\documentclass[structabstract]{aa}  

\usepackage{graphicx}
\usepackage{longtable,lscape}
\usepackage{txfonts}
\usepackage{natbib}
\bibpunct{(}{)}{;}{a}{}{,} 
\usepackage[usenames]{color}
\usepackage[utf8x]{inputenc}

\begin{document}
 
\newcommand\dotd{\hbox{$.\!\!^{\rm d}$}}
\newcommand\dotm{\hbox{$.\!\!^{\rm m}$}}
\newcommand\doth{\hbox{$.\!\!^{\rm h}$}}
\newcommand\dts{\hbox{$.\!\!^{\rm s}$}}
\newcommand\dtas{\hbox{$.\!\!^{\rm "}$}}
\newcommand \eg{{{\it e.g.},\ }}
\newcommand \etal{{\it et al.\ }}
\newcommand \etc{{\it etc.\ }}
\newcommand \cf{{\it cf.\ }}
\newcommand \ia{{{\it inter alia},\ }}
\newcommand \ie{{{\it i.e.},\ }}
\newcommand \via{{\it via\ }}
\newcommand \viz{{\it viz.\ }}
\newcommand \vs{{\it vs.\ }}
\newcommand \Teff{{$T_{\rm {ef\!f}} $}}
\newcommand \teff{{T_{\rm {ef\!f}} }}
\newcommand \Lo{{$L_\odot $}}
\newcommand \Mo{{$M_\odot $}}
\newcommand \Log{{\rm Log}\thi}
\newcommand \approxgt{\,\raise2pt \hbox{$>$}\kern-8pt\lower2.pt\hbox{$\sim$}\,}
\newcommand \approxlt{\,\raise2pt \hbox{$<$}\kern-8pt\lower2.pt\hbox{$\sim$}\,}
\newcommand \dd{{$^d$}}
\newcommand \ML{{$M$-$L$\ }}
\newcommand \aanda{A\&A}
\newcommand \RV{$V_r$}

\title{Revisiting CoRoT RR\,Lyrae stars: detection of period doubling and temporal variation of additional frequencies
 \thanks{The CoRoT space mission was developed and operated by the French space agency
CNES, with participation of ESA's RSSD and Science Programmes, Austria, Belgium, Brazil,
Germany and Spain.}' \thanks{Follow-up observations were obtained at Piszk\'es-tet\H o, the 
Mountain Station of Konkoly Observatory.}' \thanks{Table~3 is only available in electronic form at the CDS via anonymous ftp to cdsarc.u-strasbg.fr (130.79.128.5)
or via http://cdsweb.u-strasbg.fr/cgi-bin/qcat?J/A+A/}}

   \subtitle{}
\titlerunning{A review of CoRoT RR\,Lyrae observations}

   \author{R. Szab\'o \inst{1}\and
	J.~M. Benk\H o\inst{1} \and   
	M. Papar\'o\inst{1}\and	
	E. Chapellier\inst{2}\and
	E. Poretti\inst{3} \and
        A. Baglin\inst{4} \and
        W.~W. Weiss\inst{5} \and 
        K. Kolenberg\inst{6,7} \and 
        E. Guggenberger\inst{8,9} \and
        J.-F. Le Borgne\inst{10,11}
          }

   \institute{Konkoly Observatory, MTA CSFK,  
             Konkoly Thege Mikl\'os \'ut 15-17. H-1121 Budapest, Hungary\\
              \email{rszabo@konkoly.hu}\and
	     Laboratoire Lagrange, Universit\'e Nice Sophia-Antipolis, UMR 7293, Observatoire de la C\^ote d'Azur 06300, Nice, France\and
             INAF - Osservatorio Astronomico di Brera, via E. Bianchi 46, 23807 Merate (LC), Italy\and
              LESIA, Universit\'e Pierre et Marie Curie, Universit\'e Denis Diderot, Observatoire de Paris, 92195 Meudon Cedex, France\and 
	     Institute of Astronomy, University of Vienna, T\"urkenschanzstrasse 17, 1180 Vienna, Austria\and 
	     Harvard-Smithsonian Center for Astrophysics, 60 Garden Street, Cambridge MA 02138, USA\and 
Instituut voor Sterrenkunde, K.U. Leuven, Celestijnenlaan 200D, B-3001 Heverlee, Belgium\and 
Max Planck Institute for Solar System Research, Justus-von-Liebig-Weg 3, 37077 Göttingen, Germany\and
Stellar Astrophysics Centre, Department of Physics and Astronomy, Aarhus
University, Ny Munkegade 120, 8000 Aarhus C, Denmark\and
Universit\'e de Toulouse, UPS-OMP, IRAP, Toulouse, France\and
CNRS, IRAP, 14, avenue Edouard Belin, F-31400 Toulouse, France}

   \date{received; accepted}

 
  \abstract
   {High-precision, space-based photometric missions like CoRoT and $Kepler$ revealed new and surprising phenomena in classical variable stars. Such discoveries were the period doubling in RR\,Lyrae stars and the frequent occurrence of additional periodicities some of which can be explained by radial overtone modes, but others are discordant with the radial eigenfrequency spectrum.}
   {We search for signs of period doubling in CoRoT RR\,Lyrae stars. The occurrence of this dynamical effect in modulated RR\,Lyrae stars might help us to gain more information about the mysterious Blazhko effect. The temporal variability of the additional frequencies in representatives of all subtypes of RR\,Lyrae stars is also investigated.}
   {We pre-process CoRoT light curves by applying trend and jump correction and outlier removal. Standard Fourier technique is used to analyze the frequency content of our targets and follow the time dependent phenomena.}
   {The most comprehensive collection of CoRoT RR\,Lyrae stars, including new discoveries is presented and analyzed. We found alternating maxima and in some cases half-integer frequencies in four CoRoT Blazhko RR\,Lyrae stars, as clear signs of the presence of period doubling. This reinforces that period doubling is an important ingredient to understand the Blazhko effect -- a premise we derived previously from the $Kepler$ RR\,Lyrae sample. As expected, period doubling is detectable only for short time intervals in most modulated RRab stars. Our results show that the temporal variability of the additional frequencies in all RR\,Lyrae sub-types is ubiquitous. 
 The ephemeral nature and the highly variable amplitude of these variations suggest a complex underlying dynamics of and an intricate interplay between radial and possibly nonradial modes in RR\,Lyrae stars. The omnipresence of additional modes in all types of RR\,Lyrae -- except in non-modulated RRab stars -- implies that asteroseismology of these objects should be feasible in the near future.}
   {}


   \maketitle
%

\section{Introduction}

\begin{table*}
\caption{Basic parameters of the CoRoT RR Lyrae stars. The columns contain the CoRoT ID number, the coordinates, {\it V} magnitude, the CoRoT run, RR\, Lyrae sub-type, the amplitude of the dominant mode(s), 
the Blazhko period in case of modulated stars, the pulsation period and references.}
\begin{tabular}{ccccclcccc}
\hline\hline
CoRoT ID & R.A. (2000) & Dec. (2000) & {\it V} [mag]& Run & Type & {${\rm f_{0,1}}$ ampl. [mag]} & ${\rm P}_{\rm Bl}$ [days] & {puls.\ \ period} [days] & Ref.\\
\hline
0101370{\bf 131} & 19 28 14.40 & +0 06 02.27 &15.60 & LRc01 & RRab     &  $0.257$                & -         &	0.619332  & (1)\\
0101315{\bf 488} & 19 27 47.40 & +0 58 35.62 &16.15 & LRc01 & RRab     &  $0.029^{b}$  & -         & 0.4853033 & this work\\
0103800{\bf 818} & 18 31 23.80 & +9 10 10.45 &13.70 & LRc04 & RRab     & $0.345^{c}$	  & -         & 0.4659348 & this work\\
0104315{\bf 804} & 18 34 28.92 & +8 57 00.43 &15.94 & LRc04 & RRab     &$0.130$	   & -         & 0.7218221 & this work\\
0100689{\bf 962} & 19 24 00.10 & +1 41 48.70 &14.96 & LRc01 & RRab Bl  & 0.201        & 26.88 & 0.3559966 & (2) \\		    
0101128{\bf 793} & 19 26 37.32 & +1 13 34.90 &15.93 & LRc01 & RRab Bl  & 0.235        & 17.86 & 0.4719296 & (3) \\		    
0100881{\bf 648} & 19 25 05.43 & +1 39 23.83 &16.16 & LRc01 & RRab Bl  & $0.033^{b}$  & 59.77 & 0.607186  & this work\\  	    
0101503{\bf 544} & 19 29 10.12 & +0 43 47.14 &14.79 & LRc01 & RRab Bl  & $0.009^{b,c}$& 25.60 & 0.605087  & this work\\  	    
0105288{\bf 363} & 18 39 30.86 & +7 26 53.95 &15.32 & LRc02 & RRab Bl  &$0.179$	   & 35.6  &	0.5674412 & (4,5,6)\\	     
0103922{\bf 434} & 18 32 08.55 & +8 32 40.78 &15.84 & LRc04 & RRab Bl  &$0.280$	   & 54.5  & 0.5413828 & (7) \\
0105036{\bf 241} & 18 38 09.60 & +7 43 56.68 &15.58 & LRc02 & RRc      &$0.196$	   & -         & 0.372921 & this work\\
0105735{\bf 652} & 18 42 10.13 & +6 33 05.15 &15.01 & LRc02 & RRc      &$0.204$	   & -         &	0.2791596 & this work\\
0101368{\bf 812} & 19 28 13.61 & +0 40 42.46 &15.86 & LRc01 & RRd ${f_0}$ &$0.053$ & -         & 0.4880408 & (8)\\
                 &             &             &      &       & RRd ${f_1}$ &$0.143$ & -         & 0.3636016 & \\
\hline
\end{tabular}
\tablefoot{ The superscript 'b' denotes blended pulsators. Superscript 'c' denotes a 
CoRoT target with colors. {\it V} magnitude values were taken from the ExoDat catalog. 
No good candidates classified as RR\,Lyrae stars were found in the third runs.
References: {{\bf(1)}:  \cite{paparo2009}, \bf (2)}: \cite{chadid2010}, {\bf(3)}: \cite{poretti2010}, {\bf(4)}: \cite{guggenberger2011}, {\bf(5)}: \cite{chadid2011}, 
{\bf(6)}: \cite{guggenberger2012}, {\bf(7)}: Poretti et al., in prep., 
{\bf(8)}: \cite{chadid2012}}
\label{tab1}
\end{table*}

The advent of space photometry has opened up new vistas in investigating stellar pulsations 
and oscillations. Pulsating variable stars located in the classical instability strip are no 
exception. RR Lyrae stars in particular have benefited from the continuous and ultra-precise 
space photometric data delivered by MOST \citep{walker2003}, CoRoT \citep{baglin2006} and 
{\it Kepler} \citep{borucki2010}.

One of the surprising findings was the period doubling phenomenon (hereafter PD) 
in the  ultra-precise $Kepler$ RR Lyrae light curves \citep{kolenberg2010, szabo2010}, which manifests as the alternating height of maximum brightness and alternating light curve shape from cycle to cycle in the photometric light curve and half-integer frequencies (HIFs) between the dominant pulsation mode and its harmonics in the frequency domain. So far only the $Kepler$ stars were investigated to describe the phenomenon, such a detailed analysis is still missing for CoRoT RR\,Lyrae stars. This prompted us to reexamine the published CoRoT RR Lyrae light curves and extend the study by adding new ones, as well. 

PD is found only in case of modulated (Blazhko) RR Lyrae stars, though a thorough search was done on non-modulated $Kepler$ RRab stars, too \citep{szabo2010, nemec2011}. Therefore, it is plausible to assume that either PD as a nonlinear dynamical phenomenon plays an important role in causing the Blazhko-modulation (see \citealt{smolec2012} for a hydrodynamic example of PD causing modulation in BL Herculis stars) or at least it shows up as a frequent companion effect of the modulation. In either case, it is important to find well-documented cases, establish the frequency of the PD occurrence, investigate its temporal behavior and other characteristics. In addition, the physical explanation of PD itself was unambiguously traced back to a 9:2 resonance between the fundamental pulsational mode and a high-order radial (strange) overtone \citep{kollath2011} which opens a new way to study the dynamics of these  high-amplitude variable stars belonging to the horizontal branch. 

In recent years a new picture has started to emerge regarding the frequency spectrum of RR\,Lyrae stars. Thanks to dedicated telescopes \citep{jurcsik2009a} and space-based photometry, additional frequencies were found in many stars that were observed with at least millimagnitude precision: AQ Leo (RRd) \citep{gruberbauer2007}, MW Lyr \citep{jurcsik2008}, CoRoT \citep{poretti2010,chadid2010} and $Kepler$ stars \citep{benko2010}.

These additional periodicities do not fit the series of radial pulsational modes (nor the occasional half-integer series due to PD) and present a low-amplitude variability in each case. Presently, the best explanation for their presence is the excitation of nonradial modes. If this turns out to be true, we may have another handle on the interior of horizontal branch stars to understand better the stability, pulsation and evolution of these standard candles. In some cases we unexpectedly found frequencies with low-amplitudes at or near the radial overtones (e.g. \citealt{benko2010}) that might correspond to radial modes excited by resonances or nonradial modes with frequencies being in 1:1 resonance with the corresponding radial mode \citep{dziembowski2004, vanhoolst1998}. The era of exploiting the power of nonlinear seismology using radial modes in RR\,Lyrae stars is imminent \citep{molnar2012}.

In this work we embarked on investigating all the known CoRoT Blazhko RR\,Lyrae stars up to LRc04 in order to detect period-doubling. The detection or non-detection would help to find out how frequent this phenomenon is, and further strengthen its role in and connection to the mysterious Blazhko-effect. Our second aim is to investigate the recently found new periodicities in all types of RR\,Lyrae stars. This kind of research is only possible with the photometric precision of CoRoT for these stars. We are especially interested in the temporal stability of these additional frequencies. 

The structure of this paper is the following: in Sec.~\ref{obs} we introduce the original space-borne and ground-based follow-up data we use in this analysis. In Sec.~\ref{methods} we discuss the methods to find period doubling and to detect the variations of low amplitude periodicities. In Sec.~\ref{results} we present our results, then in Sec.~\ref{disc}  we discuss them and draw our conclusions followed by a short summary in Sec.~\ref{sum}. Hitherto unpublished frequency tables and other complementary results can be found in the Appendices. 

Throughout this work we use the following notations: by ${f_0}$, ${f_1}$, ${f_2}$, ... we denote the frequency of the radial pulsation modes (fundamental, first overtone, second overtone, etc.), while ${f'}$, ${f''}$ stand for additional, {\it independent} frequencies. We refer to the CoRoT stars with the last three digits of their ID number for short notation, which provides a unique identification, \eg {\tt 962} for CoRoT\,0100689{\bf 962}.

\section{Observations}\label{obs}

\begin{center}
\begin{table}
\begin{flushleft}
\caption[]{Observations of the CoRoT LRc01 RR\,Lyrae stars at the 
Konkoly Observatory that were used in this paper. 
The CoRoT ID, the night of the observations, total number of
the scientific frames (N$_{\mathrm f}$), the ID of the
primary comparison stars from the USNO-A2.0 are given.
}\label{table_obslog}
\begin{tabular}{rcrc}
\hline \hline
\noalign{\smallskip}
CoRoT~ID        & DATE & N$_{\mathrm f}$ &
Comp. \\
\noalign{\smallskip}
\hline
\noalign{\smallskip}
100881648       &  2008-07-28/29  &  55  &  0900-14969350 \\
                &  2008-07-29/30  &  30  &   \\
101503544       &  2008-06-20/21  &  49  &  0900-15291694  \\
                &  2008-06-21/22  &  68  &  \\
100689962       &  2008-06-09/10  &  57  &  0900-14903871 \\
                &  2008-06-19/20  &  58 & \\
                &  2008-07-08/09  &  38 & \\
                &  2008-07-11/12  &  57 & \\
101370131       &  2008-06-22/23  &  110 & 0900-15209129 \\
                &  2008-06-25/26  &  33 & \\
                &  2008-07-10/11  &  85 & \\
                &  2008-07-11/12  &  36 & \\
                &  2008-07-12/13  &  38 & \\ 
101128793       &  2008-07-27/28  &  63 &  0900-15077950 \\
                &  2009-06-29/30  &  47 &      \\            
\noalign{\smallskip}
\hline
\end{tabular}
\end{flushleft}
\end{table}
\end{center}

\subsection{CoRoT observations}
CoRoT has conducted long-duration, continuous, very-high-precision relative photometry on a few very bright stars, and a large number of faint ones \citep{baglin2006}. 
As an attempt to review the CoRoT RR\,Lyrae stars we investigate 13 stars listed in Table\,\ref{tab1}. Table\,\ref{tab1} gives the CoRoT ID, the coordinates, the {\it V} magnitude, the CoRoT run in which the object was observed and discovered, its pulsation and modulation parameters (if applicable), as well as references if the star was already analyzed. 

The brightness of the targets that were observed in the exo-field of CoRoT are measured through a prism which creates images that contain red, green and blue fluxes. 
The color fluxes are not available for all stars, in some cases only the sum of them (white light) is retained. Though these fluxes are not related to any photometric systems, they are still useful for example to compare the amplitudes of variability in some of our cases. The interested reader can find an example for the use of CoRoT color fluxes in \citet{paparo2011}. 

The CoRoT RR Lyrae group analyzed those stars that were classified as such by the 
CoRoT Variability Classifier (CVC) \citep{debosscher2009} up to the fourth pointing (LRc04). We note that one of these stars, CoRoT\,1027817{\bf 750}, observed during LRa01, originally classified as an RR\,Lyrae star with a 67\% probability is found to be non-RR\,Lyrae by \citet{paparo2011}. We took those targets that were classified as RRab pulsators with non-zero probability and found eight bona fide RRab stars out of 14. In addition to that sample we found CoRoT RR\,Lyrae star {\tt 648} early in the mission. 
Furthermore we added one more object to the classified RR Lyrae stars
which was found by \citet{affer2012} while investigating stellar rotation periods in CoRoT light curves in runs LRc01 and LRa01. Upon inspecting the published light curves manually, we confirmed that {\tt 488} is a new RRab variable. Light curve characteristics, frequencies, epoch and more details on this target can be found in Appendix~\ref{app488}. 

RRc classification is more problematic, as eclipsing binaries (\eg W\,UMa stars) can exhibit similar light curve shapes, thus the contamination of a sample originating from  automated classification is much higher. Therefore we chose the ${\rm >80\%}$ probability level assigned by CVC and found only two genuine RRc stars out of 17 candidates. We note that decreasing the probability limit to ${\rm 50\%}$ or ${\rm 10\%}$ did not help to find more RRc stars, but increased the number of eclipsing binaries and other types enormously. 

Interestingly, we have RR Lyrae stars only in the direction of the Galactic Center. This obviously might be a selection effect due to the fact that that we observe mostly disc-population stars when the satellite points toward the anticenter direction. The length of observations varies from 158 days (LRc01) and 150 days (LRc02) to 88 days (LRc04), in all cases the duty cycle being over 90\%. Time is given in Heliocentric Julian Date throughout the paper, however when showing light curves we use CoRoT Julian Date (CJD). The relation between the Heliocentric Julian Date and the CJD is 
\begin{equation}
${\rm HJD = CJD} + 2451545.0$
\end{equation}

We used the N2 level calibrated CoRoT light curves. We applied trend and jump filtering and outlier removal as described in detail in \citet{chadid2010}. As we usually deal with relatively faint targets, only two RR Lyrae in our sample have  CoRoT colors ({\tt 544} and {\tt 818}), but we use only the integrated (white) light to increase the signal-to-noise ratio. The nominal sampling is 512 sec for all the stars, except the above mentioned two targets, where the initial part of the light curves were observed with the 512 sec mode while the rest of the light curve was observed with the much denser short cadence mode (32 sec).

\subsection{Ground-based follow-up observations}

Ground-based multi-color observations on some of the CoRoT RR\,Lyrae 
stars discussed in this paper were collected in 2008 and 2009 with 
the 1-m Ritchey-Chr\'etien-Coud\'e (RCC) telescope mounted at Piszk\'es-tet\H{o} Mountain Station
of the Konkoly Observatory. These observations were especially helpful 
in determining the true pulsational nature in the cases where our targets 
are heavily blended in the CoRoT apertures.

A Versarray 1300B camera with an UV-enhancement coating constructed by Princeton Instruments was used. This device contains a back illuminated EEV CCD36-40 1340$\times$1300 chip that corresponds to a $6\farcm{6}\times 6\farcm{8}$ field of view (FOV) with 0.303$\arcsec$/pixel scale. Standard Johnson {\it BV}\/ and Kron-Cousins
{\it R$_{\mathrm C}$}\/ filters were used in the observations. The typical exposure times were 300, 180, 100~s for bands {\it B}, {\it V}\/ and {\it R$_{\mathrm C}$}, respectively. Each night dome flats, bias and dark frames were taken as main calibration images.

We used the {\sc iraf/ccdred}\footnote{IRAF is distributed by the National Optical Astronomy Observatory, which is operated by the Association of Universities for Research in Astronomy (AURA) under cooperative agreement with the National Science Foundation.}
package for the standard reduction procedures: bias, dark and flat field correction.
The brightness of stars was determined by using the aperture photometry task {\sc daophot/phot} of {\sc iraf}. We carried out differential photometry with carefully selected comparison stars. These were found to be constant and fit the target stars both in their brightness and colours. The telescope constants were obtained using the standard stars of the open cluster M67 \citep{chevalier1991}. The typical errors of the individual observations were between 0.01–-0.02\,mag in {\it BVR$_{\mathrm C}$}. The log of the observations is presented in Table~\ref{table_obslog}, while the ground-based {\it BVR$_{\mathrm C}$} light curves are available in electronic form via the website of the journal. A sample of this file is shown in Table~\ref{table_online1} for guidance.

\begin{center}
\begin{table}
\caption[]{Ground-based multi-color observations of CoRoT RR\,Lyrae stars 
taken with the 1m RCC telescope of the Piszk\'es-tet\H o Mountain Station of the 
Konkoly Observatory. This table is published in its entirety in the electronic edition of the journal. A portion is shown here for guidance regarding its form and content.}\label{table_online1}
\begin{tabular}{cccc}
\hline\hline
\noalign{\smallskip}
CoRoT~ID        & HJD & magnitude &  filter \\
\noalign{\smallskip}
\hline
\noalign{\smallskip}
100881648   &    2454676.34320 &  16.903    &     {\it B} \\
100881648   &    2454676.35783 &  16.919    &     {\it B} \\
100881648   &    2454676.36861 &  16.913    &     {\it B} \\
100881648   &    2454676.37939 &  16.933    &     {\it B} \\
100881648   &    2454676.39017 &  16.913    &     {\it B} \\
... & ... & ... & ... \\
\noalign{\smallskip}
\hline
\end{tabular}
\end{table}
\end{center}
 
\begin{figure*}                      
\includegraphics[height=9.0cm,angle=270]{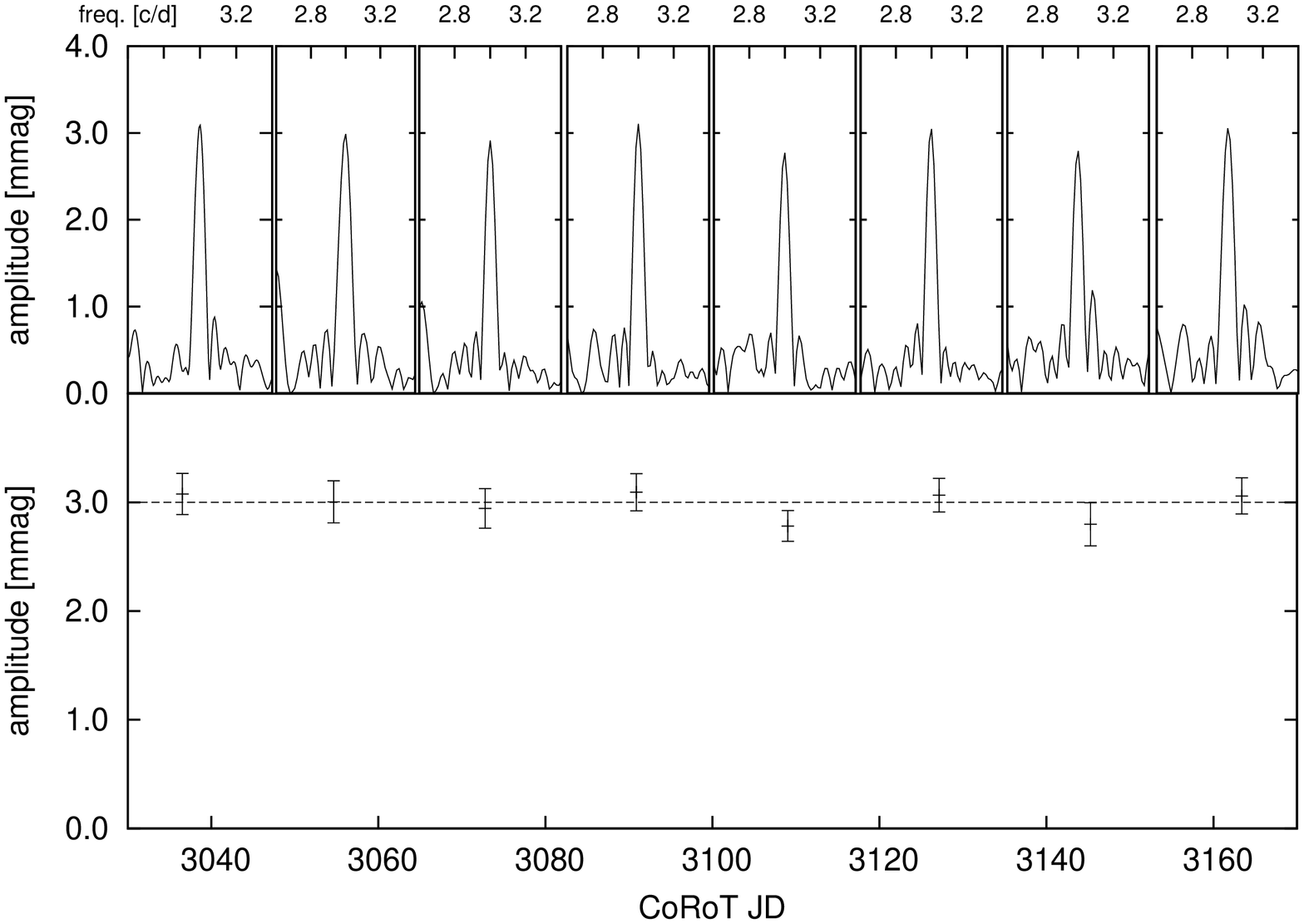}
\includegraphics[height=9.0cm,angle=270]{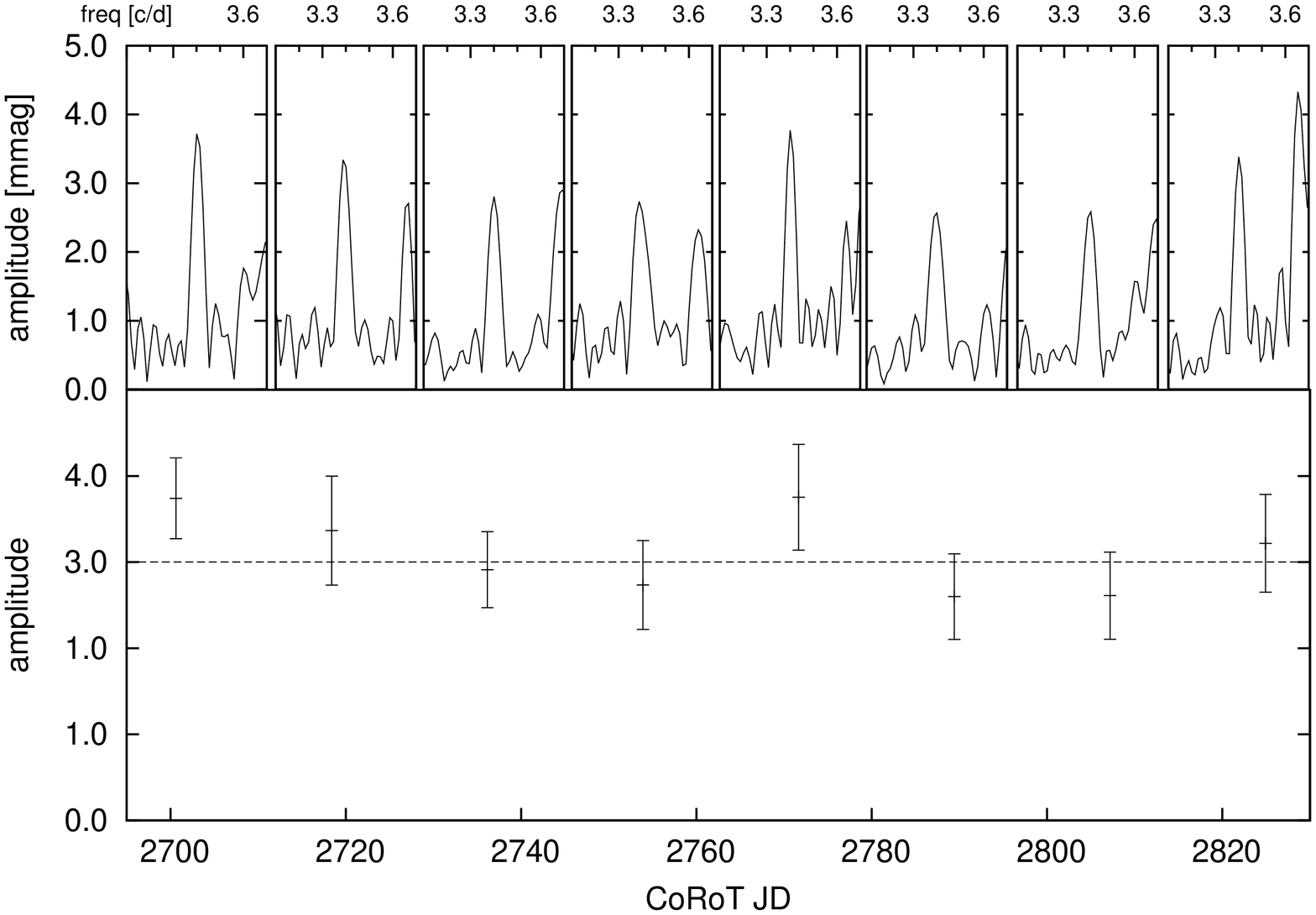}
\caption{Test results for the detection of the temporal variation of low amplitude frequencies.   
{\bf Left panel}: The  amplitude of the constant, 3 mmag amplitude frequency injected at 
${f_{\rm test}=3.0\,d^{-1}}$ into the light curve of {\tt 241}, a CoRoT RRc 
star (see Sec.~\ref{rrc}). 
{\bf Right panel}: A more complex case: we added a constant (3 mmag) amplitude frequency to the light curve of the strongly modulated Blazhko RR\,Lyrae, {\tt 962}. 
We chose ${f_{\rm test} ={\rm 3.4\,d^{-1}}}$. Horizontal lines show the amplitude of the injected signal, points with error bars come from our method. The upper row shows the shape and the vicinity of the recovered frequency in the applied eight bins.}
\label{test}
\end{figure*}

\section{Methodology}\label{methods}

We applied standard packages such as {\sc MuFrAn} \citep{kollath1990} and {\sc Period04} \citep{lenz2005} to perform Fourier-analysis. 

\subsection{Detection of period doubling}\label{dpd}
 
There is no obvious, strongly alternating pattern in the CoRoT RR\,Lyrae maxima 
that could have triggered a thorough analysis earlier, though in some cases a 
detailed inspection do show period doubling persisting through a few pulsational 
cycles as we show later in this work. 

CoRoT light curves are shorter than the {\it Kepler} ones and are of lower precision at 
the same apparent brightness. The typical error of the individual data points for a 
${\rm 16^{th}}$ magnitude RR\,Lyrae is 0.006 -- 0.012\,mag for the CoRoT observations, and 0.0008 -- 0.003\,mag for an RR\,Lyrae of the same brightness observed by {\it Kepler}. In addition, since period doubling is usually only temporarily noticeable in the light curves (but see \citealt{leborgne2014}), we do not expect to see PD signs in both the light curve and the frequency spectrum in each cases. The strength of PD is not correlated with the Blazhko phase, as we saw earlier on the example of RR\,Lyrae itself and other modulated RRab stars. Therefore here we need to relax our strict criteria applied in \citet{szabo2010}, where we required the presence of PD signs in both the time and the frequency domain. Consequently, in this work we investigate both the Fourier-spectrum between the dominant mode and its harmonics aiming at finding HIFs at exactly between two consecutive harmonics and the light curves themselves looking for suspicious alternating pattern. We report tentative detection in those cases where PD is present in either or both domains. 

For the detection of PD in the Fourier spectrum we set a conservative limit, \ie 3\,$\sigma$ detection for the HIFs.  The rms for half-integer frequencies ($(2{\it  k}+1)/2 \cdot f_0$) was computed in the intervals between consecutive harmonics of the dominant pulsational mode [$k\cdot f_0; (k+1)\cdot f_0$] after pre-whitening with significant additional frequencies and the Blazhko side peaks. The [$f_0$;$2f_0$] frequency interval is the most interesting one, because all the known stars undergoing the period doubling bifurcation found in the $Kepler$ sample show the highest half-integer frequency amplitude in this particular interval \citep{szabo2010}. 

In order to investigate the alternating nature of the pulsational cycles, the photometric maxima were fitted with 7th or 9th order polynomials \citep{chadid2010} to get rid of the detrimental effects of missing photometric points or (in the case of blended Blazhko stars) large scatter. For more details of this process we refer to \cite{chadid2010}.

\begin{figure*}                      
\includegraphics[height=6.2cm,angle=270]{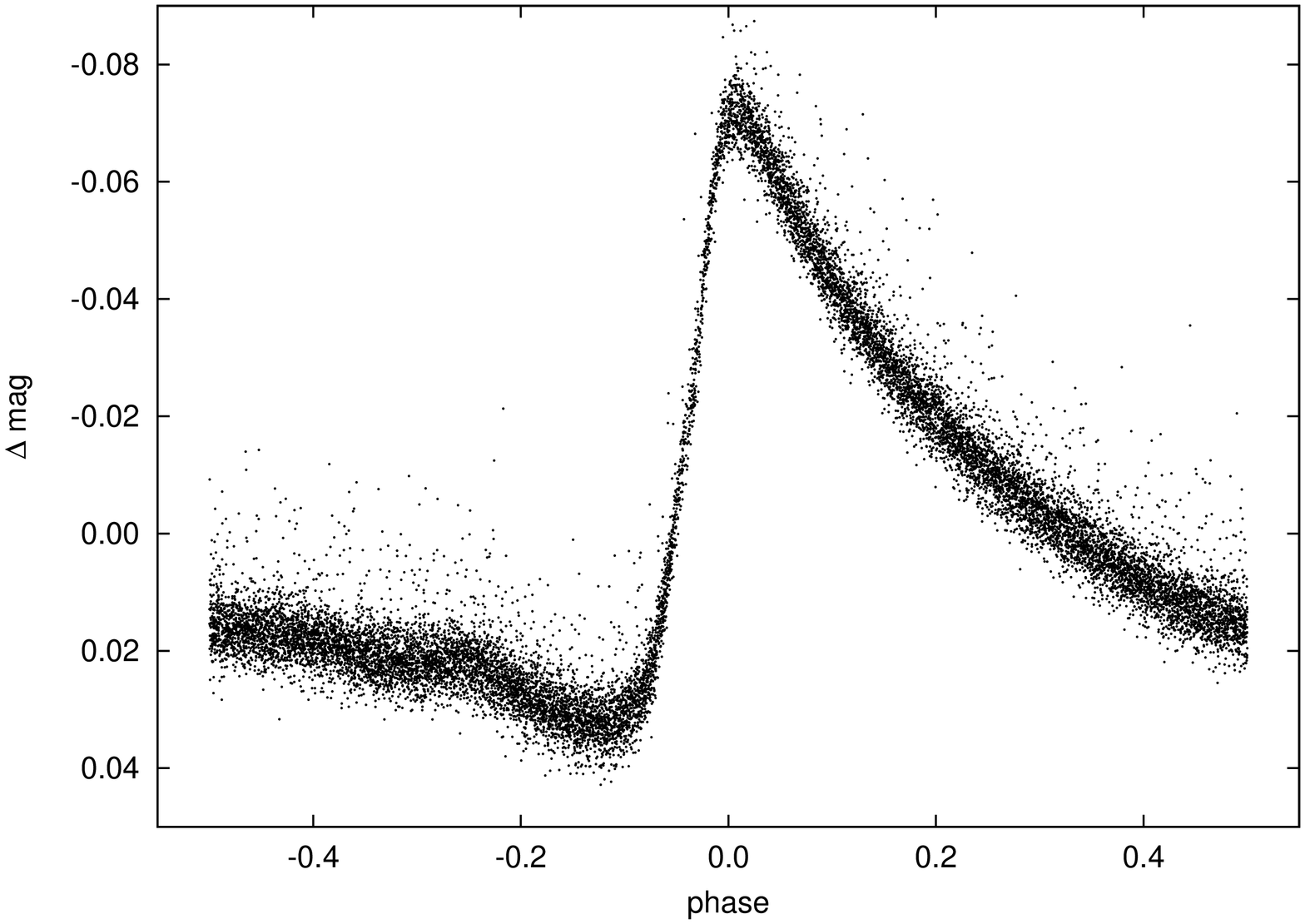}
\includegraphics[height=6.2cm,angle=270]{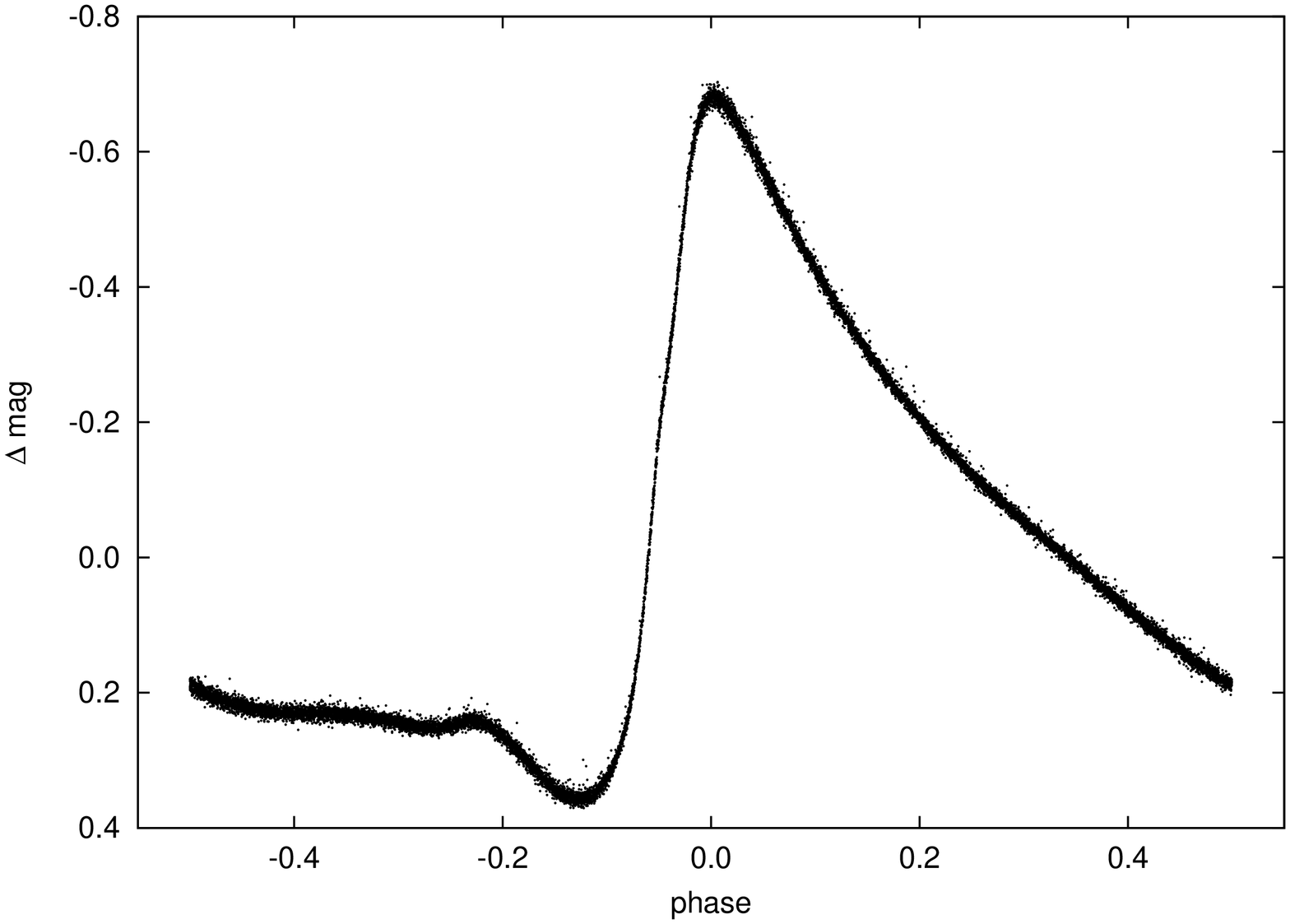}
\includegraphics[height=6.2cm,angle=270]{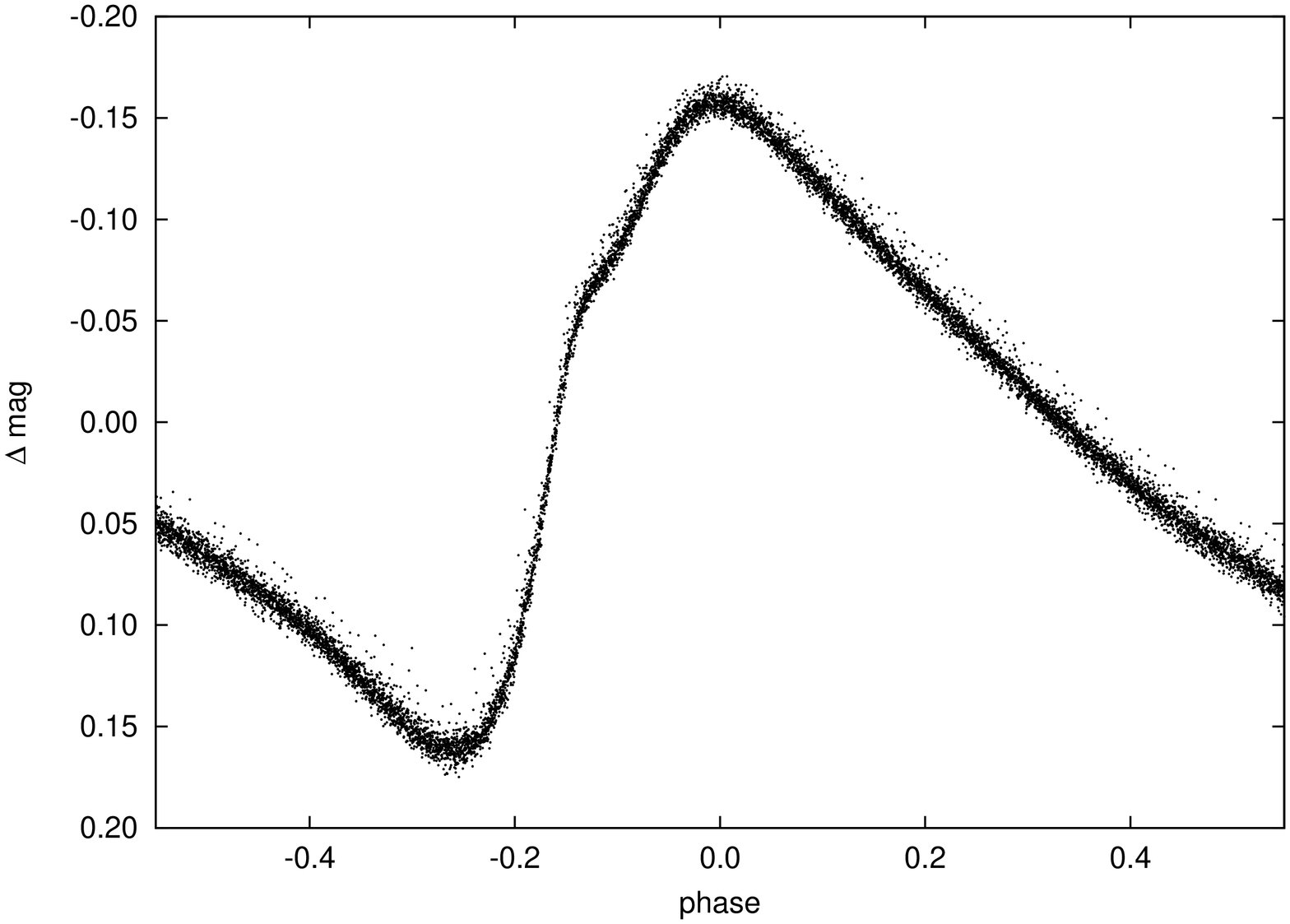}
\caption{Jump- and trend filtered phased light curves of CoRoT non-Blazhko RRab stars. 
{\bf Left panel}: The newly found RR Lyrae, {\tt 488}. Note the small amplitude 
and the relatively large scatter, indicating a blended CoRoT target. 
{\bf Middle panel:} The new, non-modulated RR Lyrae, {\tt 818}. Having over 177,000 data points, we plot only every $10^{\rm th}$ data points for visualization purposes. 
{\bf Right panel:} Light curve of the new, non-modulated RR Lyrae, {\tt 804}.}
\label{488}
\end{figure*}

\subsection{Frequency analysis and time-dependent frequencies}

To get the time-dependence of the properties of a frequency peak we cut the photometric data into equal-length chunks, and the respective amplitudes and frequencies and their uncertainties were obtained from these shorter data sections  with {\sc Period04}. We tried to cut the light curve into two-four-eight etc. pieces,  and stopped where the uncertainties (primarily due to the shortening of the data sections) precludes the derivation of any meaningful information. In most of the cases, for uniformity, eight bins were used. This is a compromise between time resolution of the variabilities and the frequency resolution permitted by the lengths of the data chunks, which could not be deteriorated arbitrarily, because often close-by frequency peaks occur. We performed simulations to make sure that any arbitrary time shift (\ie the exact starting epoch of the subsections) does not have significant effect on the amplitudes of the studied frequencies.

In all cases the main pulsation mode(s), their harmonics, the Blazhko frequency, its harmonics and the modulation side peaks (where applicable) were subtracted before we embarked on computing the time variable amplitude of the additional peaks. Error bars were derived by a Monte Carlo method available in {\sc Period04}.

In order to check whether the successive pre-whitening, the strong Blazhko modulation and other time-dependent features have an appreciable effect on our method used to detect temporal variation of low amplitude frequency peaks, we designed a series of tests. First we added a 3.0 mmag, constant amplitude sinusoidal signal with a frequency of ${\rm 3.0\,d^{-1}}$ to the light curve of {\tt 241}, an RRc star (see Sec.~\ref{rrc}). The vicinity of ${\rm 3.0\,d^{-1}}$ in the Fourier spectrum is relatively clean, \ie free of contaminating frequencies. This light curve was subjected to the same 
procedure as all the CoRoT targets, namely the light curve was cut into eight bins, and the main pulsation frequency and its harmonics were subtracted in each bin. As we see in the left panel of Fig.~\ref{test}, the injected constant amplitude signal was preserved, the error introduced by the process is a few percent, and does not exceed 10\% in any case. This means that in a sparse frequency spectrum we can confidently detect the time-dependence of low-amplitude signals  without introducing amplitude variability with our methods.

The right panel of Fig.~\ref{test} shows a more complex case. This time we chose {\tt 962}, a strongly modulated Blazhko star. The frequency spectrum of this star contains more frequencies \citep{chadid2010} making the analysis more complicated. Now we squeezed an ${f_{\rm test}}={\rm 3.4\,d^{-1}}$ signal with 3-mmag constant amplitude in between other frequencies. As suspected, in this case we got a higher scatter around the constant amplitude. Ten percent deviation from the nominal value is not unusual, and even larger excursions occur in some cases. The computed error bars more or less capture this scatter, the deviation rarely exceeds the assigned sigma. Evidently, in case of a dense frequency spectrum with evidence of strong temporal variations (such as a Blazhko-modulation and period-doubling) one has to exercise extra care when interpreting amplitude variability of low-amplitude signals. It is reassuring, however, that the structure of the frequency peaks does not deform due to the procedure, as long as there are no close immediate frequency peaks in the vicinity of our test signal.

\begin{figure}                      
\includegraphics[width=4.4cm]{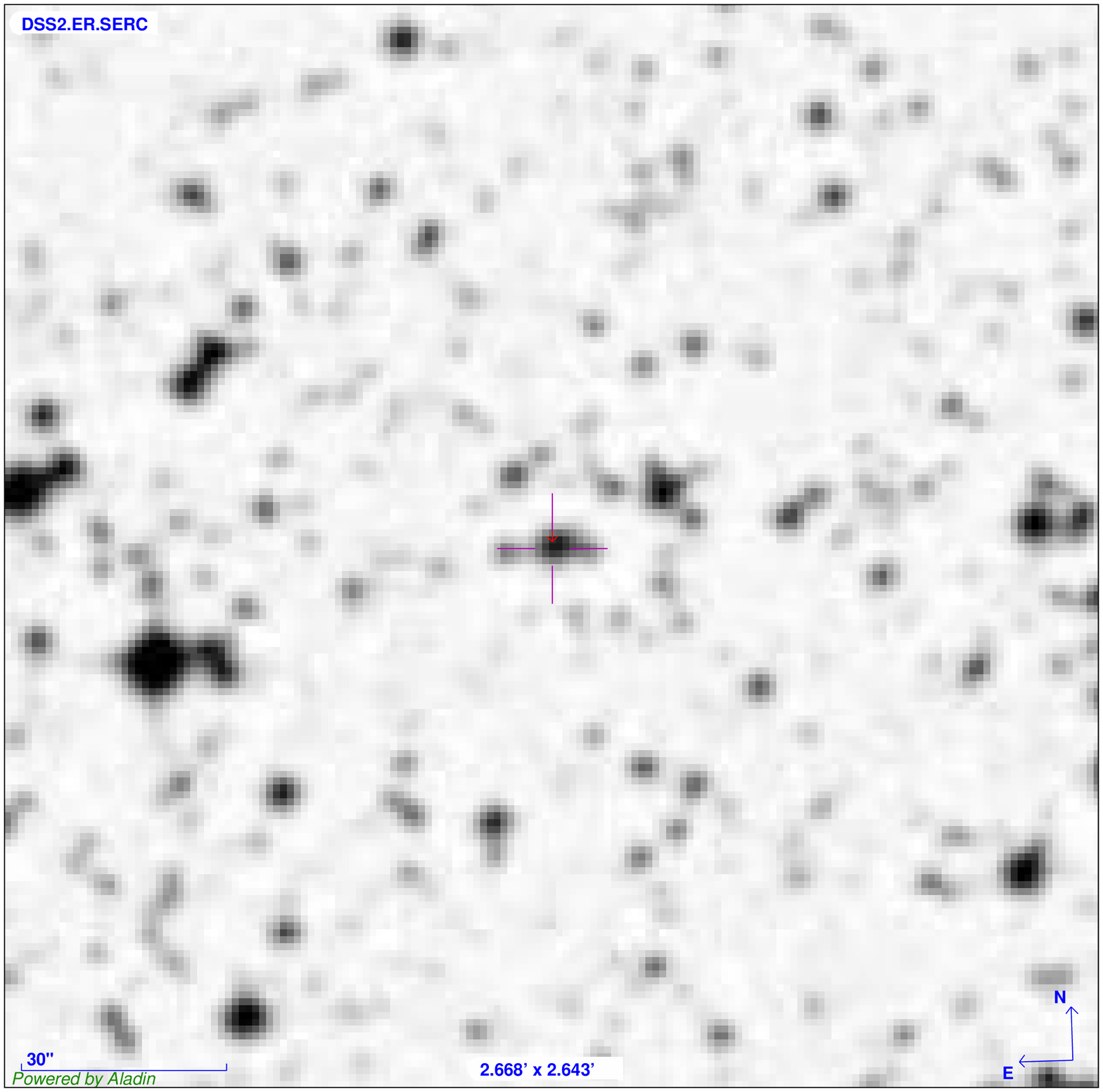}\includegraphics[width=4.5cm]{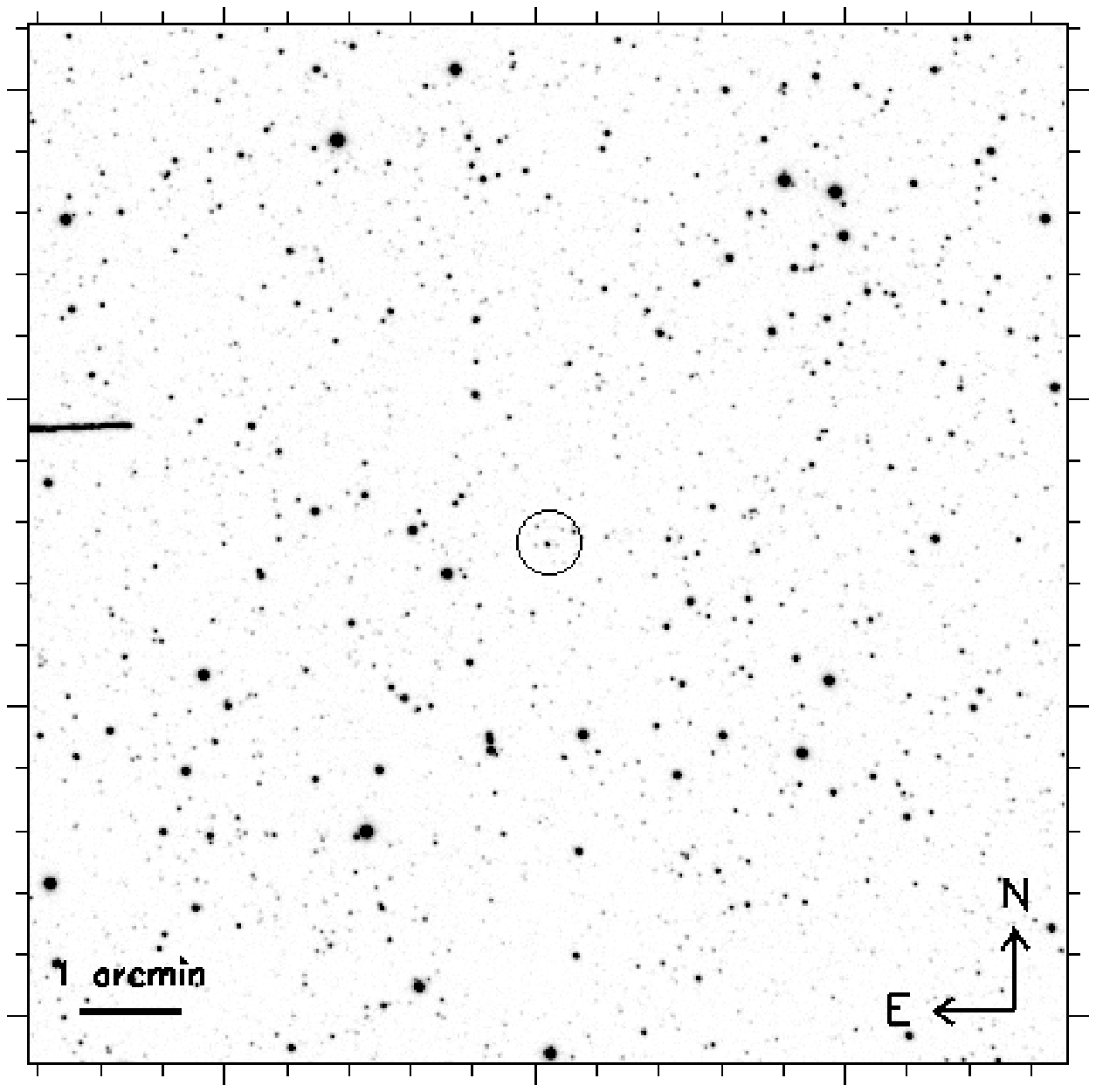}
\caption{{\bf Left:} 2.6' x 2.6' DSS image showing the neighborhood of {\tt 488}.
{\bf Right:} A larger surroundings of the star {\tt 488} showing the applied aperture  as well. Several faint stars are shown within the aperture. North is up and East is to the left in both images.}
\label{488map}
\end{figure}

\section{Results}\label{results}

In the next subsections we discuss our targets one by one focusing on those stars and features that have not been discussed earlier. In case of previously published results or those in preparation we refer to the corresponding paper(s). However, for the sake of completeness we decided to list all CoRoT objects reliably classified as RR\,Lyrae stars up to LRc04. First we describe non-modulated RRab stars followed by Blazhko stars, then RRc stars are listed, finally the only double-mode (RRd) star is discussed. 

\subsection{Non-Blazhko RRab stars}\label{nonBL}

\noindent CoRoT\,0101370{\bf 131}: This mono-periodic RRab star was analyzed in detail in \citet{paparo2009}. In the course of this work we checked again that no additional  frequencies, including signs of HIFs can be found in the frequency spectrum. We give an upper limit of 0.30\,mmag for the latter (Table~\ref{tab4}). We also checked whether the maxima show any periodicity or pattern,  as a sign of a so far hidden, low-amplitude Blazhko-effect, but failed to find any significant periodicities. 
\vskip 5mm

\noindent CoRoT\,0101315{\bf 488}: \citet{affer2012} found 169 pulsating stars while investigating stellar rotation periods and ages in CoRoT light curves of spotted stars in runs LRc01 and LRa01. Upon inspecting the published light curves manually we encountered a new RR Lyrae like object: {\tt 488}, besides finding other, already known RR\,Lyrae variables. 

As this is a new finding, we discuss the properties of this object in detail. 
The coordinates of the new variable are RA= $19^{\rm h}$ $27^{\rm m}$ $47\dts40$, 
Dec = 0\degr \thinspace 58'\thinspace $35\dtas652$. For the brightness of the main target the ExoDat database gives the following values: $B$=17.163, $V$=16.152, $R$=15.740, $I$=15.072. Based on the contamination index (0.222), blending can be considered to be severe.

The light curve shape is of a typical RRab, but the total amplitude is rather small ($0\dotm1$). Hence, we suspect that this variable star is indeed blended. The relatively high scatter also corroborates this argument (see Fig.~\ref{488}, left panel). The pulsation period is constant during the CoRoT run. 

\begin{figure*}            
\includegraphics[height=18.5cm,angle=270]{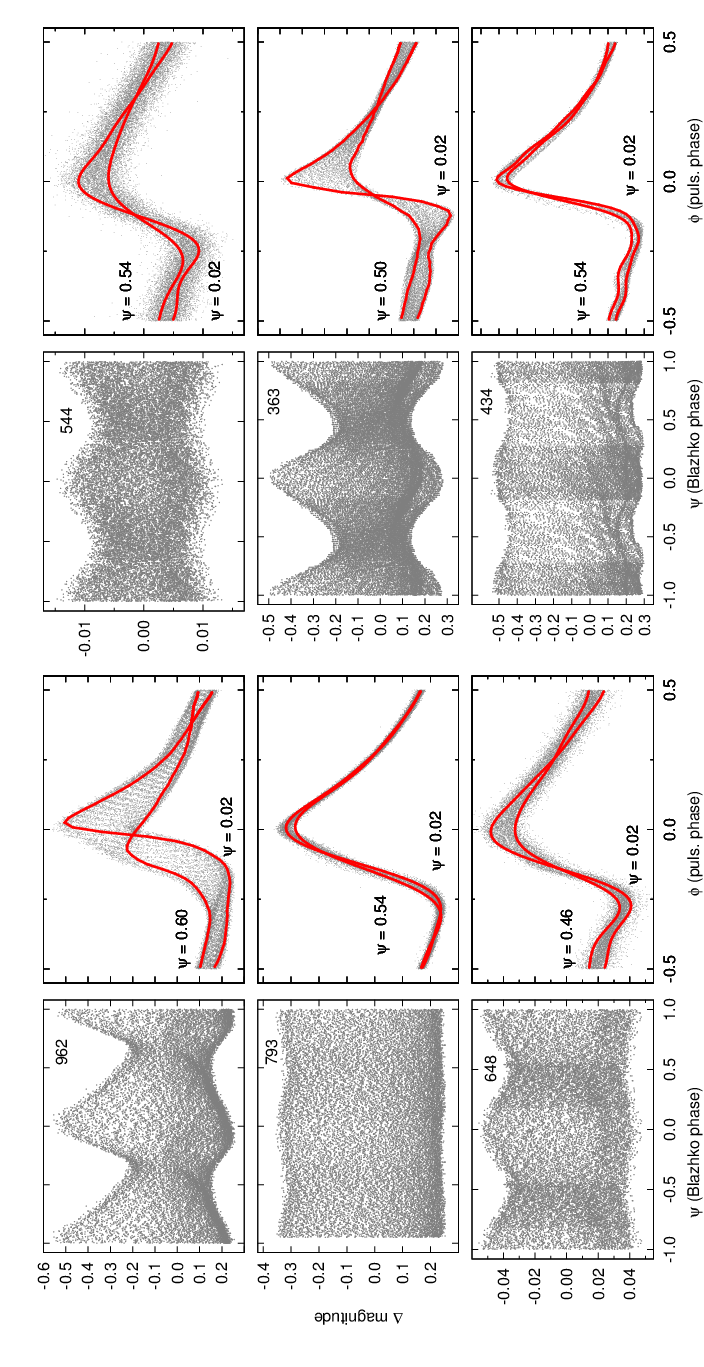} 
\caption{Phase-folded CoRoT Blazhko light curves. {\bf Left}: folded by the Blazhko-period, {\bf right}: folded by the pulsational period. Representative averaged light curves are shown in two extreme modulation phases in each case in the right panels.}
\label{962_363}
\end{figure*}

Fig.~\ref{488map} shows the DSS and the EXODAT images of our target. EXODAT lists six close-by stars of magnitude 17.5 -- 20.5, within the CoRoT aperture. It is conceivable that one of these might be the RR Lyrae, but based on the available data we cannot tag the exact source of variation. The pulsation parameters, frequency table, epoch and discussion of the frequency spectrum can be found in Appendix~\ref{app488}. We note here only that no PD was found in the spectrum with an upper limit of 0.10\,mmag for any half-integer frequencies.

\vskip 5mm

\noindent CoRoT\,0103800{\bf 818}: This star was observed by CoRoT in LRc04, and it is described here for the first time. It started to be observed with nominal cadence (512\,s) but soon after it was switched to short cadence (32\,s) observations in color mode. To enhance the signal-to-noise here we use the integrated (white) light curve only. With a period of 0.4659348 days it is a non-modulated RRab star. We did not detect period doubling with an the upper limit of 0.10\,mmag set for the half-integer frequencies.  The phase-folded light curve is shown in the middle panel of Fig.~\ref{488}. 

Due to the early switch to high cadence observations, fifty-six harmonics has been found in the frequency spectrum. So far this is the largest number we are aware of. The amplitude distribution is very similar to {\tt 131} \citep{paparo2009}, \ie there is a local minimum in the amplitudes of the harmonics around the ${\rm 15^{th}}$ harmonics (${\rm 32\,d^{-1}}$). The harmonics in the higher frequency range come with a different slope after this trough in the distribution. While we can not confirm this behavior with the other two non-modulated CoRoT RRab stars ({\tt 488} and {\tt 804}) because of the lower cadence observations, it would be interesting to see the amplitude distribution in the case of other RRab stars observed from space with high cadence. 

Apart from that, only the CoRoT orbital frequencies and some residuals around the main frequency peaks are visible in the frequency spectrum. We could identify neither regular frequency side peaks that would signal the presence of the Blazhko effect, nor half-integer or any other frequencies. The frequency table is given in Appendix~\ref{app818}.

\begin{figure*}                        
\resizebox{\hsize}{!}{\includegraphics[angle=270]{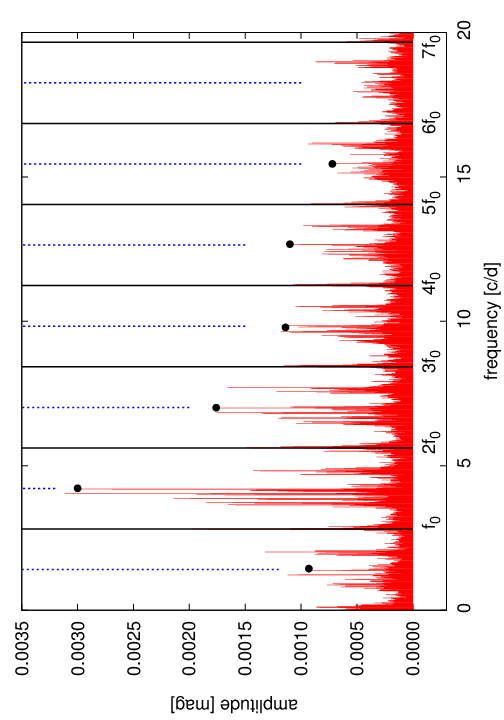}\includegraphics[angle=270]{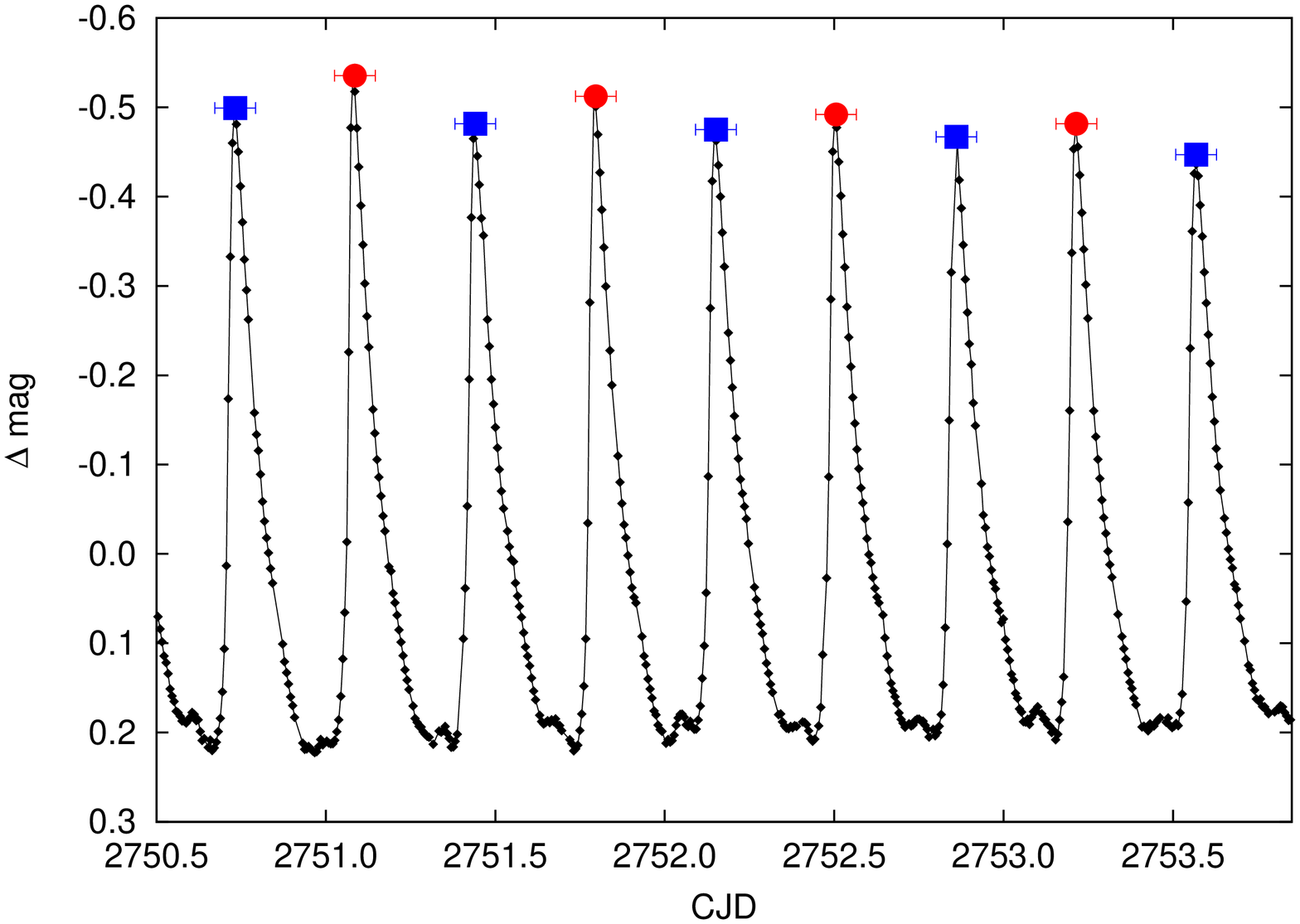}}
\caption{{\bf Left panel:} Low-frequency part of the Fourier-spectrum of {\tt 962}. The dominant pulsation frequency (${f_0}$), its harmonics and the modulation triplets were pre-whitened. Black lines show the location of ${f_0}$ and its harmonics, while the blue dash-dotted lines show the location of the half-integer frequencies. To visualize the distribution of the amplitudes of the HIFs we plotted black points on top of the frequencies originating from the PD effect. The HIFs are significant up to ${\rm 11/2} {f_0}$. {\bf Right panel:} Small part of the CoRoT light curve of {\tt 962} clearly showing alternating maxima, \ie  period doubling. Fitted maxima are plotted to guide the eye.}
\label{962_PD}
\end{figure*}

\vskip 5mm

\noindent CoRoT\,0104315{\bf 804}: This star was also observed in LRc04, and it has not been published, yet. It was monitored with nominal cadence (512s) and only in white light. It is a non-Blazhko star with a period of 0.7218221 days.  We could not detect period doubling with an upper limit of 0.30\,mmag. The frequency spectrum is also devoid of statistically significant additional periodicities. The frequency table is given in Appendix~\ref{app804}, while the phase-folded light curve is plotted in the right panel of Fig.~\ref{488}.

\begin{figure*}                        
\includegraphics[height=9.2cm,angle=270]{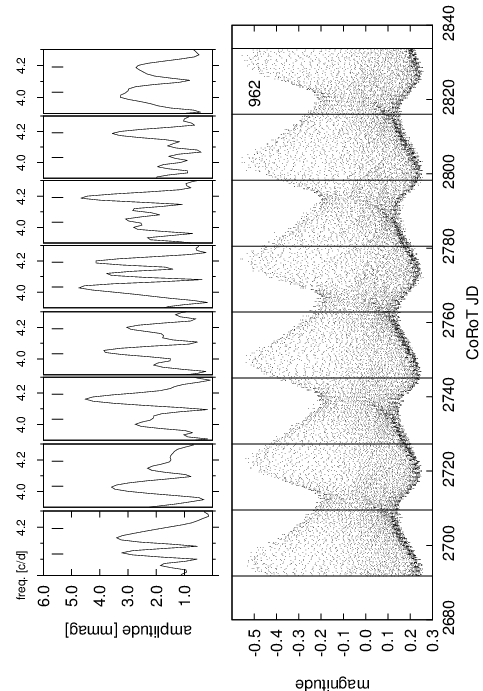}
\includegraphics[height=9.2cm,angle=270]{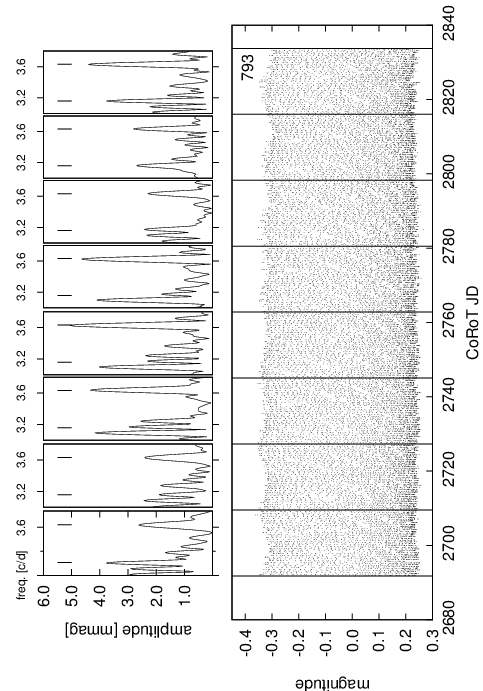} \\
\includegraphics[height=9.2cm,angle=270]{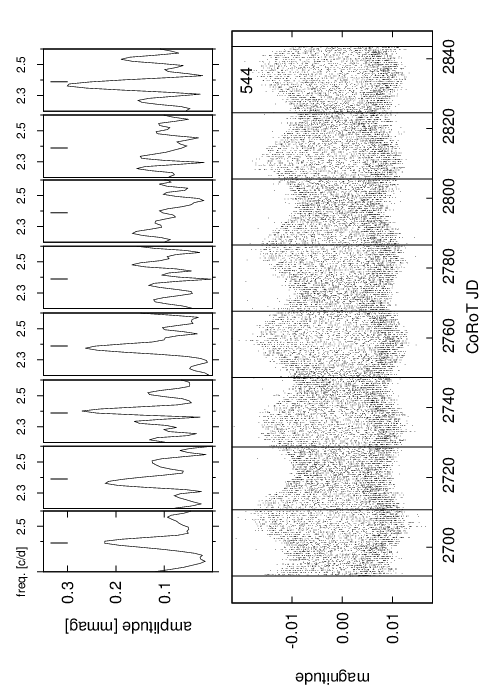}
\includegraphics[height=9.2cm,angle=270]{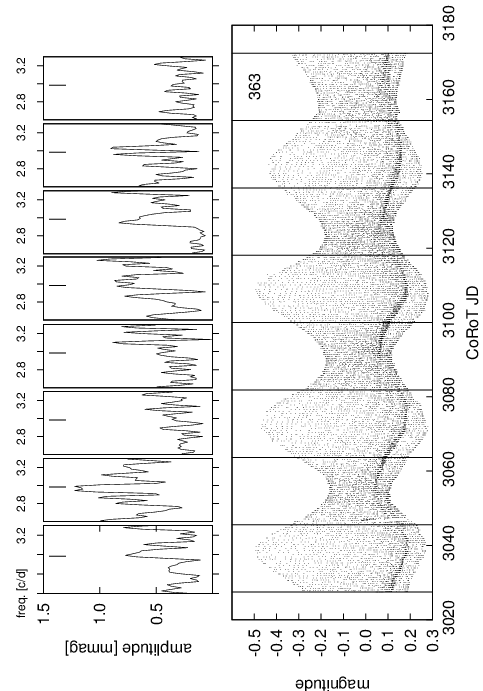}
\caption{Temporal variation of the amplitude of additional frequencies in CoRoT Blazhko RR\,Lyrae stars. 
The CoRoT photometric light curves were split into eight chunks, the dominant pulsational frequency, its harmonics and the modulation side peaks were pre-whitened. The Fourier spectrum in the vicinity of the 
additional frequencies (denoted by small verticals lines) are shown in the upper parts of the sub-figures. 
{\bf Upper left:} Temporal variation of the half-integer frequency at ${\rm 3/2} {f_0} = {\rm 4.1916\,d^{-1}}$ and the independent frequency ${f'} = {\rm 4.03265\,d^{-1}}$ in the Blazhko RR\,Lyrae {\tt 962}. 
{\bf Upper right:} The same for the CoRoT Blazhko RR\,Lyrae star {\tt 793}. The frequencies are 
${\rm 3/2} {f_0} = {\rm 3.159\,d^{-1}}$ and ${f_1} = {\rm 3.63088\,d^{-1}}$.
{\bf Lower left:} The variation of ${f'} = {\rm 2.389287\,d^{-1}}$ additional frequency in the blended CoRoT Blazhko RR\,Lyrae star {\tt 544}.
{\bf Lower right:} The same for the Blazhko star {\tt 363}, where ${f'} = {\rm 2.98400\,d^{-1}}$. }
\label{962_cut8}
\end{figure*}

\subsection{RRab stars showing the Blazhko effect}

\noindent CoRoT\,0100689{\bf 962} is identical with V1127\,Aql. 
It has a strongly modulated light curve, with a high pulsational amplitude and a 
pronounced phase modulation as is easily seen in the left panel of  Fig.~\ref{962_363}. The modulation is not sinusoidal \citep{bszp11}, and is asymmetric in the sense that the maximum of the maxima and the minimum of the minima shows considerable shift. A detailed study has been published on this object by  \cite{chadid2010}. 

We found alternating cycles in the light curve of {\tt 962} (see the right 
panel of Fig.~\ref{962_PD}.) Half-integer frequencies between the fundamental mode and its harmonics up to ${\rm 11/2} {f_0}$ are also found, demonstrated in the left panel of Fig.~\ref{962_PD}. A forest of peaks can be found around the HIFs. This may be a consequence of the excitation of additional (nonradial) modes, or due to the interplay of the time-variable PD and the modulation itself (see a simulation suggesting this scenario in \citet{szabo2010}), or both. The amplitude distribution of the HIFs is very similar to other cases (see \citealt{szabo2010}): the highest peak is at ${\rm 3/2} {f_0}$, then there is a monotone decrease with the order, however at ${\rm 9/2} {f_0}$ there is a local maximum (a standstill in the case of {\tt 962}), which is the consequence of the 9:2 resonance with the 9th overtone \citep{kollath2011}. We could identify HIFs up to ${\rm 11/2} {f_0}$ above the $3 {\rm \sigma}$ significance level. The alternating cycles in the light curve and the presence of the half-integer frequencies in the frequency domain constitute an unambiguous detection of the period doubling phenomenon in {\tt 962}. It is evident that the PD is not always present with the same strength throughout the 142-day CoRoT observations. Instead, we see it in the light curve for short time intervals, \eg CJD[2750:2754] as shown in Fig.~\ref{962_PD} or CJD[2794:2798] (not shown). 

In the frequency spectrum of {\tt 962} there is another, independent frequency peak, ${f'}={\rm 4.03265 d^{-1}}$. This is quite close to the 3/2${f_0}$ frequency value, and there is a dense region of frequencies around them (Fig.~7 of \citealt{chadid2010}), which makes the detection of the temporal variation of these frequencies extremely challenging. We made a series of tests to take into account the effects of the decreased frequency resolution due the splitting of the original data set, and we found variation in the amplitude in both the 3/2${f_0}$ HIF and ${f'}$ (upper left panel of Fig.~\ref{962_cut8}), although this later conclusion is quite weak and may be a consequence of contamination from other close-by frequencies. 

\vskip 5mm

\begin{figure}
\includegraphics[height=8.5cm,angle=270]{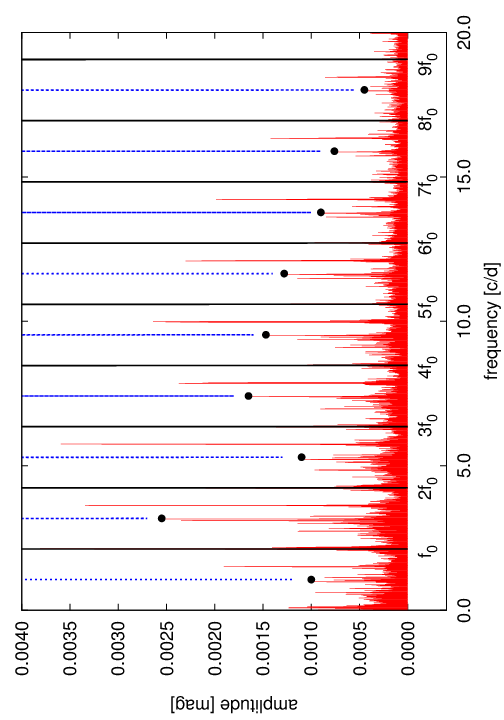}
\caption{Same as the left panel of Fig.~\ref{962_cut8}, but for CoRoT Blazhko RRab {\tt 793}. The dominant pulsation frequency (${f_0}$), its harmonics, the Blazhko-frequency (${f_B}$) and ${\rm 2}f_B$ \citep{poretti2010} were pre-whitened. The HIFs are significant up to ${\rm 17/2} f_0$.} 
\label{793_PD}
\end{figure}

\noindent CoRoT\,0101128{\bf 793} is a CoRoT RR\,Lyrae star with a low Blazhko modulation amplitude (the amplitude of the highest side peak due to the modulation is 4.6\,mmag) and a Blazhko period of 18.66\,d (see Fig.~\ref{962_363}). This star shows one of the lowest modulation amplitudes detected so far \citep{jurcsik2006,skarka2014}. \citet{poretti2010} studied this object in detail. In the course of this work we find clear sign of period doubling in the frequency spectrum in the form of half-integer frequencies (see Fig.~\ref{793_PD}). \citet{poretti2010} interpreted $f_2 = {\rm 3.159\,d^{-1}}$ and ${f_0+f_2} = {\rm 5.279\,d^{-1}}$ as combination frequencies, but in our context the period doubling interpretation seems to be more plausible, especially that it is found in all possible places from ${\rm 1/2} f_0$ to ${\rm 17/2} f_0$ \footnote{We make an  exception here and retain the original designation of ${f_1}$ and ${f_2}$ from \citet{poretti2010}}. In our interpretation these frequencies are located closer than 1\% to ${\rm 3/2} {f_0} = {\rm 3.178 \,d^{-1}}$ and ${\rm 5/2} {f_0} = {\rm 5.297\,d^{-1}}$, thus we identify them as HIFs.

We note that there is another additional frequency, ${\it f_1} = {\rm 3.63088\,d^{-1}}$, which might be the second radial overtone \citep{poretti2010}. Its combinations with the fundamental mode frequency and its harmonics ($ f_1 + k f_0, k=0,1,2...$) are clearly visible in Fig.~\ref{793_PD} with high amplitudes in between the main harmonics and the HIFs. As in the previous cases, temporal variability is easily seen for both the ${3/2 f_0}$ half-integer frequency and the additional frequency $f_1 = {\rm 3.63088\,d^{-1}}$ (upper right panel of Fig.~\ref{962_cut8}). 
\vskip 5mm

\noindent CoRoT\,0100881{\bf 648} is a Blazhko star exhibiting a typical asymmetric RR~Lyrae light variation with a pulsation period of 0.607186\,d and a Blazhko cycle of 59.77\,d \citep{szabo2009}. The modulation is sinusoidal and symmetric (Fig.~\ref{962_363}). 

\begin{figure}                      
\includegraphics[width=4.65cm]{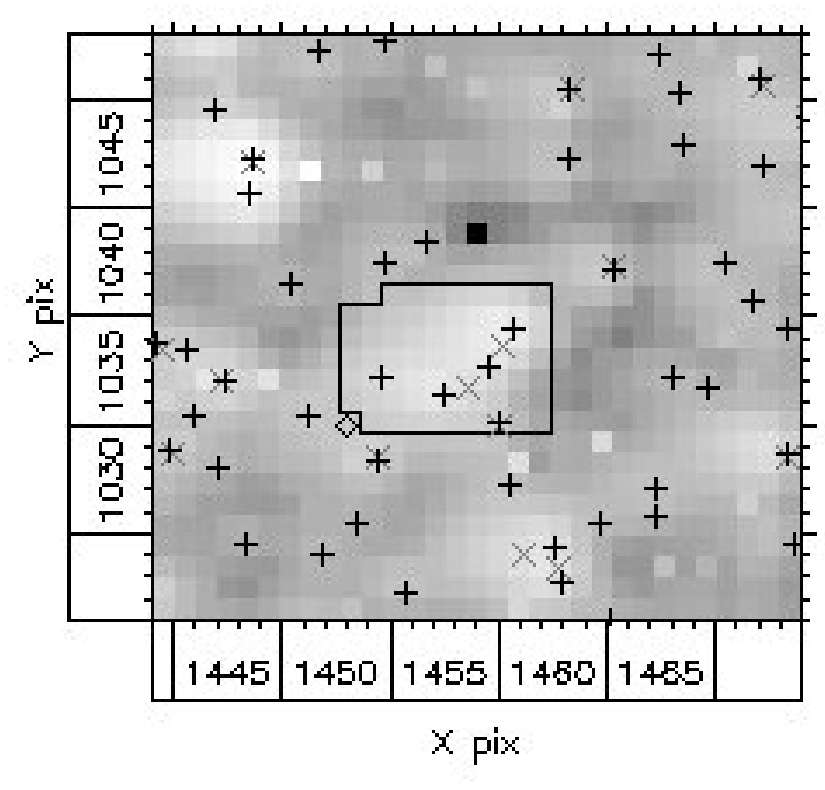}\includegraphics[width=4.5cm]{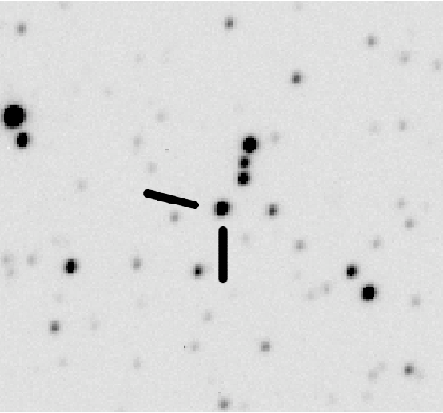}\\
\includegraphics[width=4.65cm]{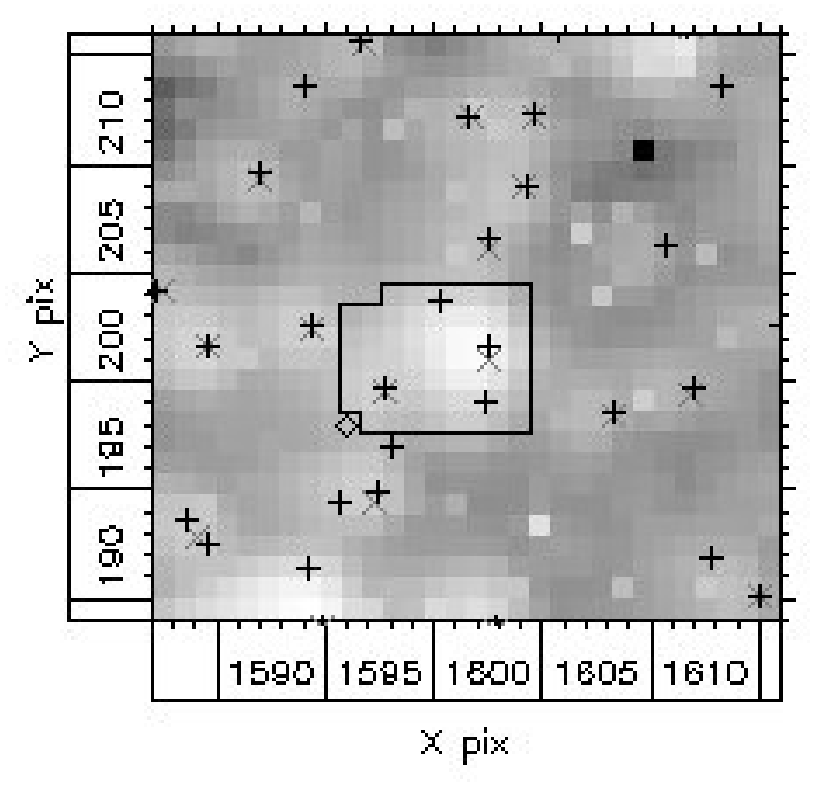}\includegraphics[width=4.5cm]{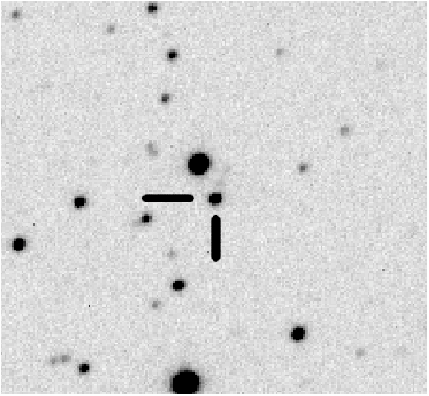}
\caption{Left: EXODAT image showing the aperture applied to {\tt 648} (upper panel) and {\tt 544} (lower panel). Right: 1' x 1' {\it V} image of the targets taken by the 1m RCC telescope at Piszk\'es-tet\H o. North is to the right and East is to the bottom. The variable star is located between the thick lines. We transformed our frames to match the EXODAT images except a slight (30 degrees) rotation. {\tt 648} was caught around pulsational minimum, while {\tt 544} around pulsational maximum. }
\label{648}
\end{figure}

The typical RRab light variation is diluted by a string of three close stellar companions to W-NW and a fainter star to the North (Fig.~\ref{648}). The flux from these stars are contained by the CoRoT aperture together with that of the star {\tt 648}, and their presence is revealed by our ground-based observations that easily resolved the near-by contaminating sources. These follow-up observations allow us to make a crude estimate on the intrinsic light variation amplitude of this Blazhko star. One {\it V} image was taken in July 2008 (HJD 2454676.44742) with the Konkoly Observatory 1m RCC telescope which corresponds to a pulsational minimum and is slightly past the Blazhko-maximum ($\psi=0.06$, where $\psi$ denotes the Blazhko phase). Based on the fluxes of the contaminating stars (assumed to be constant) and the variable in its minimum and using the apparent total amplitude we calculate the true light variation to be 0\dotm36 in the Blazhko maximum and 0\dotm30 in the Blazhko minimum. This is still a bit lower than the amplitude of the light variation of a normal, non-blended Blazhko RRab star, but to reconstruct the undiluted light variation more precisely, much longer follow-up observations would be necessary. Other fainter stars have negligible effect on our estimate.

We detected 10 harmonics of the ${\rm 1.64694 d^{-1}}$ frequency and a 
triplet around ${f_0}$ and 2${f_0}$. The right peaks are much higher and are detected around all the harmonics. No additional frequency is detected above the  3$\sigma$ level. Also, we could not detect any signs of the period doubling phenomenon. The frequency table is available at Appendix~\ref{app648}.

\vskip 5mm

\noindent CoRoT\,0101503{\bf 544} is a Blazhko star with an asymmetric light variation. Its pulsation period is 0.605087\,d and a Blazhko-variation of 25.60\,d  is clearly seen (Fig.~\ref{962_363}). The color light curves are preserved for this object. The star exhibits a typical skewed RRab light curve, with its amplitude diminished to 0\dotm021 in red, 0\dotm045 in green and 0\dotm052 in blue passbands. Based on the CoRoT data this object is most probably an intrinsically high-amplitude RR\,Lyrae star. This conclusion is supported by the relatively high modulation (e.g. 0\dotm028 in the green light compared to the overall light variation amplitude 0\dotm045). We note that the color amplitude ratios are typical for an RR Lyrae variable. The L1 (level-1) contamination value given by the ExoDat catalog is $0.13871 \pm 0.00663$, which might be a too low estimate. 

CCD frames taken by one of us (J. Benk\H o) with the Konkoly Observatory 1m RCC telescope and a Johnson {\it V} filter allowed us to reconstruct the true light variation of this heavily blended object. The contaminating star is brighter by 2\dotm0 in {\it V} compared to the brightest phase of {\tt 544}. One image was taken in June 2008 (HJD 2454639.49092) which corresponds to a pulsational maximum and is slightly past the Blazhko-maximum ($\psi=0.10$). Using a similar process that was applied in the case of {\tt 648} (and assuming that the Johnson {\it V} filter is reasonably close to the CoRoT green passband) we calculate the true light variation to be 0\dotm39 in the 
Blazhko maximum and 0\dotm23 in the Blazhko minimum in the {\it V} passband. 
The frequency content of {\tt 544} can be found in Appendix~\ref{app544}.

After a few days of normal sampling (8-min) the observations were switched to the  oversampling mode (32-sec). Before embarking on the frequency analysis we re-sampled the over-sampled data to normal sampling to get a more tractable number of points (23\,915 vs. 351\,086). We checked that no information was lost in the relevant low-frequency range which is the main focus of this study. 

High-precision space-based observations frequently show additional frequencies in RR\,Lyrae stars. This Blazhko star is no exception, it also has additional periodicities in the frequency spectrum. One of them is $f'={\rm 2.389287\,d^{-1}}$. The lower left panel of Fig.~\ref{962_cut8} shows the temporal variation of the $f'$. Again, one can easily recognize the time-variability of the appearance of this frequency peak during the CoRoT observations. 

\begin{figure*} 
\includegraphics[height=8.9cm,width=5.8cm,angle=270]{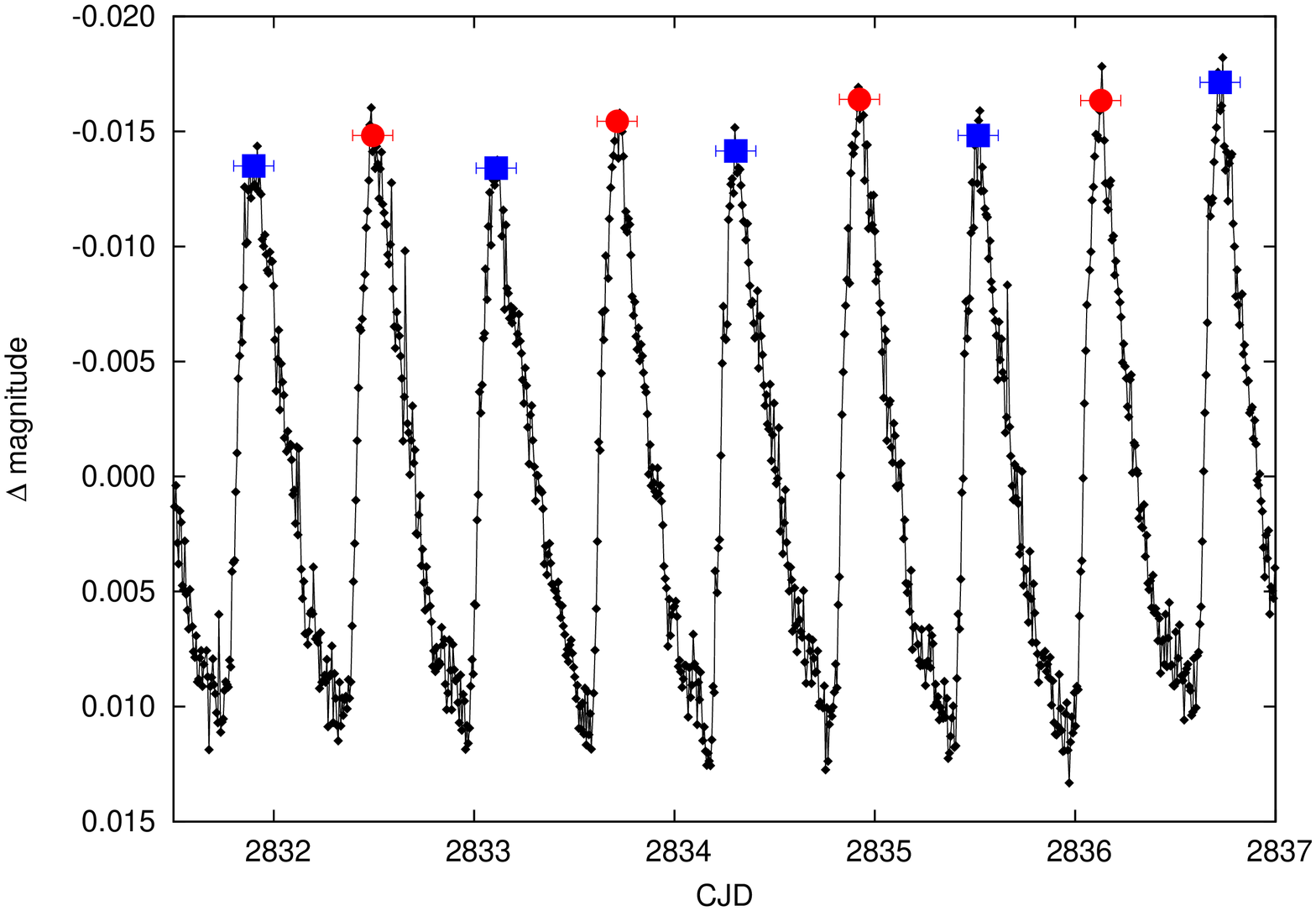}\includegraphics[height=8.9cm,width=5.8cm,angle=270]{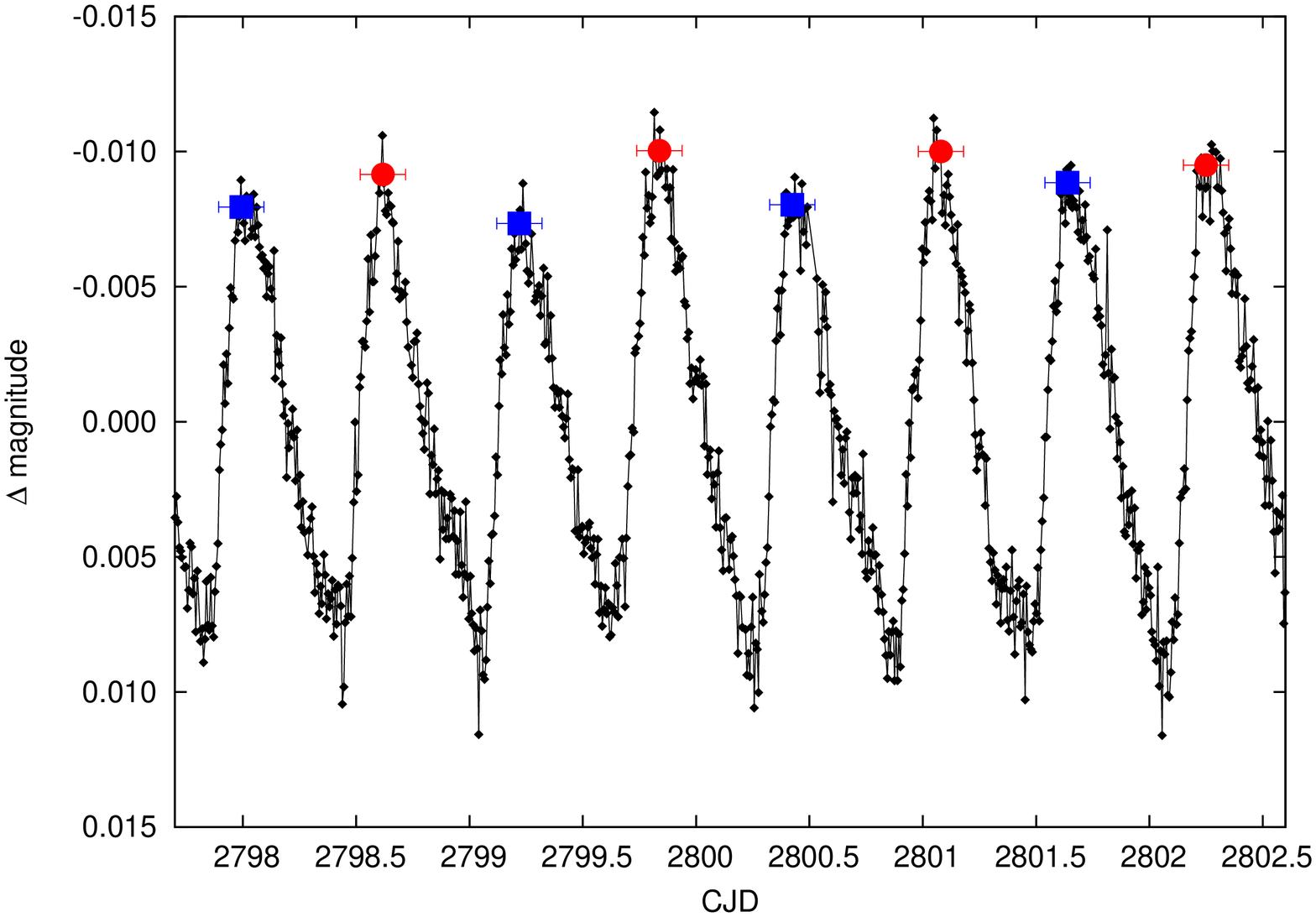}
\caption{Sections of light curve where period doubling is present in the blended Blazhko CoRoT RRab star {\tt 544}. Even and odd maxima are marked with different symbols. Original data points are denoted by small black points.}
\label{544pd}  
\end{figure*}

We have not found any signatures pointing to period doubling in the frequency spectrum neither in the white light, nor in the color observations, which is not surprising, given the blended nature of our target. However, a close inspection of the light curve reveals that there are sections where the alternating pulsating cycles are clearly visible (Fig.~\ref{544pd}). We checked that the frequency spectrum 
does not contain any other additional frequency peaks with high enough amplitude to cause the observed variation in the maxima from cycle to cycle. Clearly, the 0.15 mmag ${f'}$ can not cause fluctuation exceeding several mmag as seen in Fig.~\ref{544pd}.  Interestingly, CoRoT's detection capability allowed us to reveal a temporary nature of the period doubling phenomenon in this heavily blended CoRoT Blazhko RR\,Lyrae star. 

\vskip 5mm

\noindent CoRoT\,0105288{\bf 363} is an RR\,Lyrae star showing a strong Blazhko 
modulation with a period of 35.06\,d. The lower right panel of Fig.~\ref{962_363} shows that the maxima of the pulsational cycles do not move considerably in sharp contrast with {\tt 962}. The amplitude of the Blazhko modulation is not constant. A detailed study was presented by \citep{guggenberger2011}. After reanalyzing the CoRoT light curve, we have not found 
any sign of period doubling either in the light curve or in the frequency spectrum down to 0.35\,mmag in accordance with \citet{guggenberger2011}. There is an additional frequency in the star ($f'={\rm 2.98400\,d^{-1}}$) with a frequency ratio of 0.591, which can be identified as the second radial overtone. Despite its low amplitude and the cluttered frequency spectrum we tried to analyze its temporal variation (Fig.~\ref{962_cut8}, lower right panel). Although there are hints that the amplitude of this frequency peak varies in time as well, we can not draw firm conclusions about its temporal variability due to the difficulties mentioned. Also, there is a non-negligible chance that this periodicity comes from a different, nearby source \citep{guggenberger2011}. According to \citet{guggenberger2012}, the star also shows a frequency around its first radial overtone ($f_1 = {\rm 2.3793\,d^{-1}}$, frequency ratio 0.741) and another one which is most probably a nonradial mode\footnote{Denoted by ${f_{N}}$ in the original paper.} ($f_{nr} = {\rm 2.4422\,d^{-1}}$, $f_{0}/f_{nr}=0.722$). Due to their low amplitude we did not try to analyze their temporal behavior in this study.

\vskip 5mm

\noindent CoRoT\,0103922{\bf 434}: 
This is a new Blazhko RRab star (also known as V922\,Oph) that will be discussed in detail in Poretti et al. (in prep). It's period is 0.5413828 days and shows a Blazhko-modulation of roughly 54.5 days, thus CoRoT LRc04 observations cover two modulation cycles. 

\begin{figure*} 
\includegraphics[height=8.9cm,width=5.8cm,angle=270]{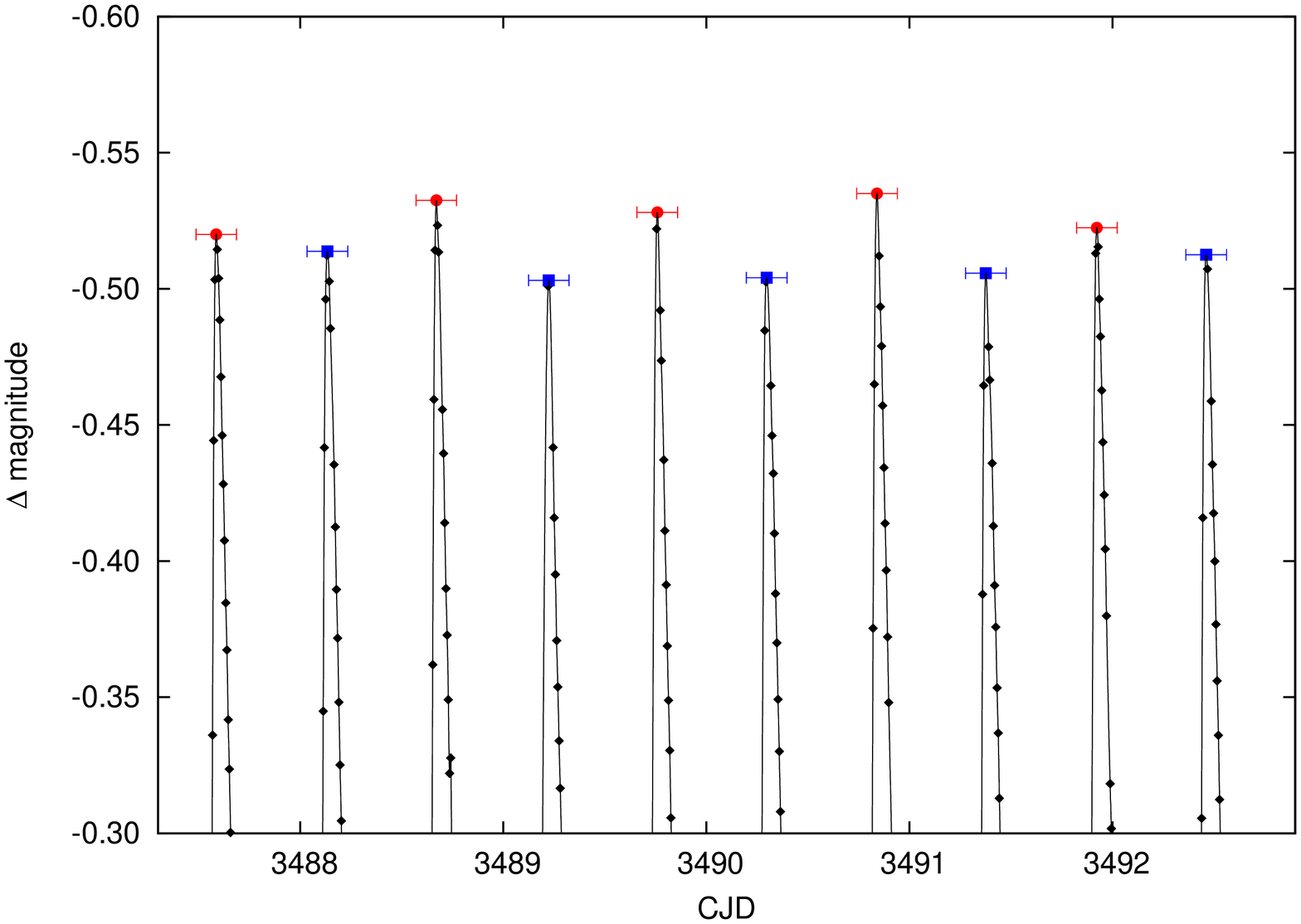}
\includegraphics[height=8.9cm,width=5.8cm,angle=270]{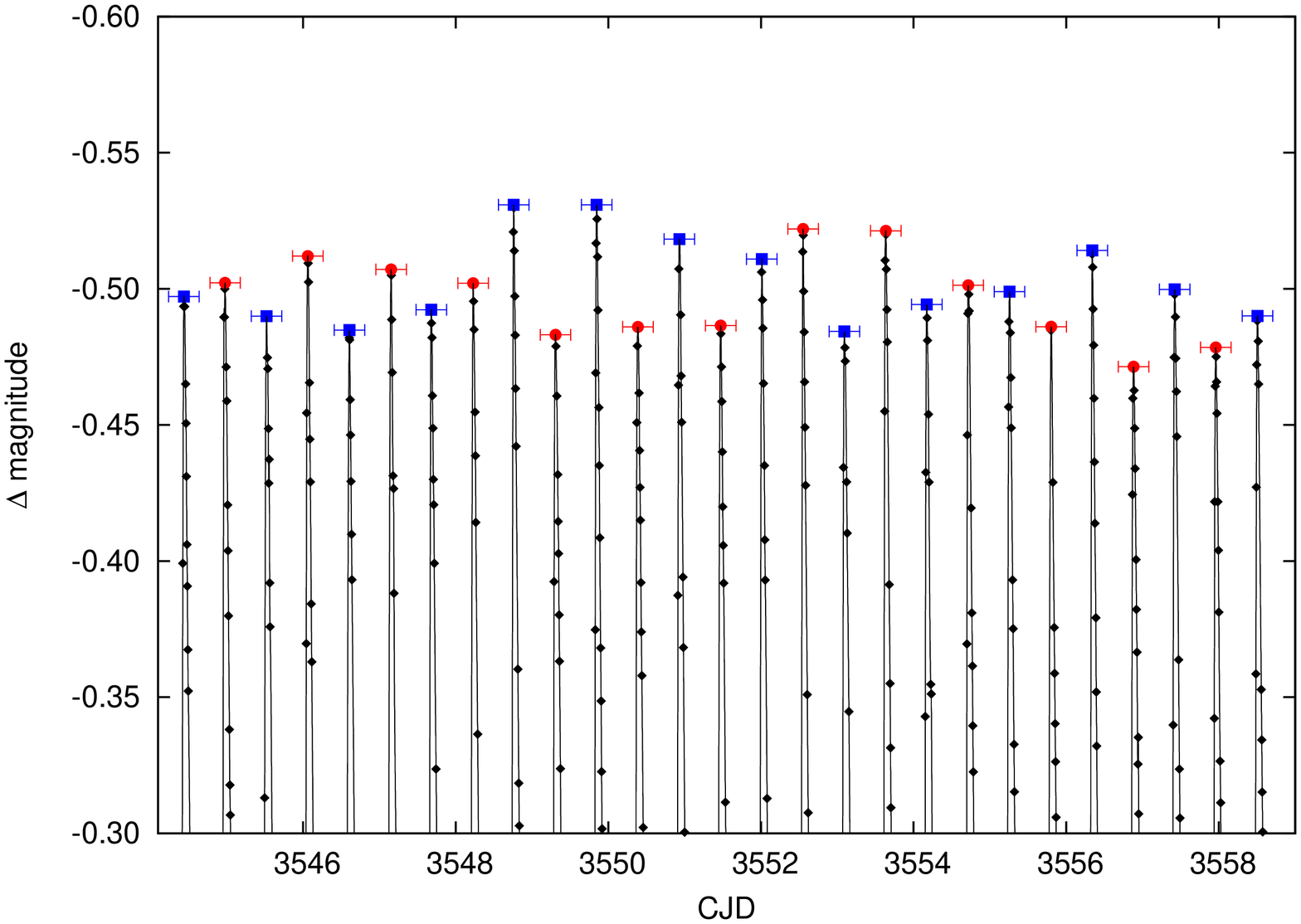}
\caption{Same as Fig.~\ref{544pd}, but for the Blazhko CoRoT RRab star {\tt 434}. The switch back and forth between the two branches in the right panel is a clear signal points to more complex dynamical behavior.}
\label{434pd}  
\end{figure*}

Here we only note that despite the fact that we do not see half-integer frequencies in the frequency spectrum of this star, there are hints that the period doubling might be present temporarily. Namely, we find 6 -- 10 consecutive pulsational cycles in the light curve that are alternating (Fig.~\ref{434pd}). One is found around CJD 3490 and lasts only for 10 pulsational cycles. The other one is at the end of the CoRoT observations (starting from CJD 3548). Such regularity can not be seen in any of the other RRab stars that do not show PD. The amplitudes of the additional frequencies are not sufficient to cause as large as many hundredths of a magnitude alternation from maximum to maximum even in the most favorable (constructive interference) case. In addition, finding alternating maxima by chance more than once has a very low probability.\footnote{To be more specific: in a very simplistic approach, the probability of ${n}$ jump (up or down) is ${1 / ({2^n-1})}$, since the random variable $X$ of the height of the pulsational maximum follows a binomial distribution assuming an equal probability (${p=1/2}$) of an up or down jump. Thus, the chance configuration of $n=8$ switches  in the left panel of Fig.~\ref{434pd} is $1/127 \approx 0.0079$.} We also note that in RR\,Lyrae and other {\it Kepler} PD-stars the temporal nature of the period doubling is also clearly demonstrated \citep{szabo2010}. Thus, we conclude that {\tt 434} is a Blazhko RRab star showing the period doubling temporarily. One can notice that there are switches between the different branches connecting even and odd maxima (Fig.~\ref{434pd}). The switch between the two branches of the 
lower and higher maximum strings is a definite sign of a more complex dynamical behavior besides the high-order resonance causing the period doubling. It points to the presence of additional resonances (see e.g. \cite{molnar2012}) or low-dimensional chaos \citep{plachy2013}.

\begin{figure*} 
\includegraphics[height=8.9cm,width=5.8cm,angle=270]{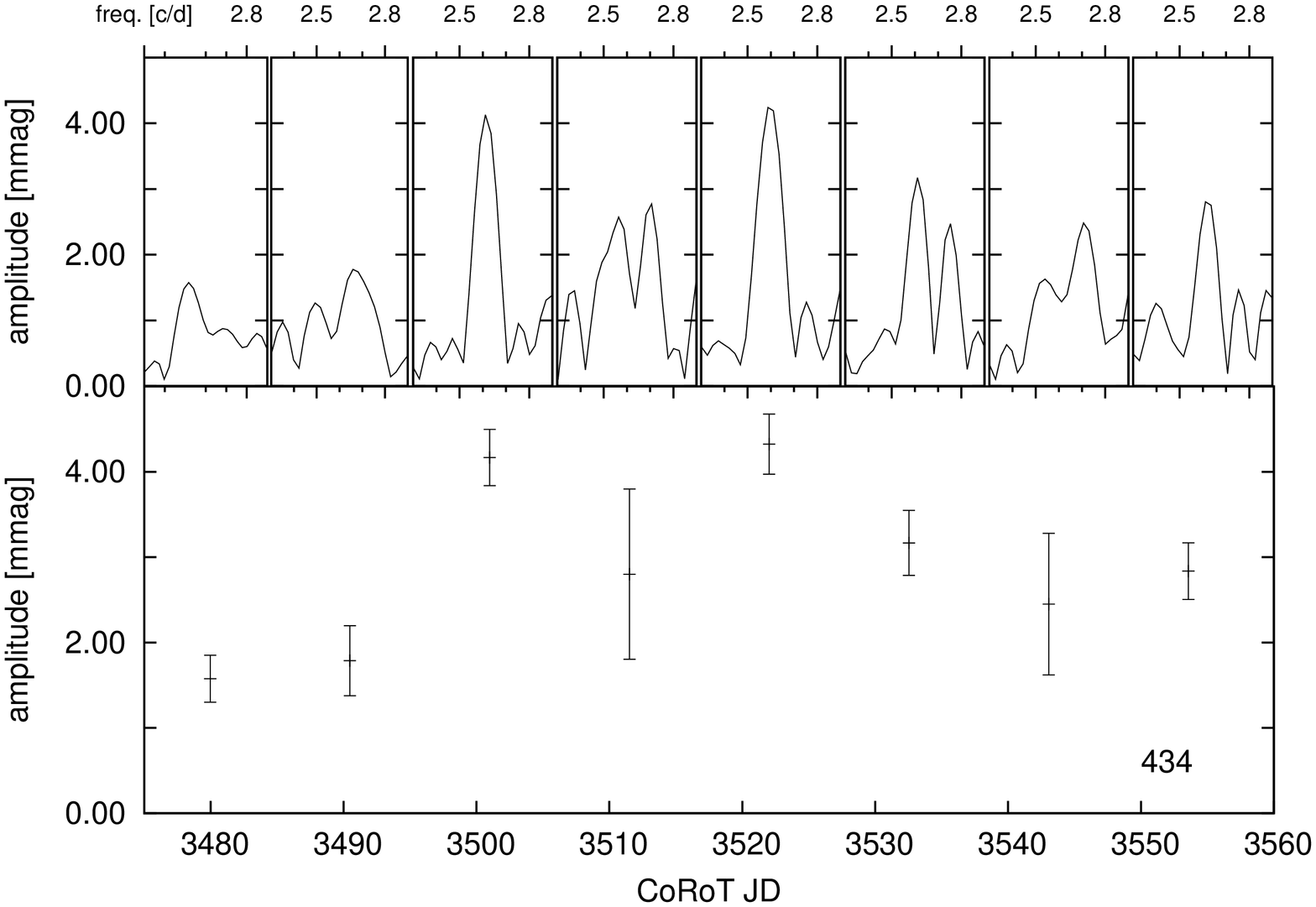}
\includegraphics[height=8.9cm,width=5.8cm,angle=270]{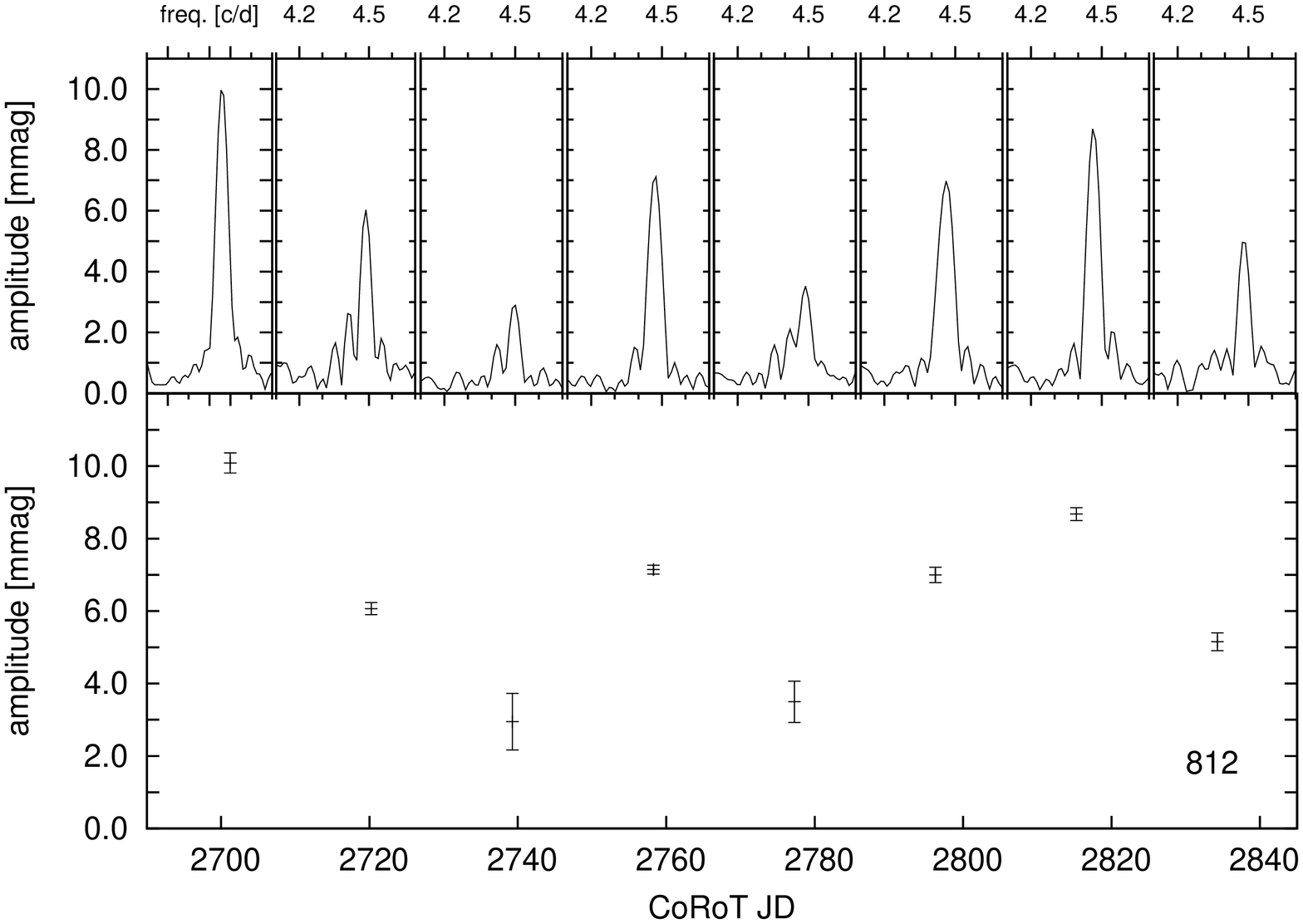}\\
\includegraphics[height=8.9cm,width=5.8cm,angle=270]{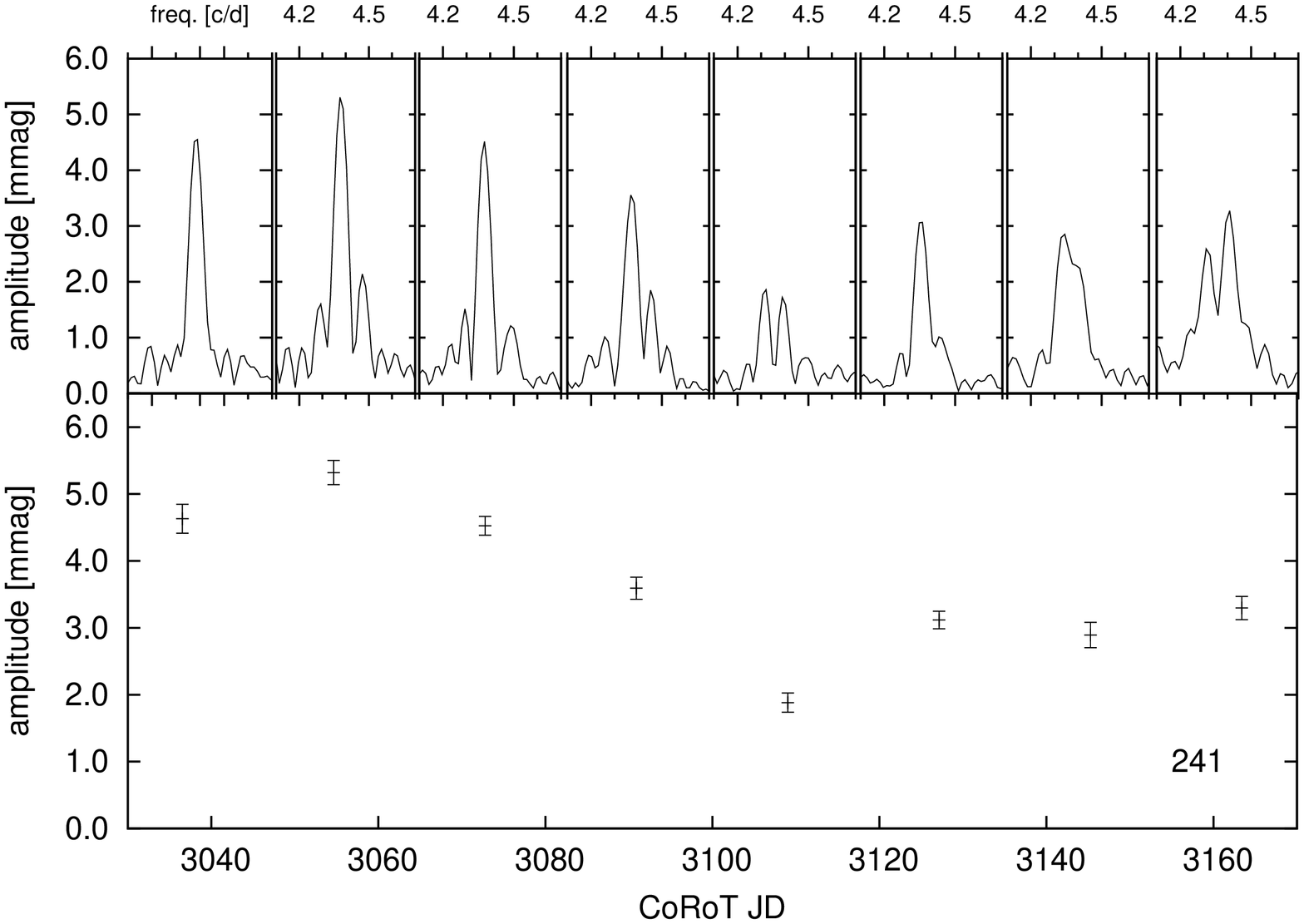}
\includegraphics[height=8.9cm,width=5.8cm,angle=270]{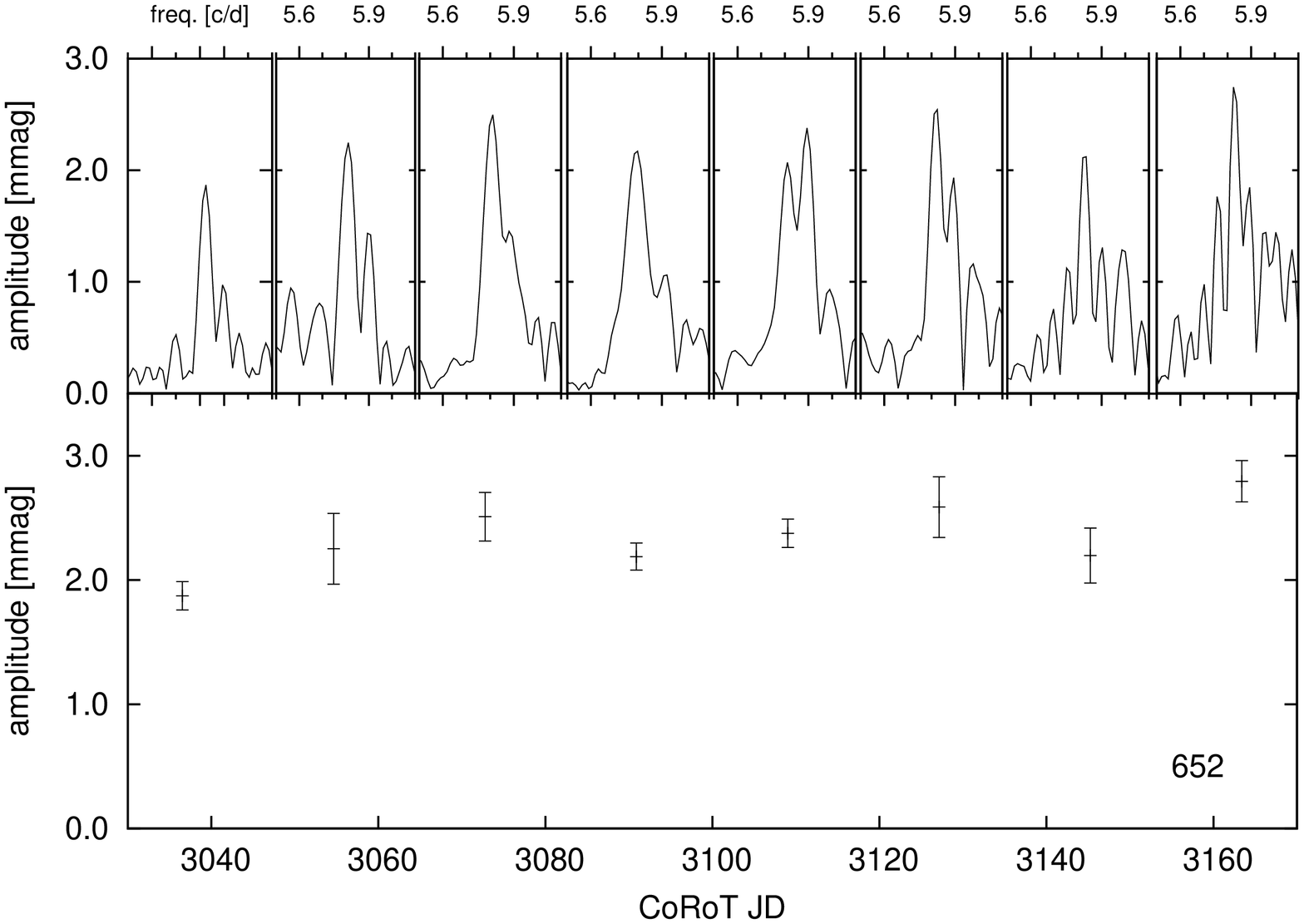}
\caption{Temporal variability of additional frequencies in RRab, RRd and RRc stars. 
{\bf Upper left:} Lower panel: Amplitude variation of the additional $f'= {\rm 2.61219586\,d^{-1}}$ frequency of CoRoT Blazhko RR Lyrae star {\tt 434}. The fundamental mode frequency, its harmonics 
and the side peaks related to the modulation (and the modulation frequency itself) were 
removed. Upper panels: the vicinity of  ${f'}$ in the frequency spectrum, lower panels: amplitude variation.
{\bf Upper right:} The same for the temporal variability of $f' = {\rm 4.4786159\,d^{-1}}$
CoRoT RRd star {\tt 812}. The first overtone frequency and its harmonics were removed.
{\bf Lower left:} Temporal variation of the $f'= {\rm 4.37783\,d^{-1}}$ frequency of CoRoT RRc star {\tt 241}. The first overtone frequency and its harmonics were removed. 
{\bf Lower right:} The same as for the RRc star {\tt 241}, but for {\tt 652}. Note that the amplitudes are smaller than in {\tt 241}, but the variation of the structure of the  $f' = {\rm 5.82484\,d^{-1}}$ peak is strikingly similar.}
\label{f2var}  
\end{figure*}

Although there are indications that the second radial overtone is excited in this star with $f_2 = {\rm 3.165859\,d^{-1}}$, we examined only the strongest additional frequency $f'={\rm 2.612196\,d^{-1}}$ and the result is presented in the upper right panel of Fig.\ref{f2var}. There is a clear amplitude variation (lower panel). The figure was made after pre-whitening with the fundamental mode ${f_0}$, its harmonics and the modulation side peaks. Large number of combination frequencies (${kf_0 + nf_m}$) are present as well, suggesting that these signals  come from the star itself. The frequency ratio is ${f_0 / f'} = {\rm 0.707091}$, thus ${f'}$ can not be a radial mode. 

\subsection{RRc stars}\label{rrc}

\begin{figure*}         
\includegraphics[height=8.9cm,width=5.2cm,angle=270]{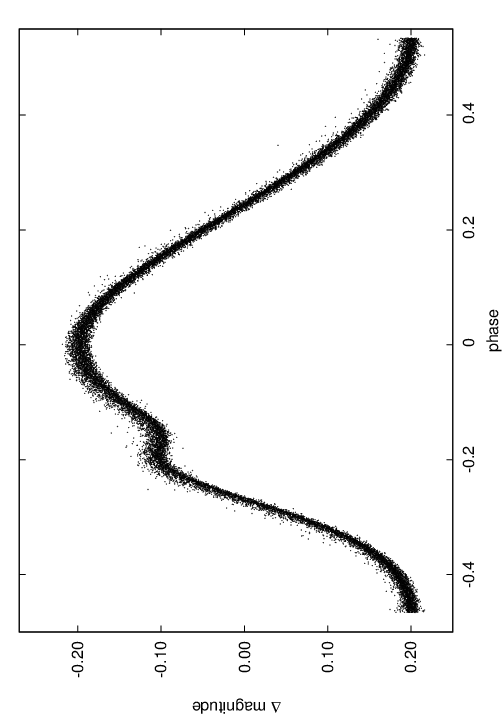}
\includegraphics[height=8.9cm,width=5.2cm,angle=270]{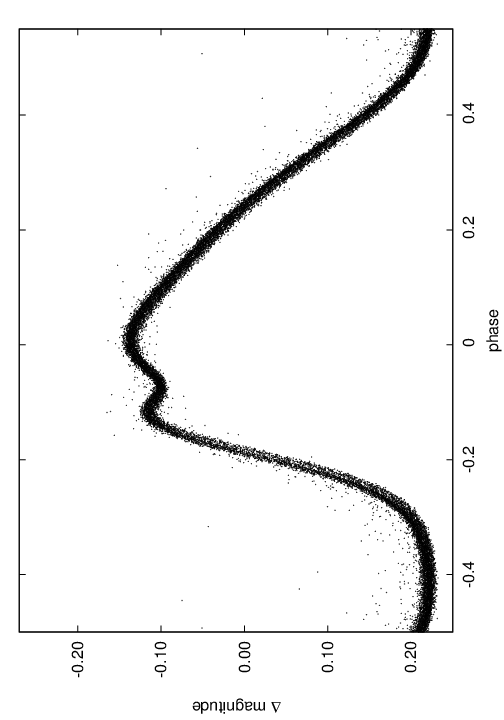}
\caption{Phased light curves of two CoRoT RRc stars. 
{\bf Left:} {\tt 241}. {\bf Right:} {\tt 652}.}
\label{rrcfigs}        
\end{figure*}

\noindent CoRoT\,0105036{\bf 241}: We have found two RR\,Lyrae stars predominantly pulsating in the first radial overtone (RRc) by CoRoT, one of them is {\tt 241}. The phased light curve can be seen in the left panel of Fig.~\ref{rrcfigs}. 
Besides the first overtone mode ($f_1 = {\rm 2.68153 \,d^{-1}}$), this RRc star has an additional frequency $f' = {\rm 4.37783\,d^{-1}}$ with a frequency ratio ${f_1/f'}$=0.613. This frequency ratio occurs surprisingly often in overtone RR Lyrae and Cepheid stars, see \eg \cite{moskalik2013}. {\tt 241} represents another member of this class of classical pulsators. If -- as it is conceivable -- ${f_1}$ belongs to the first radial overtone then ${f'}$ can not be a radial mode, it is most probably a nonradial one. 

\begin{figure*}                      
\includegraphics[height=16.0cm,angle=270]{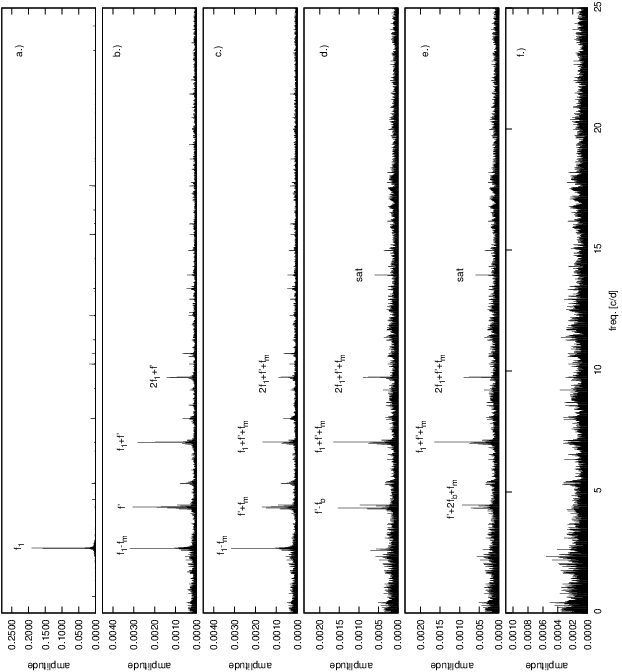}
\caption{Frequency spectrum of the star {\tt 241} with sequential pre-whitening. Some 
characteristic frequencies are labeled with their identification. {\it Sat} refers to frequencies connected to the orbit of the CoRoT satellite. {\bf a.)} The original spectrum dominated by the first overtone pulsation mode (${f_1}$). {\bf b.)} After pre-whitening with ${f_1}$ and its harmonics. {\bf c.)} The result of removing ${f'}$ and all combination terms involving ${f_1}$ and ${f'}$. {\bf d.)} Pre-whitening with all ${f_1 \pm k*f_m}$ and ${f' \pm k*f_m}$ frequencies. {\bf e.)} After removing all significant ${f_1 \pm k*f_b}$ and ${f' \pm k*f_b}$ frequencies. {\bf f.)} Removing another 13 frequencies including those originating from the satellite and third order 
linear combinations of \{${f_1, f', f_m, f_b}$\}.}
\label{241_frsp}
\end{figure*}

The Fourier spectrum and sequential pre-whitening with the highest amplitude frequencies in {\tt 241} are displayed in Fig.~\ref{241_frsp}. After successively removing the dominant pulsation frequency, its harmonics, and the additional ${f'}$ frequency, two sets of side peaks appear. We denote the corresponding frequency differences with $f_{\rm m}$ and $f_{\rm b}$ in Fig.~\ref{241_frsp} and in Appendix~\ref{app241}. The first one ($f_{\rm m}$) may be a long-period (longer than the length of our data set) Blazhko modulation, although more data would be necessary to establish its exact nature, while the second one ($f_{\rm b}$) most probably appears because a clear amplitude variability is seen in the additional frequency, ${f'}$ (see below). 

After removing all the previously mentioned frequencies there are still a lot of remaining peaks in the spectrum, the highest among them is $f'' = {\rm 2.345174\,d^{-1}}$ with an amplitude of 0.57 mmag. However, instead of (over)interpreting the data we prefer to stop at this level. The average value of the 'grass' of the residual frequency spectrum is 0.09 millimagnitude. A detailed discussion of the frequency spectrum can be found in Appendix~\ref{app241}. 

We computed the temporal variation of the amplitude of ${f'}$ by using eight bins. 
The lower left panel of Fig.~\ref{f2var} shows that the amplitude of ${f'}$ very clearly varies on long time-scales. In the same figure we also plotted the close neighborhood of ${f'}$ in the frequency spectrum corresponding to the bins for which the amplitude was calculated (upper panel). It is interesting to see the structural variation of the peak. Sometimes the frequency peak is split into two separate peaks. The variation over the 145 day-interval is striking.

\vskip 5mm

\noindent CoRoT\,0105735{\bf 652} is similar to {\tt 241} in many aspects. Its  phased light curve is shown in the right panel of Fig.~\ref{rrcfigs}. However, this star has a B5 II spectrum in the CoRoT observation log Release 13 \citep{corot2014}. This and the presence of many low amplitude frequencies (that may be nonradial modes) might imply a $\beta$\,Cep scenario. On the other hand, the light curve shape resembles that of an RRc star with the characteristic bump just before maximum light, though some $\beta$\,Cep variables show a similar feature \citep{sterken1987}. The CoRoT classification algorithm \citep{debosscher2009} assigns a probability of more than $80$\% for an RRc variable, while gives less than $10$\% chance for the $\beta$\,Cep variation. In addition, the frequency ratio ${f_1/f'}$= 0.615 is seen frequently in recent space-based photometric observations (e.g. \citealt{moskalik2013}), which makes it likely that the star is a first-overtone RR~Lyrae star, hence we stick to this scenario.

The temporal variation of the amplitude of the ${f'}$ frequency is plotted in lower right panel of Fig.~\ref{f2var}. The ${f'}$ frequency shows similar structural variations as its sibling, the other CoRoT RRc star, {\tt 241} that we discussed above. Although this star does not show as large amplitude variation as {\tt 241}, the change in the structure of this secondary frequency is obvious. Note that while the timescale of the variation is similar, the variations in the two stars are not identical, and other frequencies do not show the same effect, so we conclude that they can not come from instrumental effects. We also checked the well-behaved spectral window to see whether the first side lobe is able to cause any trouble in combination with ${f'}$, but this possibility can be safely excluded. The frequency table of the star is shown in Appendix~\ref{app652}.

\subsection{CoRoT RRd star}

\noindent CoRoT\,0101368{\bf 812}: While the original {\it Kepler} field does not 
contain any classical double-mode RR\,Lyrae stars, this RRd star has been found in CoRoT's LRc01 run and analyzed by \cite{chadid2012}. The period of the fundamental mode is 0.4880408\,d, while that of the first overtone is 0.3636016\,d. The period ratio is 0.7450 and the amplitude ratio ${A_1/A_0}$ is 2.7055, the first overtone being the dominant mode. We note in passing that RR\,Lyrae, the eponym of its class is an example of a modulated RR\,Lyrae pulsating in the fundamental mode and exhibiting the first overtone mode in addition \citep{molnar2012}, with an exceedingly tiny amplitude (non-classical double-mode RR\,Lyrae).

It is of interest to investigate whether classical double-mode RR\,Lyrae show the period doubling, because of the lack of RRds in the original $Kepler$ field. Based on the CoRoT data we see no half-integer frequencies in the Fourier-spectrum of {\tt 812} down to 0.07\,mmag. We mention here that MOST observations of AQ\,Leo, another classical RRd star do not show PD frequencies, either \citep{gruberbauer2007}.

We examined the additional frequency $f' = {\rm 4.4786159\,d^{-1}}$ \citep{chadid2012} (${f_2}$ in the original paper) and the result is presented in the upper right panel of Fig.\ref{f2var}. The figure was made after pre-whitening with the fundamental mode ${f_0}$ (and its harmonics) the first overtone ${f_1}$ (along with all harmonics) and all the visible linear combinations of these. The ${f'}$ periodicity shows a clear amplitude variation. We tested that the same variation is present in the original (un-prewhitened) spectrum as well, so the effect is not caused by the pre-whitening process. We also checked that the observed change is not caused by any variation of the ${f_0}$ or ${f_1}$ themselves, or an interaction with the spectral window side-lobes.

\subsection{Brief summary of the results}

After discussing individual objects it is worth summarizing the most important results:

We discovered signs of period doubling in four CoRoT RR\,Lyrae stars by re-analyzing their light curves. These are {\tt 962}, {\tt 793}, {\tt 544} and {\tt 434}. Although these stars were known before, the presence of period doubling has not been recognized. These objects are all Blazhko RRab stars showing either characteristic alternating maxima (or pulsating cycles) in the time domain, or half-integer frequencies in between the dominant pulsation mode and its harmonics \citep{szabo2010}, or both (see Table~\ref{tab4} for a summary). 

In Tab.~\ref{tab4} we list additional frequencies reported here for the first time, as well as a few previously found ones. These frequencies can be interpreted as 
\begin{itemize}
{\item the second radial overtone in Blazhko RRab stars (or at least frequencies around its expected location with a frequency ratio ${\rm \approx 0.58}$) in case of {\tt 962}, {\tt 793} and {\tt 434}.
Here we investigated only the one belonging to {\tt 793} in detail,}
{\item the well-known frequency with a frequency ratio of $0.61$ most probably corresponding to a nonradial mode in RRc and RRd stars: {\tt 241}, {\tt 652}, {\tt 812} (we use the notation ${f_X}$ throughout this work) , and}
{\item other nonradial modes in  {\tt 962}, {\tt 544} and {\tt 434}.}
\end{itemize}
We emphasize that the list in Tab.~\ref{tab4} is by no means exhaustive or complete, since we do not investigate and discuss all the significant additional frequencies here, only the most prominent ones. Interestingly, based on this work and \cite{nemec2011}, non-Blazhko RRab stars do not show additional periodicities down to the exquisitely low amplitude limits provided of CoRoT and {\it Kepler}. 

Another interesting conclusion of this work is that in many cases the amplitudes of additional frequencies vary over time on shorter time scales than the typical length of the CoRoT runs, irrespective of their nature (HIF, second radial overtone, or hypothesized nonradial modes). We found clear temporal variability of the amplitudes of a frequency peak close to the second radial overtone of {\tt 793}, the probable nonradial modes with a frequency ratio of 0.61 in {\tt 241} (RRc) and {\tt 812} (RRd), and other possibly nonradial modes in {\tt 544} and {\tt 434}. Variability is also obvious in the structure of the frequency peak $f'= {\rm 5.82484\,d^{-1}}$ in {\tt 652} (another RRc, $f'/f= {\rm 0.615}$). Only marginal evidence is found for temporal variability of the amplitudes of a possibly nonradial mode and the second radial overtone in the Blazhko-modulated RRab stars {\tt 962} and {\tt 363}, respectively.

\section{Discussion and conclusions}\label{disc}

\subsection{Period doubling}

\begin{table*}
\begin{center}
\caption{Summary of the main results: period doubling search and additional frequencies that were investigated in detail along with their most probable identification.  '$<$' denotes upper amplitude limit. For the only RRd star in the sample half-integer frequencies 
corresponding to both radial pulsation modes were searched for.}
\begin{tabular}{ccclc|lccc}
\hline\hline
CoRoT ID & Ampl. of ${\rm 1.5}f_0$ & alternating & Type & PD & ${f_0}$ or ${f_1}$ & additional freq. & frequency ratio & identification \\
         & [mmag]                 & maxima      &  & yes / no  & [${\rm d^{-1}}$]    &  ${f'}$ [${\rm d^{-1}}$]     & ${f/f'}$ &      \\
\hline
0101370{\bf 131} & $<0.30$ & n & RRab    & n & ${\rm 1.61464}$  & & \\
0101315{\bf 488} & $<0.10$ & n & RRab    & n & ${\rm 2.06057}$  & & \\
0103800{\bf 818} & $<0.07$ & n & RRab    & n & ${\rm 2.14622}$  & & \\
0104315{\bf 804} & $<0.15$ & n & RRab    & n & ${\rm 1.38538}$  & &  \\
0100689{\bf 962} & $ 3.21$ & y & RRab Bl & {\bf y} & ${\rm 2.80902}$  & 4.03265   & 0.697 & nr \\
0101128{\bf 793}& $ 2.44$ & y & RRab Bl & {\bf y} & ${\rm 2.11895}$  & 3.63088   & 0.584 & O2 \\
0100881{\bf 648}& $<0.10$ & n & RRab Bl & ${\rm n^{b}}$ & ${\rm 1.64694}$ & & \\
0101503{\bf 544}& $<0.05$ & y & RRab Bl & ${\rm {\bf y}^{b}}$ & ${\rm 1.65266}$  & 2.38929*  & 0.692 & nr	\\	    
0105288{\bf 363}& $<0.35$ & n & RRab Bl & n & ${\rm 1.76230}$  & 2.98400   & 0.591 & O2 	\\	     
0103922{\bf 434}& $<0.30$ & y & RRab Bl & {\bf y} & ${\rm 1.84712}$  & 2.61220*  & 0.707 & nr 	\\
0105036{\bf 241}& $<0.20$ & n & RRc     & n & ${\rm 2.68153}$  & 4.37783*  & 0.613 & ${\rm f_X}$\\
0105735{\bf 652}& $<0.13$ & n & RRc     & n & ${\rm 3.58218}$  & 5.82484*  & 0.615 & ${\rm f_X}$	\\
0101368{\bf 812}& $<0.20$ & n & RRd ${\rm f_0}$    & n & ${\rm 2.04901}$  &     -     & - & -	\\
                & $<0.15$ & n & RRd ${\rm f_1}$    & n & ${\rm 2.75026}$  & 4.47862   & 0.614 & ${\rm f_X}$	\\
\hline
\end{tabular} 
\tablefoot{ The superscript 'b' denotes blended pulsators, a configuration that  
prevents detection of low amplitude features. Identifications: O2: second radial overtone, nr: nonradial mode, fx: ubiquitous frequency in RRc stars with a frequency ratio $\approx$ 0.61 \citep{moskalik2014}. '*' denotes new discoveries, other frequencies were reported in previous publications, see Tab.~\ref{tab1} for the relevant references.}
\label{tab4}
\end{center}
\end{table*}

We confirm earlier results \citep{szabo2010} that period doubling occurs only in 
Blazhko-modulated RR\,Lyrae stars, and none of the non-modulated RRab stars shows this phenomenon. In this work we added four non-modulated RRab stars to the list of RR\,Lyrae not showing PD with CoRoT's precision. 

Although the amplitude of the HIFs is empirically expected to be highest in  the [${f_0}$;${2f_0}$] frequency interval, we nevertheless checked the whole frequency spectrum of each objects when searching for the PD. Indeed, the HIFs are significant up to ${\rm 11/2} f_0$ in case of {\tt 962} and ${\rm 17/2} f_0$ in {\tt 793}. In two cases ({\tt 544} and {\tt 434}) only the alternating maxima betray the presence of PD. In {\tt 434} the reason may be twofold: the PD is rather weak and it is present only for a very brief time period. {\tt 544} is heavily blended, and we had not expected to see anything in the frequency spectrum beyond a few harmonics of the fundamental mode and some modulation side peaks. On the contrary, with CoRoT data we were able to uncover PD in this blended Blazhko\,RR Lyrae star (its total amplitude is only about $0\dotm02$), nicely demonstrating the capabilities of high duty-cycle, extreme-precision space photometry. We checked the frequency spectrum of {\tt 544} and {\tt 434} to see whether any single additional frequency or a combination of those can cause the conspicuous alternation of consecutive maxima. It turned out that the amplitudes of these frequencies are much lower than what would be required to cause the observed effect, thus we conclude that the hypothesis of the presence of PD is plausible in these cases. 

As seen in the $Kepler$ RR\,Lyrae sample, the strength of period doubling is changing with time in the CoRoT PD stars as well, and in most cases it crops up only temporarily for a few pulsational  cycles. For the estimation of the occurrence rate of the period doubling effect in Blazhko and possible non-Blazhko stars it is very important to monitor RR\,Lyrae stars with high-precision photometry (preferably from space) for a long time (several Blazhko-modulation periods, meaning several months or years) uninterrupted because of the time variability of the phenomenon. CoRoT and {\it Kepler} data sets are ideal for this purpose, because both missions delivered many months- and years-long high-precision, uninterrupted observations, respectively. The four stars with PD out of six modulated RRab targets is entirely consistent with the results of \citep{benko2014} who found PD in nine out of 15 Blazhko-modulated $Kepler$ stars. This can be augmented to 10/16, if RR\,Lyrae, the prototype is accounted for \citep{szabo2010}. This means that roughly two Blazhko-modulated RRab stars out of three show the period-doubling phenomenon. One is tempted to speculate that because of the finite length of space data (60-150 days in case of CoRoT, 4 years for $Kepler$) and the temporary presence of PD it is possible that the occurrence rate can be higher. To take this argument to the extreme, it is possible that all Blazhko-modulated stars would show period doubling if long enough times series data were taken. 

The connection between the Blazhko modulation and the period doubling is obvious. 
The high occurrence rate of PD in Blazhko-modulated stars stresses the possible underlying 
physical connection of the two dynamical phenomena. Indeed, the 9:2 resonance between the fundamental mode and the ${\rm 9^{th}}$ overtone which explains the period doubling \citep{szabo2010,kollath2011}, might be the culprit for causing the Blazhko modulation itself, as well \citep{bk11}. Given that the resonance paradigm is partly based on hydrodynamic models and partly on the successful and simple amplitude equation formalism, in our opinion currently this is the most plausible explanation for the century-old Blazhko enigma. It is backed up by the results of this paper. 

Interestingly, alternating cycles are found in other pulsating star types, such as RV\,Tauri of which the alternating deep and shallow minima are characteristic and also in BL\,Her stars \citep{smolec2012}. Surprisingly, white dwarfs may show alternating cycles, as well \citep{paparo2013}. Although in some cases this may be a result of an interaction of several independent modes in the frequency spectrum, at least in one case a period doubling bifurcation event seems to be well-documented \citep{goupil1988}, showing close resemblance of RR\,Lyrae star pulsational dynamics \citep{kollath2011}. 

\subsection{Time-dependent additional frequencies}

With the thorough analysis of CoRoT RR\,Lyrae data we confirm earlier emerging trends that additional periodicities are ubiquitous in RR\,Lyrae stars. RRd stars were the first type of objects where frequencies not fitting in the radial eigenspectrum were found \citep{gruberbauer2007}, but soon - with the advent of regular space photometric observations - Blazhko-modulated RRab  \citep{chadid2010,guggenberger2012} and RRc stars \citep{moskalik2013,moskalik2014} followed. It is especially interesting that most of the RRc stars that were observed from space show periodicities with a frequency ratio of ${f_1/f'} {\rm \approx 0.61}$ with the first radial overtone mode \citep{moskalik2014}. These additional frequencies seem to be present in all stars in our sample showing normal amplitude first overtone pulsation, \ie RRc and RRd stars (see Tab.~\ref{tab4}, where we identified them with ${f_X}$). The most plausible explanation for their origin is nonradial modes \citep{dziembowski2012}. Interestingly, modulated RRab stars also show various, low-amplitude additional periodicities \citep{benko2010,chadid2010,molnar2012,guggenberger2012}.

In this work we attempted to unveil the temporal behavior of the ubiquitously seen additional, low-amplitude frequencies that are usually attributed to nonradial oscillations by the virtue of the extended coverage and high duty cycle provided by CoRoT. We found that in almost all cases, where the brightness of the star, the data coverage and the crowdedness of the frequency spectrum allowed a detailed analysis, these frequencies showed amplitude variation over time, followed by a variation of the shape of the frequency peak. The structure of the peaks already suggests temporal variability of the amplitude and/or frequencies, since amplitude and/or frequency modulation manifests itself as side peaks around the corresponding frequency peaks \citep{bszp11}. This is exactly what we see around the HIFs in many cases \citep{szabo2010}, see also Fig.~\ref{962_PD} in this work. Maybe this same variation occur in other additional frequencies, as well. It is possible that the amplitude and the structure of these additional frequency peaks vary because there are close, unresolved frequencies around them. We consider this possibility unlikely based on our experiences with the $Kepler$ RR\,Lyrae data that have better frequency resolution.

Finding a physical explanation for ubiquitous time variability of the additional frequencies, which at the same time does not affect the dominant pulsational mode is challenging. We are tempted to think that the additional frequencies might be  nonradial modes. {\bf (i)} Maybe these modes are not self-excited, and are only present when they experience resonances. That would explain nonstationarity, if resonance conditions governed by the stellar structure are not always met. This might be the case in Blazhko-modulated stars, where the period and stellar structure also varies (quasi)periodically \citep{jurcsik2009b,sodor2009,szabo2010,kollath2011}. The typical timescale of the variation we found is several tens of days, although the variation may be seriously under-sampled because of the attainable frequency resolution. This might be compared to typical theoretical mode growth rates or timescale of interactions, as we see in stars showing period doubling \citep{kollath2011}. Anyhow, the variation is  most probably a consequence of the 
interaction with the large-scale, long-term modulation, which may be a plausible explanation for the observed phenomena. However, this explanation is challenged by the fact that the amplitude and hence non-linearity in RRc stars is much lower than in RRab stars, still the same variation occur in these stars, as well. {\bf (ii)} If not only one, but several closely-spaced nonradial modes are excited \citep{vanhoolst1998}, they can interact causing a complex behavior in the frequency space. In this case not only the time scale of the excitation, but also that of the nonlinear interaction becomes relevant. While the found variations are quite diverse, the magnitude and the timescale of the newly found variations of the additional frequencies suggest a common origin. {\bf (iii)} Rotational splitting is less likely, since we would not expect temporal variation unless other mechanism is at work. Invoking some other modes might come to the rescue if further nonlinear interactions are assumed, hence creating a complex, unresolvable (even with continuous CoRoT observations) pattern in the frequency domain. Detailed investigation of these mechanisms is beyond the scope of this paper, but we may conclude that the discovery of these variations leads us to an unexplored territory of fine details of RR\,Lyrae pulsation. 

\subsection{On the interpretation of the frequency spectra of Blazhko RRab stars}

In \citet{chadid2010} and \citet{poretti2010} the frequency spectrum of the Blazhko-modulated {\tt 962} and {\tt 793} were interpreted as the result of the presence of independent, additional frequencies (presumably nonradial modes) and their combinations besides the well-known pulsational frequency, harmonics, modulation multiplets, the modulation frequency itself and its harmonics. Here, we offer a simpler explanation, 
since the presence of half-integer frequencies, as a result of the period doubling 
removes one independent frequency together with all the related combination 
terms.

In many Blazhko RR\,Lyrae stars, peaks appear around the expected frequencies of the second (and first) radial overtone \citep{benko2010}. This is the case for {\tt 962} and {\tt 793}, as well. 
Here we identify the frequencies with a frequency ratio of 0.58 with the radial second overtone, while keeping in mind that the excitation of nonradial modes in the vicinity of radial overtones is also predicted and plausible \citep{dziembowski1977,vanhoolst1998}. 

Other significant frequencies were found by previous studies and this work (see in Tab.~\ref{tab4}. denoted by `nr') which do not fit the above described picture. Their frequency ratio (${f_0/f'}$) are about 0.7. As we mentioned it in \cite{BSz2014} these frequencies can also be interpreted as linear combination: ${\rm 2}(f_2-f_0)$. Indeed, the components and the simplest combination (${f_0}$, ${f_2}$ and ${f_2-f_0}$) of these frequencies are detectable for all Blazhko RR\,Lyrae stars except for the heavily blended {\tt 544} and {\tt 648} . 

The linear combination assumption simplifies the mathematical description, but is it 
a plausible physical explanation? Do these frequencies belong to radial modes? 
Not necessarily, since the amplitudes seem to contradict this scenario. For example, the amplitudes of ${f_2-f_0}$ and ${f_2}$ in {\tt 962} are about three times smaller than the amplitude of ${\rm 2}(f_2-f_0)$. Such a behaviour is highly unlikely for simple linear combinations. It is possible however, that ${f_2}$ and ${f_0}$ excite a nonradial mode (${f_{nr}}$) through a three-mode resonance ${f_{nr} {\rm \approx 2}(f_2-f_0)}$, in which case the amplitude of the excited mode can be higher. 
Similar effect have been detected for $\delta$\,Scuti stars, B,A,F stars and a peculiar 
roAp star \citep{balona2013,breger2014,bm2014}. All of these papers suggest the presence of coupled nonradial modes showing quasi-periodic amplitude and frequency variations similarly to our presented results. Further developments of multi-dimension hydrocodes, such as \citet{geroux2011} and \citet{mundprecht2013} will eventually make possible to test this scenario. 
\section{Summary}\label{sum}

The main results of this work can be summarized as follows:

\begin{itemize}
\item{The most comprehensive collection of CoRoT RR\,Lyrae variables is presented to date, including new discoveries. The sample consists of all RR\,Lyrae sub-types: Blazhko and non-modulated RRab stars, two RRc and one RRd star. We publish epochs, periods, frequency tables and phased light curves for those object that have not been analyzed earlier.}

\item{Thanks to the CoRoT high cadence observations, we could detect 56 harmonics of the pulsation frequency corresponding the fundamental mode in the case of non-modulated {\tt 818} RRab star. To our knowledge this is the highest number of observed Fourier-terms describing an RR\,Lyrae light curve, hence the most precise one that will serve as a benchmark for model computations. }

\item{Period doubling is detected in CoRoT Blazhko RRab stars for the first time. We discovered brief sections of alternating maxima typical of PD effect in 4 CoRoT RR\,Lyrae stars. It means that two out of three modulated RR\,Lyrae shows this dynamical phenomenon at least temporarily, in accordance with the $Kepler$ RR\,Lyrae statistics. Given the usually short time intervals where PD is detectable, the percentage can be even higher. The strong correlation of the PD occurrence with the Blazhko phenomenon and the fact that no PD was detected in non-modulated CoRoT and $Kepler$ RR\,Lyrae suggests a causal relation, such as the 'resonance paradigm' proposed by \citet{bk11}. In some cases the presence of PD offers a simpler explanation of the frequency spectrum of Blazhko-modulated RR\,Lyrae.}

\item{Our work corroborates those recently found trends, that additional frequencies (most probably higher radial overtones and nonradial modes) are ubiquitous in all sub-types of RR\,Lyrae stars (RRc, RRd, Blazhko RRab), except the non-modulated RRab pulsators. This is an extremely strict rule, since none of the non-Blazhko RRab stars observed by {\it Kepler} and CoRoT shows any additional frequency peaks beyond the dominant pulsational mode and its harmonics, while all the other types - except some blended objects - with high-precision space-based photometric observations do show this feature. If the additional frequencies proved to be nonradial modes, then we anticipate that asteroseismology of RR\,Lyrae stars should be feasible in the near future.}

\item{We analyzed the temporal variability of additional frequencies for the first time in all sub-types of RR\,Lyrae stars based on the CoRoT sample. The amplitude or the shape of these frequencies vary in time in most cases where we could draw firm conclusions. This variability can be connected to the Blazhko-cycle in modulated stars, much like the half-integer frequencies signaling the period doubling are strongly variable. A physical explanation in this case may be the changes of the stellar structure during the Blazhko cycle, and the consequent close or near-miss encounters with different resonances between radial and/or nonradial modes. That would explain the temporal excitation of nonradial modes \citep{kollath2011}. In non-modulated stars, such as RRd and RRc stars, however, a different mechanism should be at work, and it is not clear at this point whether a common mechanism can explain the temporal variability of additional frequencies in all RR\,Lyrae sub-types. Through investigations similar to this work we may get closer to the understanding of the excitation mechanism and origin of these periodicities.} 

\end{itemize}

Future high-precision photometric missions will multiply the number of interesting 
RR\,Lyrae stars to be investigated in detail. NASA's continuing Kepler Mission (dubbed K2), \citep{howell2014}, TESS \citep{ricker2014}, and PLATO \citep{rauer2013} will provide hundreds-to-thousands continuous {\bf of} RR\,Lyrae light curves spanning from a few weeks (TESS), to couple of months (K2), to several years (PLATO) coverage. Prospective data sets from upcoming missions will shed new light on the occurrence of the Blazhko effect, period doubling, additional radial and nonradial modes and other dynamical phenomena as a function of a broad range of stellar parameters. In the light of these prospects, we are entering a golden era of classical variable stars, and our observational data presented in this work will provide ample examples for detailed theoretical analysis. 

\begin{acknowledgements}
This research has made use of the ExoDat database, operated
at LAM-OAMP, Marseille, France, on behalf of the CoRoT/Exoplanet
programme. RSz, MP, and JMB acknowledge the support of the ESA PECS project 
No.~4000103541/11/NL/KML. This project has been supported by the 
Hungarian OTKA grant K83790 and the European Community's Seventh 
Framework Programme (FP7/2007–2013) under grant agreements no. 312844 
(SPACEINN), no. 269194 (IRSES/ASK) and ERC grant agreement no. 338251 (StellarAges). RSz wishes to thank the support from the J\'anos Bolyai Research Scholarship of the Hungarian Academy of Sciences. WW was supported by the Austrian Science Fonds (FWF P22691-N16). The authors thank \'Akos Gy\H orffy, P\'eter P\'apics and L\'aszl\'o Moln\'ar for their help with the ground-based observations. 

\end{acknowledgements}

\bibliographystyle{aa} 
\bibliography{rszabo}

\begin{thebibliography}{55}
\expandafter\ifx\csname natexlab\endcsname\relax\def\natexlab#1{#1}\fi

\bibitem[{{Affer} {et~al.}(2012){Affer}, {Micela}, {Favata}, \&
  {Flaccomio}}]{affer2012}
{Affer}, L., {Micela}, G., {Favata}, F., \& {Flaccomio}, E. 2012, \mnras, 424,
  11

\bibitem[{{Baglin} {et~al.}(2006){Baglin}, {Auvergne}, {Boisnard}, {Lam-Trong},
  {Barge}, {Catala}, {Deleuil}, {Michel}, \& {Weiss}}]{baglin2006}
{Baglin}, A., {Auvergne}, M., {Boisnard}, L., {et~al.} 2006, in COSPAR Meeting,
  Vol.~36, 36th COSPAR Scientific Assembly, 3749

\bibitem[{{Balona} {et~al.}(2013){Balona}, {Catanzaro}, {Crause}, {Cunha},
  {Gandolfi}, {Hatzes}, {Kabath}, {Uytterhoeven}, \& {De Cat}}]{balona2013}
{Balona}, L.~A., {Catanzaro}, G., {Crause}, L., {et~al.} 2013, \mnras, 432,
  2808

\bibitem[{{Benk{\H o}} {et~al.}(2010){Benk{\H o}}, {Kolenberg}, {Szab{\'o}},
  {Kurtz}, {Bryson}, {Bregman}, {Still}, {Smolec}, {Nuspl}, {Nemec},
  {Moskalik}, {Kopacki}, {Koll{\'a}th}, {Guggenberger}, {di Criscienzo},
  {Christensen-Dalsgaard}, {Kjeldsen}, {Borucki}, {Koch}, {Jenkins}, \& {van
  Cleve}}]{benko2010}
{Benk{\H o}}, J.~M., {Kolenberg}, K., {Szab{\'o}}, R., {et~al.} 2010, \mnras,
  409, 1585

\bibitem[{{Benk{\H o}} {et~al.}(2014){Benk{\H o}}, {Plachy}, {Szab{\'o}},
  {Moln{\'a}r}, \& {Koll{\'a}th}}]{benko2014}
{Benk{\H o}}, J.~M., {Plachy}, E., {Szab{\'o}}, R., {Moln{\'a}r}, L., \&
  {Koll{\'a}th}, Z. 2014, \apjs, 213, 31

\bibitem[{{Benk{\H o}} \& {Szab{\'o}}(2014)}]{BSz2014}
{Benk{\H o}}, J.~M. \& {Szab{\'o}}, R. 2014, in IAU Symposium, Vol. 301, IAU
  Symposium, ed. J.~A. {Guzik}, W.~J. {Chaplin}, G.~{Handler}, \&
  A.~{Pigulski}, 383--384

\bibitem[{{Benk{\H o}} {et~al.}(2011){Benk{\H o}}, {Szab{\'o}}, \&
  {Papar{\'o}}}]{bszp11}
{Benk{\H o}}, J.~M., {Szab{\'o}}, R., \& {Papar{\'o}}, M. 2011, \mnras, 417,
  974

\bibitem[{{Borucki} {et~al.}(2010){Borucki}, {Koch}, {Basri}, {Batalha},
  {Brown}, {Caldwell}, {Caldwell}, {Christensen-Dalsgaard}, {Cochran},
  {DeVore}, {Dunham}, {Dupree}, {Gautier}, {Geary}, {Gilliland}, {Gould},
  {Howell}, {Jenkins}, {Kondo}, {Latham}, {Marcy}, {Meibom}, {Kjeldsen},
  {Lissauer}, {Monet}, {Morrison}, {Sasselov}, {Tarter}, {Boss}, {Brownlee},
  {Owen}, {Buzasi}, {Charbonneau}, {Doyle}, {Fortney}, {Ford}, {Holman},
  {Seager}, {Steffen}, {Welsh}, {Rowe}, {Anderson}, {Buchhave}, {Ciardi},
  {Walkowicz}, {Sherry}, {Horch}, {Isaacson}, {Everett}, {Fischer}, {Torres},
  {Johnson}, {Endl}, {MacQueen}, {Bryson}, {Dotson}, {Haas}, {Kolodziejczak},
  {Van Cleve}, {Chandrasekaran}, {Twicken}, {Quintana}, {Clarke}, {Allen},
  {Li}, {Wu}, {Tenenbaum}, {Verner}, {Bruhweiler}, {Barnes}, \&
  {Prsa}}]{borucki2010}
{Borucki}, W.~J., {Koch}, D., {Basri}, G., {et~al.} 2010, Science, 327, 977

\bibitem[{{Breger}(2014)}]{breger2014}
{Breger}, M. 2014, in IAU Symposium, Vol. 301, IAU Symposium, ed. J.~A.
  {Guzik}, W.~J. {Chaplin}, G.~{Handler}, \& A.~{Pigulski}, 93--100

\bibitem[{{Breger} \& {Montgomery}(2014)}]{bm2014}
{Breger}, M. \& {Montgomery}, M.~H. 2014, \apj, 783, 89

\bibitem[{{Buchler} \& {Koll{\'a}th}(2011)}]{bk11}
{Buchler}, J.~R. \& {Koll{\'a}th}, Z. 2011, \apj, 731, 24

\bibitem[{{Chadid}(2012)}]{chadid2012}
{Chadid}, M. 2012, \aap, 540, A68

\bibitem[{{Chadid} {et~al.}(2010){Chadid}, {Benk{\H o}}, {Szab{\'o}},
  {Papar{\'o}}, {Chapellier}, {Kolenberg}, {Poretti}, {Bono}, {Le Borgne},
  {Trinquet}, {Artemenko}, {Auvergne}, {Baglin}, {Debosscher}, {Grankin},
  {Guggenberger}, \& {Weiss}}]{chadid2010}
{Chadid}, M., {Benk{\H o}}, J.~M., {Szab{\'o}}, R., {et~al.} 2010, \aap, 510,
  A39

\bibitem[{{Chadid} {et~al.}(2011){Chadid}, {Perini}, {Bono}, {Auvergne},
  {Baglin}, {Weiss}, \& {Deboscher}}]{chadid2011}
{Chadid}, M., {Perini}, C., {Bono}, G., {et~al.} 2011, \aap, 527, A146

\bibitem[{{Chevalier} \& {Ilovaisky}(1991)}]{chevalier1991}
{Chevalier}, C. \& {Ilovaisky}, S.~A. 1991, \aaps, 90, 225

\bibitem[{{COROT Team}(2014)}]{corot2014}
{COROT Team}. 2014, VizieR Online Data Catalog, 1, 2028

\bibitem[{{Debosscher} {et~al.}(2009){Debosscher}, {Sarro}, {L{\'o}pez},
  {Deleuil}, {Aerts}, {Auvergne}, {Baglin}, {Baudin}, {Chadid}, {Charpinet},
  {Cuypers}, {De Ridder}, {Garrido}, {Hubert}, {Janot-Pacheco}, {Jorda},
  {Kaiser}, {Kallinger}, {Kollath}, {Maceroni}, {Mathias}, {Michel}, {Moutou},
  {Neiner}, {Ollivier}, {Samadi}, {Solano}, {Surace}, {Vandenbussche}, \&
  {Weiss}}]{debosscher2009}
{Debosscher}, J., {Sarro}, L.~M., {L{\'o}pez}, M., {et~al.} 2009, \aap, 506,
  519

\bibitem[{{Dziembowski}(1977)}]{dziembowski1977}
{Dziembowski}, W. 1977, \actaa, 27, 95

\bibitem[{{Dziembowski}(2012)}]{dziembowski2012}
{Dziembowski}, W.~A. 2012, \actaa, 62, 323

\bibitem[{{Dziembowski} \& {Mizerski}(2004)}]{dziembowski2004}
{Dziembowski}, W.~A. \& {Mizerski}, T. 2004, \actaa, 54, 363

\bibitem[{{Geroux} \& {Deupree}(2011)}]{geroux2011}
{Geroux}, C.~M. \& {Deupree}, R.~G. 2011, \apj, 731, 18

\bibitem[{{Goupil} {et~al.}(1988){Goupil}, {Auvergne}, \&
  {Baglin}}]{goupil1988}
{Goupil}, M.~J., {Auvergne}, M., \& {Baglin}, A. 1988, \aap, 196, L13

\bibitem[{{Gruberbauer} {et~al.}(2007){Gruberbauer}, {Kolenberg}, {Rowe},
  {Huber}, {Matthews}, {Reegen}, {Kuschnig}, {Cameron}, {Kallinger}, {Weiss},
  {Guenther}, {Moffat}, {Rucinski}, {Sasselov}, \& {Walker}}]{gruberbauer2007}
{Gruberbauer}, M., {Kolenberg}, K., {Rowe}, J.~F., {et~al.} 2007, \mnras, 379,
  1498

\bibitem[{{Guggenberger} {et~al.}(2011){Guggenberger}, {Kolenberg},
  {Chapellier}, {Poretti}, {Szab{\'o}}, {Benk{\H o}}, \&
  {Papar{\'o}}}]{guggenberger2011}
{Guggenberger}, E., {Kolenberg}, K., {Chapellier}, E., {et~al.} 2011, \mnras,
  415, 1577

\bibitem[{{Guggenberger} {et~al.}(2012){Guggenberger}, {Kolenberg}, {Nemec},
  {Smolec}, {Benk{\H o}}, {Ngeow}, {Cohen}, {Sesar}, {Szab{\'o}}, {Catelan},
  {Moskalik}, {Kinemuchi}, {Seader}, {Smith}, {Tenenbaum}, \&
  {Kjeldsen}}]{guggenberger2012}
{Guggenberger}, E., {Kolenberg}, K., {Nemec}, J.~M., {et~al.} 2012, \mnras,
  424, 649

\bibitem[{{Howell} {et~al.}(2014){Howell}, {Sobeck}, {Haas}, {Still},
  {Barclay}, {Mullally}, {Troeltzsch}, {Aigrain}, {Bryson}, {Caldwell},
  {Chaplin}, {Cochran}, {Huber}, {Marcy}, {Miglio}, {Najita}, {Smith},
  {Twicken}, \& {Fortney}}]{howell2014}
{Howell}, S.~B., {Sobeck}, C., {Haas}, M., {et~al.} 2014, ArXiv e-prints

\bibitem[{{Jurcsik} {et~al.}(2008){Jurcsik}, {S{\'o}dor}, {Hurta},
  {V{\'a}radi}, {Szeidl}, {Smith}, {Henden}, {D{\'e}k{\'a}ny}, {Nagy},
  {Posztob{\'a}nyi}, {Szing}, {Vida}, \& {Vityi}}]{jurcsik2008}
{Jurcsik}, J., {S{\'o}dor}, {\'A}., {Hurta}, Z., {et~al.} 2008, \mnras, 391,
  164

\bibitem[{{Jurcsik} {et~al.}(2009{\natexlab{a}}){Jurcsik}, {S{\'o}dor},
  {Szeidl}, {Hurta}, {V{\'a}radi}, {Posztob{\'a}nyi}, {Vida}, {Hajdu}, {K{\H
  o}v{\'a}ri}, {Nagy}, {Moln{\'a}r}, \& {Belucz}}]{jurcsik2009a}
{Jurcsik}, J., {S{\'o}dor}, {\'A}., {Szeidl}, B., {et~al.} 2009{\natexlab{a}},
  \mnras, 400, 1006

\bibitem[{{Jurcsik} {et~al.}(2009{\natexlab{b}}){Jurcsik}, {S{\'o}dor},
  {Szeidl}, {Koll{\'a}th}, {Smith}, {Hurta}, {V{\'a}radi}, {Henden},
  {D{\'e}k{\'a}ny}, {Nagy}, {Posztob{\'a}nyi}, {Szing}, {Vida}, \&
  {Vityi}}]{jurcsik2009b}
{Jurcsik}, J., {S{\'o}dor}, {\'A}., {Szeidl}, B., {et~al.} 2009{\natexlab{b}},
  \mnras, 393, 1553

\bibitem[{{Jurcsik} {et~al.}(2006){Jurcsik}, {Szeidl}, {S{\'o}dor},
  {D{\'e}k{\'a}ny}, {Hurta}, {Posztob{\'a}nyi}, {Vida}, {V{\'a}radi}, \&
  {Szing}}]{jurcsik2006}
{Jurcsik}, J., {Szeidl}, B., {S{\'o}dor}, {\'A}., {et~al.} 2006, \aj, 132, 61

\bibitem[{{Kolenberg} {et~al.}(2010){Kolenberg}, {Szab{\'o}}, {Kurtz},
  {Gilliland}, {Christensen-Dalsgaard}, {Kjeldsen}, {Brown}, {Benk{\H o}},
  {Chadid}, {Derekas}, {Di Criscienzo}, {Guggenberger}, {Kinemuchi}, {Kunder},
  {Koll{\'a}th}, {Kopacki}, {Moskalik}, {Nemec}, {Nuspl}, {Silvotti}, {Suran},
  {Borucki}, {Koch}, \& {Jenkins}}]{kolenberg2010}
{Kolenberg}, K., {Szab{\'o}}, R., {Kurtz}, D.~W., {et~al.} 2010, \apjl, 713,
  L198

\bibitem[{{Koll\'ath}(1990)}]{kollath1990}
{Koll\'ath}, Z. 1990, Konkoly Observatory Occasional Technical Notes, 1, 1

\bibitem[{{Koll{\'a}th} {et~al.}(2011){Koll{\'a}th}, {Moln{\'a}r}, \&
  {Szab{\'o}}}]{kollath2011}
{Koll{\'a}th}, Z., {Moln{\'a}r}, L., \& {Szab{\'o}}, R. 2011, \mnras, 414, 1111

\bibitem[{{Le Borgne} {et~al.}(2014){Le Borgne}, {Poretti}, {Klotz}, {Denoux},
  {Smith}, {Kolenberg}, {Szab{\'o}}, {Bryson}, {Audejean}, {Buil}, {Caron},
  {Conseil}, {Corp}, {Drillaud}, {de France}, {Graham}, {Hirosawa}, {Klotz},
  {Kugel}, {Loughney}, {Menzies}, {Rodr{\'{\i}}guez}, \&
  {Ruscitti}}]{leborgne2014}
{Le Borgne}, J.~F., {Poretti}, E., {Klotz}, A., {et~al.} 2014, \mnras, 441,
  1435

\bibitem[{{Lenz} \& {Breger}(2005)}]{lenz2005}
{Lenz}, P. \& {Breger}, M. 2005, Communications in Asteroseismology, 146, 53

\bibitem[{{Moln{\'a}r} {et~al.}(2012){Moln{\'a}r}, {Koll{\'a}th}, {Szab{\'o}},
  {Bryson}, {Kolenberg}, {Mullally}, \& {Thompson}}]{molnar2012}
{Moln{\'a}r}, L., {Koll{\'a}th}, Z., {Szab{\'o}}, R., {et~al.} 2012, \apjl,
  757, L13

\bibitem[{{Moskalik}(2013)}]{moskalik2013}
{Moskalik}, P. 2013, in Astrophysics and Space Science Proceedings, Vol.~31,
  Stellar Pulsations: Impact of New Instrumentation and New Insights, ed. J.~C.
  {Su{\'a}rez}, R.~{Garrido}, L.~A. {Balona}, \& J.~{Christensen-Dalsgaard},
  103

\bibitem[{{Moskalik}(2014)}]{moskalik2014}
{Moskalik}, P. 2014, in IAU Symposium, Vol. 301, IAU Symposium, ed. J.~A.
  {Guzik}, W.~J. {Chaplin}, G.~{Handler}, \& A.~{Pigulski}, 249--256

\bibitem[{{Mundprecht} {et~al.}(2013){Mundprecht}, {Muthsam}, \&
  {Kupka}}]{mundprecht2013}
{Mundprecht}, E., {Muthsam}, H.~J., \& {Kupka}, F. 2013, \mnras, 435, 3191

\bibitem[{{Nemec} {et~al.}(2011){Nemec}, {Smolec}, {Benk{\H o}}, {Moskalik},
  {Kolenberg}, {Szab{\'o}}, {Kurtz}, {Bryson}, {Guggenberger}, {Chadid},
  {Jeon}, {Kunder}, {Layden}, {Kinemuchi}, {Kiss}, {Poretti},
  {Christensen-Dalsgaard}, {Kjeldsen}, {Caldwell}, {Ripepi}, {Derekas},
  {Nuspl}, {Mullally}, {Thompson}, \& {Borucki}}]{nemec2011}
{Nemec}, J.~M., {Smolec}, R., {Benk{\H o}}, J.~M., {et~al.} 2011, \mnras, 417,
  1022

\bibitem[{{Papar{\'o}} {et~al.}(2013){Papar{\'o}}, {Bogn{\'a}r}, {Plachy},
  {Moln{\'a}r}, \& {Bradley}}]{paparo2013}
{Papar{\'o}}, M., {Bogn{\'a}r}, Z., {Plachy}, E., {Moln{\'a}r}, L., \&
  {Bradley}, P.~A. 2013, \mnras, 432, 598

\bibitem[{{Papar{\'o}} {et~al.}(2011){Papar{\'o}}, {Chadid}, {Chapellier},
  {Benk{\H o}}, {Szab{\'o}}, {Kolenberg}, {Guggenberger}, {Reg{\'a}ly},
  {Auvergne}, {Baglin}, \& {Weiss}}]{paparo2011}
{Papar{\'o}}, M., {Chadid}, M., {Chapellier}, E., {et~al.} 2011, \aap, 531,
  A135

\bibitem[{{Papar{\'o}} {et~al.}(2009){Papar{\'o}}, {Szab{\'o}}, {Benk{\H o}},
  {Chadid}, {Poretti}, {Kolenberg}, {Guggenberger}, \&
  {Chapellier}}]{paparo2009}
{Papar{\'o}}, M., {Szab{\'o}}, R., {Benk{\H o}}, J.~M., {et~al.} 2009, in
  American Institute of Physics Conference Series, Vol. 1170, American
  Institute of Physics Conference Series, ed. J.~A. {Guzik} \& P.~A. {Bradley},
  240--244

\bibitem[{{Plachy} {et~al.}(2013){Plachy}, {Koll{\'a}th}, \&
  {Moln{\'a}r}}]{plachy2013}
{Plachy}, E., {Koll{\'a}th}, Z., \& {Moln{\'a}r}, L. 2013, \mnras, 433, 3590

\bibitem[{{Poretti} {et~al.}(2010){Poretti}, {Papar{\'o}}, {Deleuil}, {Chadid},
  {Kolenberg}, {Szab{\'o}}, {Benk{\H o}}, {Chapellier}, {Guggenberger}, {Le
  Borgne}, {Rostagni}, {Trinquet}, {Auvergne}, {Baglin}, {Sarro}, \&
  {Weiss}}]{poretti2010}
{Poretti}, E., {Papar{\'o}}, M., {Deleuil}, M., {et~al.} 2010, \aap, 520, A108

\bibitem[{{Rauer} {et~al.}(2013){Rauer}, {Catala}, {Aerts}, {Appourchaux},
  {Benz}, {Brandeker}, {Christensen-Dalsgaard}, {Deleuil}, {Gizon}, {Goupil},
  {G{\"u}del}, {Janot-Pacheco}, {Mas-Hesse}, {Pagano}, {Piotto}, {Pollacco},
  {Santos}, {Smith}, {-C.}, {Su{\'a}rez}, {Szab{\'o}}, {Udry}, {Adibekyan},
  {Alibert}, {Almenara}, {Amaro-Seoane}, {Ammler-von Eiff}, {Asplund},
  {Antonello}, {Ball}, {Barnes}, {Baudin}, {Belkacem}, {Bergemann}, {Bihain},
  {Birch}, {Bonfils}, {Boisse}, {Bonomo}, {Borsa}, {Brand{\~a}o}, {Brocato},
  {Brun}, {Burleigh}, {Burston}, {Cabrera}, {Cassisi}, {Chaplin}, {Charpinet},
  {Chiappini}, {Church}, {Csizmadia}, {Cunha}, {Damasso}, {Davies}, {Deeg},
  {D{\'{i}}az}, {Dreizler}, {Dreyer}, {Eggenberger}, {Ehrenreich},
  {Eigm{\"u}ller}, {Erikson}, {Farmer}, {Feltzing}, {de Oliveira Fialho},
  {Figueira}, {Forveille}, {Fridlund}, {Garc{\'{\i}}a}, {Giommi}, {Giuffrida},
  {Godolt}, {Gomes da Silva}, {Granzer}, {Grenfell}, {Grotsch-Noels},
  {G{\"u}nther}, {Haswell}, {Hatzes}, {H{\'e}brard}, {Hekker}, {Helled},
  {Heng}, {Jenkins}, {Johansen}, {Khodachenko}, {Kislyakova}, {Kley}, {Kolb},
  {Krivova}, {Kupka}, {Lammer}, {Lanza}, {Lebreton}, {Magrin}, {Marcos-Arenal},
  {Marrese}, {Marques}, {Martins}, {Mathis}, {Mathur}, {Messina}, {Miglio},
  {Montalban}, {Montalto}, {Monteiro}, {Moradi}, {Moravveji}, {Mordasini},
  {Morel}, {Mortier}, {Nascimbeni}, {Nelson}, {Nielsen}, {Noack}, {Norton},
  {Ofir}, {Oshagh}, {Ouazzani}, {P{\'a}pics}, {Parro}, {Petit}, {Plez},
  {Poretti}, {Quirrenbach}, {Ragazzoni}, {Raimondo}, {Rainer}, {Reese},
  {Redmer}, {Reffert}, {Rojas-Ayala}, {Roxburgh}, {Salmon}, {Santerne},
  {Schneider}, {Schou}, {Schuh}, {Schunker}, {Silva-Valio}, {Silvotti},
  {Skillen}, {Snellen}, {Sohl}, {Sousa}, {Sozzetti}, {Stello}, {Strassmeier},
  {{\v S}vanda}, {Szab{\'o}}, {Tkachenko}, {Valencia}, {van Grootel},
  {Vauclair}, {Ventura}, {Wagner}, {Walton}, {Weingrill}, {Werner}, {Wheatley},
  \& {Zwintz}}]{rauer2013}
{Rauer}, H., {Catala}, C., {Aerts}, C., {et~al.} 2013, ArXiv e-prints

\bibitem[{{Ricker} {et~al.}(2014){Ricker}, {Winn}, {Vanderspek}, {Latham},
  {Bakos}, {Bean}, {Berta-Thompson}, {Brown}, {Buchhave}, {Butler}, {Butler},
  {Chaplin}, {Charbonneau}, {Christensen-Dalsgaard}, {Clampin}, {Deming},
  {Doty}, {De Lee}, {Dressing}, {Dunham}, {Endl}, {Fressin}, {Ge}, {Henning},
  {Holman}, {Howard}, {Ida}, {Jenkins}, {Jernigan}, {Johnson}, {Kaltenegger},
  {Kawai}, {Kjeldsen}, {Laughlin}, {Levine}, {Lin}, {Lissauer}, {MacQueen},
  {Marcy}, {McCullough}, {Morton}, {Narita}, {Paegert}, {Palle}, {Pepe},
  {Pepper}, {Quirrenbach}, {Rinehart}, {Sasselov}, {Sato}, {Seager},
  {Sozzetti}, {Stassun}, {Sullivan}, {Szentgyorgyi}, {Torres}, {Udry}, \&
  {Villasenor}}]{ricker2014}
{Ricker}, G.~R., {Winn}, J.~N., {Vanderspek}, R., {et~al.} 2014, ArXiv e-prints

\bibitem[{{Skarka}(2014)}]{skarka2014}
{Skarka}, M. 2014, \aap, 562, A90

\bibitem[{{Smolec} {et~al.}(2012){Smolec}, {Soszy{\'n}ski}, {Moskalik},
  {Udalski}, {Szyma{\'n}ski}, {Kubiak}, {Pietrzy{\'n}ski}, {Wyrzykowski},
  {Ulaczyk}, {Poleski}, {Koz{\l}owski}, \& {Pietrukowicz}}]{smolec2012}
{Smolec}, R., {Soszy{\'n}ski}, I., {Moskalik}, P., {et~al.} 2012, \mnras, 419,
  2407

\bibitem[{{S{\'o}dor} {et~al.}(2009){S{\'o}dor}, {Jurcsik}, \&
  {Szeidl}}]{sodor2009}
{S{\'o}dor}, {\'A}., {Jurcsik}, J., \& {Szeidl}, B. 2009, \mnras, 394, 261

\bibitem[{{Sterken} {et~al.}(1987){Sterken}, {Young}, \&
  {Furenlid}}]{sterken1987}
{Sterken}, C., {Young}, A., \& {Furenlid}, I. 1987, \aap, 177, 150

\bibitem[{{Szab{\'o}} {et~al.}(2010){Szab{\'o}}, {Koll{\'a}th}, {Moln{\'a}r},
  {Kolenberg}, {Kurtz}, {Bryson}, {Benk{\H o}}, {Christensen-Dalsgaard},
  {Kjeldsen}, {Borucki}, {Koch}, {Twicken}, {Chadid}, {di Criscienzo}, {Jeon},
  {Moskalik}, {Nemec}, \& {Nuspl}}]{szabo2010}
{Szab{\'o}}, R., {Koll{\'a}th}, Z., {Moln{\'a}r}, L., {et~al.} 2010, \mnras,
  409, 1244

\bibitem[{{Szab{\'o}} {et~al.}(2009){Szab{\'o}}, {Papar{\'o}}, {Benk{\H o}},
  {Chadid}, {Kolenberg}, \& {Poretti}}]{szabo2009}
{Szab{\'o}}, R., {Papar{\'o}}, M., {Benk{\H o}}, J.~M., {et~al.} 2009, in
  American Institute of Physics Conference Series, Vol. 1170, American
  Institute of Physics Conference Series, ed. J.~A. {Guzik} \& P.~A. {Bradley},
  291--293

\bibitem[{{Van Hoolst} {et~al.}(1998){Van Hoolst}, {Dziembowski}, \&
  {Kawaler}}]{vanhoolst1998}
{Van Hoolst}, T., {Dziembowski}, W.~A., \& {Kawaler}, S.~D. 1998, \mnras, 297,
  536

\bibitem[{{Walker} {et~al.}(2003){Walker}, {Matthews}, {Kuschnig}, {Johnson},
  {Rucinski}, {Pazder}, {Burley}, {Walker}, {Skaret}, {Zee}, {Grocott},
  {Carroll}, {Sinclair}, {Sturgeon}, \& {Harron}}]{walker2003}
{Walker}, G., {Matthews}, J., {Kuschnig}, R., {et~al.} 2003, \pasp, 115, 1023

\end{thebibliography}

\begin{appendix}

\section{Frequency content of the new CoRoT RR Lyrae star: 0101315{\bf 488}}\label{app488}

\begin{center}
\begin{table}
\begin{flushleft}
\caption[]{Frequency table of the newly discovered CoRoT RRab star, {\tt 488}.}
\label{table_488}
\begin{tabular}{rrcr}
\hline \hline
\noalign{\smallskip}
ID        & freq. & ampl. & phase   \\
\hline
          & [$\rm d^{-1}$] & [mag] & [rad]   \\
\noalign{\smallskip}
\hline
\noalign{\smallskip}
${f_0}$       &  2.060567  & 0.02902   &    5.282 \\
${2f_0}$      &  4.121135  & 0.01676   &    0.301 \\
${3f_0}$      &  6.181702  & 0.01121   &    1.921 \\
${4f_0}$      &  8.242270  & 0.00783   &    3.485 \\
${5f_0}$      & 10.302837  & 0.00521   &    5.147 \\
${6f_0}$      & 12.363404  & 0.00340   &    0.345 \\
${7f_0}$      & 14.423972  & 0.00258   &    1.832 \\
${8f_0}$      & 16.484539  & 0.00195   &    3.453 \\
${9f_0}$      & 18.545106  & 0.00132   &    5.020 \\
${10f_0}$     & 20.605674  & 0.00082   &    0.236 \\
\noalign{\smallskip}
\hline
\end{tabular}
\end{flushleft}
\end{table}
\end{center}

Table~\ref{table_488} contains the frequencies that we found during the frequency analysis. The star pulsates with a period of 0.485299 days. The following epoch for maxima was found:  

\begin{equation}
$$2454236.8709 \ {\rm HJD} + 0\dotd4853033(15) \cdot {\rm E} $$
\end{equation}

Digits in parentheses denote the uncertainties. Nine harmonics can be found in the frequency spectrum. No additional frequencies were found. 


\section{Frequency table of the new CoRoT RRab star: 0103800{\bf818}}\label{app818}

\begin{center}
\begin{table}
\begin{flushleft}
\caption[]{Frequency table of the non-modulated CoRoT RRab star, {\tt 818}. We refrain from listing the CoRoT orbital frequencies and a large number of residual frequencies around the harmonics.}
\label{table_818}
\begin{tabular}{rrcr}
\hline \hline
\noalign{\smallskip}
ID        & freq. & ampl. & phase   \\
\hline
          & [$\rm d^{-1}$] & [mag] & [rad]   \\
\noalign{\smallskip}
\hline
\noalign{\smallskip}
${f_0}$       &  2.146223  & 0.34509  &    3.195 \\
${2f_0}$      &  4.292446  & 0.17020  &    2.449 \\
${3f_0}$      &  6.438669  & 0.12905  &    1.966 \\ 
${4f_0}$      &  8.584893  & 0.08415  &    1.563 \\   
${5f_0}$      & 10.731116  & 0.06027  &    1.171 \\
${6f_0}$      & 12.877339  & 0.03956  &    0.803 \\
${7f_0}$      & 15.023562  & 0.02415  &    0.333 \\    
${8f_0}$      & 17.169785  & 0.01659  &    6.033 \\    
${9f_0}$      & 19.316008  & 0.01289  &    5.518 \\     
${10f_0}$     & 21.462232  & 0.00980  &    5.128 \\    
${11f_0}$     & 23.608468  & 0.00639  &    4.784 \\  
${12f_0}$     & 25.754666  & 0.00353  &    4.380 \\
${13f_0}$     & 27.900991  & 0.00184  &    3.722 \\
${14f_0}$     & 30.047044  & 0.00102  &    3.146 \\  
${15f_0}$     & 32.193767  & 0.00030  &    2.846 \\
${16f_0}$     & 34.339850  & 0.00051  &    4.970 \\
${17f_0}$     & 36.485549  & 0.00110  &    4.983 \\    
${18f_0}$     & 38.631887  & 0.00139  &    4.690 \\    
${19f_0}$     & 40.777791  & 0.00147  &    4.351 \\    
${20f_0}$     & 42.924600  & 0.00152  &    3.732 \\   
${21f_0}$     & 45.070642  & 0.00158  &    3.280 \\    
${22f_0}$     & 47.216924  & 0.00163  &    2.834 \\   
${23f_0}$     & 49.363091  & 0.00161  &    2.429 \\    
${24f_0}$     & 51.509352  & 0.00156  &    1.989 \\   
${25f_0}$     & 53.655410  & 0.00146  &    1.558 \\
${26f_0}$     & 55.801814  & 0.00139  &    1.085 \\
${27f_0}$     & 57.948115  & 0.00132  &    0.649 \\
${28f_0}$     & 60.094089  & 0.00120  &    0.284 \\
${29f_0}$     & 62.240271  & 0.00110  &    6.150 \\
${30f_0}$     & 64.386660  & 0.00103  &    5.672 \\
${31f_0}$     & 66.532893  & 0.00093  &    5.188 \\
${32f_0}$     & 68.679352  & 0.00086  &    4.709 \\
${33f_0}$     & 70.825334  & 0.00075  &    4.368 \\
${34f_0}$     & 72.971610  & 0.00072  &    3.945 \\
${35f_0}$     & 75.117668  & 0.00063  &    3.607 \\
${36f_0}$     & 77.264339  & 0.00056  &    2.980 \\
${37f_0}$     & 79.410227  & 0.00052  &    2.709 \\
${38f_0}$     & 81.556746  & 0.00048  &    2.243 \\
${39f_0}$     & 83.703001  & 0.00041  &    1.847 \\
${40f_0}$     & 85.849075  & 0.00039  &    1.373 \\
${41f_0}$     & 87.995504  & 0.00034  &    0.962 \\
${42f_0}$     & 90.141755  & 0.00031  &    0.570 \\
${43f_0}$     & 92.287719  & 0.00030  &    0.102 \\
${44f_0}$     & 94.433877  & 0.00027  &    5.992 \\
${45f_0}$     & 96.580393  & 0.00025  &    5.741 \\
${46f_0}$     & 98.727051  & 0.00022  &    4.954 \\
${47f_0}$     & 100.872457 & 0.00020  &    4.753 \\
${48f_0}$     & 103.020019 & 0.00016  &    3.993 \\
${49f_0}$     & 105.164947 & 0.00015  &    4.051 \\
${50f_0}$     & 107.311097 & 0.00015  &    3.692 \\
${51f_0}$     & 109.456036 & 0.00013  &    3.666 \\
${52f_0}$     & 111.604225 & 0.00011  &    2.343 \\
${53f_0}$     & 113.749783 & 0.00008  &    2.633 \\
${54f_0}$     & 115.896231 & 0.00011  &    1.962 \\
${55f_0}$     & 118.041414 & 0.00010  &    1.920 \\
${56f_0}$     & 120.189737 & 0.00010  &    0.893 \\
${57f_0}$     & 122.334629 & 0.00007  &    1.414 \\
\noalign{\smallskip}
\hline
\end{tabular}
\end{flushleft}
\end{table}
\end{center}

Table~\ref{table_818} enumerates the frequencies of the non-modulated RR Lyrae, {\tt 818}. The period of this star is 0.4659348 days. No modulation was found in this RRab star. We found the following epoch for maxima: 

\begin{equation}
$$2455029.3049 \thinspace {\rm HJD} + 0\dotd4659348(6) \cdot {\rm E} $$
\end{equation}

After pre-whitening, 'forests' of peaks remain around the harmonics. These do not show any obvious modulation pattern, so we decided not to list them in Table~\ref{table_818}.

\section{Frequency table of the new CoRoT RR Lyrae star: 0104315{\bf804}}\label{app804}

In Table~\ref{table_804} we give the frequency content of {\tt 804}. The star pulsates with a period of 0.7218221 days. No signs of modulation or additional frequencies were  found in this object. The following epoch for maxima was obtained:  

\begin{equation}
$$2455019.8524 \thinspace {\rm HJD} + 0\dotd7218221(36) \cdot {\rm E} $$
\end{equation}

\begin{center}
\begin{table}
\begin{flushleft}
\caption[]{Frequency table of the non-modulated CoRoT RRab star, {\tt 804}. We list the 
dominant fundamental mode pulsation frequency and its harmonics only.}
\label{table_804}
\begin{tabular}{rrcr}
\hline \hline
\noalign{\smallskip}
ID        & freq. & ampl. & phase   \\
\hline
          & [$\rm d^{-1}$] & [mag] & [rad]   \\
\noalign{\smallskip}
\hline
\noalign{\smallskip}
${f_0}$       &  1.385383    & 0.12961    &   2.944 \\
${2f_0}$      &  2.770767    & 0.05177    &   2.608 \\
${3f_0}$      &  4.156150    & 0.02263    &   2.573 \\
${4f_0}$      &  5.541534    & 0.00878    &   2.853 \\
${5f_0}$      &  6.926917    & 0.00592    &   3.223 \\
${6f_0}$      &  8.312300    & 0.00524    &   3.217 \\
${7f_0}$      &  9.697684    & 0.00423    &   3.031 \\
${8f_0}$      & 11.083067    & 0.00330    &   2.748 \\
${9f_0}$      & 12.468451    & 0.00242    &   2.489 \\
${10f_0}$     & 13.853834    & 0.00166    &   2.207 \\
${11f_0}$     & 15.239217    & 0.00139    &   1.930 \\
${12f_0}$     & 16.624601    & 0.00105    &   1.759 \\
${13f_0}$     & 18.009984    & 0.00070    &   1.513 \\
${14f_0}$     & 19.395368    & 0.00055    &   1.244 \\
${15f_0}$     & 20.780751    & 0.00042    &   0.810 \\
${16f_0}$     & 22.166134    & 0.00030    &   0.915 \\
${17f_0}$     & 23.551518    & 0.00033    &   0.438 \\
${18f_0}$     & 24.936901    & 0.00019    &   0.808 \\
${19f_0}$     & 26.322285    & 0.00021    &   6.203 \\
\noalign{\smallskip}
\hline
\end{tabular}
\end{flushleft}
\end{table}
\end{center}


\section{Frequencies of 0100881648}\label{app648}

In Table.~\ref{table_648} we present the result of the Fourier analysis of the blended CoRoT Blazhko RRab star, {\tt 648}. The following epoch for maxima was found:  

\begin{equation}
$$2454236.9472 \thinspace {\rm HJD} + 0\dotd6071863(48) \cdot {\rm E} $$
\end{equation}

\begin{center}
\begin{table}
\begin{flushleft}
\caption[]{Frequency table of the blended Blazhko RRab star, {\tt 648}.}
\label{table_648}
\begin{tabular}{rrcr}
\hline \hline
\noalign{\smallskip}
ID        & freq. & ampl. & phase   \\
\hline
          & [$\rm d^{-1}$] & [mag] & [rad]   \\
\noalign{\smallskip}
\hline
\noalign{\smallskip}
${f_0}$       &  1.646941  & 0.03280   &  0.911  \\
${2f_0}$      &  3.293434  & 0.01208   &  6.160  \\
${3f_0}$      &  4.940375  & 0.00504   &  3.919  \\
${4f_0}$      &  6.586869  & 0.00163   &  3.594  \\
${5f_0}$      &  8.233809  & 0.00094   &  2.063  \\
${6f_0}$      &  9.880303  & 0.00084   &  1.410  \\
${7f_0}$      & 11.527244  & 0.00067   &  5.489  \\
${8f_0}$      & 13.173737  & 0.00048   &  4.474  \\
${9f_0}$      & 14.821125  & 0.00032   &  0.821  \\
${10f_0}$     & 16.467619  & 0.00020   &  6.190  \\

${f_0 - f_{\rm m}}$ &  1.633088  & 0.00065   &  0.490  \\
${f_0 + f_{\rm m}}$ &  1.663030  & 0.00437   &  0.759  \\
${2f_0 - f_{\rm m}}$&  3.275110  & 0.00033   &  3.656  \\
${2f_0 + f_{\rm m}}$&  3.309971  & 0.00215   &  3.767  \\
${3f_0 + f_{\rm m}}$&  4.956911  & 0.00168   &  1.290  \\
${4f_0 + f_{\rm m}}$&  6.603405  & 0.00062   &  0.828  \\
${5f_0 + f_{\rm m}}$&  8.250346  & 0.00021   &  5.675  \\
${6f_0 + f_{\rm m}}$&  9.898180  & 0.00029   &  0.681  \\
${7f_0 + f_{\rm m}}$& 11.543333  & 0.00028   &  4.654  \\
${8f_0 + f_{\rm m}}$& 13.190274  & 0.00027   &  2.231  \\
${9f_0 + f_{\rm m}}$& 14.836320  & 0.00019   &  2.538  \\
${10f_0 + f_{\rm m}}$&16.482814  & 0.00014   &  1.552  \\

\noalign{\smallskip}
\hline
\end{tabular}
\end{flushleft}
\end{table}
\end{center}

This is a heavily blended Blazhko RRab star. The spectrum shows the frequency corresponding 
to the fundamental mode pulsation (${f_0}$) and nine harmonics. In addition, the 
right components of the modulation triplets are seen prominently around the harmonics 
after pre-whitening. In a few cases the left side peaks are also detected. The remaining spectrum consists of the known orbital frequencies of CoRoT, their linear combinations with the sideral day, many peaks due to low-frequency variations in the frequency interval ${\rm 0.1-0.7\,d^{-1}}$, some residuals around the main frequency and around the low-order harmonics. The remaining grass is at the level of 73 $\mu$mag. The amplitude and frequency variation of the star due to the Blazhko-modulation was already presented in \citet{szabo2009}. 

\section{Frequencies of the blended Blazhko RRab star 0101370544}\label{app544}

Table.~\ref{table_544} lists the result of the Fourier analysis of the second blended CoRoT Blazhko RRab star, {\tt 544}. This object was observed in color mode of CoRoT, but we chose to present the co-added (white) light frequencies, because this data set is superior compared to the individual color observations. The following epoch for 
maxima was found:

\begin{equation}
$$2454237.1010 \thinspace {\rm HJD} + 0\dotd6050870(43) \cdot {\rm E} $$
\end{equation}

\begin{center}
\begin{table}
\begin{flushleft}
\caption[]{Frequency table of the white band light curve of the blended Blazhko RRab star, {\tt 544}.}
\label{table_544}
\begin{tabular}{rrcr}
\hline \hline
\noalign{\smallskip}
ID        & freq. & ampl. & phase   \\
\hline
          & [$\rm d^{-1}$] & [mag] & [rad]   \\
\noalign{\smallskip}
\hline
\noalign{\smallskip}
${f_0}$       &  1.652655  & 0.00930   &  5.137  \\
${2f_0}$      &  3.305310  & 0.00323   &  0.437  \\
${3f_0}$      &  4.957965  & 0.00162   &  2.316  \\
${4f_0}$      &  6.610620  & 0.00051   &  4.400  \\
${5f_0}$      &  8.263275  & 0.00016   &  0.517  \\
${6f_0}$      &  9.915930  & 0.00017   &  3.025  \\
${7f_0}$      & 11.568585  & 0.00013   &  5.008  \\
${8f_0}$      & 13.221240  & 0.00013   &  0.456  \\
${9f_0}$      & 14.873895  & 0.00010   &  2.244  \\
${10f_0}$     & 16.526550  & 0.00007   &  4.090  \\
${11f_0}$     & 18.179205  & 0.00007   &  5.635  \\
 
${f_0+f_{\rm m}}$    &    1.691938 &   0.00163 &  0.917 \\
${2f_0+f{\rm _m}}$   &    3.344600 &   0.00083 &  1.996 \\
${3f_0+f{\rm _m}}$   &    4.997249 &   0.00067 &  3.691 \\
${4f_0+f{\rm _m}}$   &    6.649602 &   0.00033 &  5.857 \\
${5f_0+f{\rm _m}}$   &    8.302697 &   0.00014 &  1.593 \\
${6f_0+f{\rm _m}}$   &    9.955316 &   0.00009 &  4.047 \\
${7f_0+f{\rm _m}}$   &   11.608022 &   0.00013 &  6.247 \\
${8f_0+f{\rm _m}}$   &   13.260818 &   0.00011 &  1.749 \\
${9f_0+f{\rm _m}}$   &   14.913797 &   0.00010 &  3.478 \\
${10f_0+f{\rm _m}}$  &   16.566175 &   0.00008 &  5.297 \\
${11f_0+f{\rm _m}}$  &   18.219099 &   0.00006 &  0.945 \\

${f_0-f{\rm _m}}$    &    1.613977  &   0.00028  &  2.873 \\
${2f_0-f{\rm _m}}$   &    3.266438  &   0.00015  &  4.474 \\
${3f_0-f{\rm _m}}$   &    4.919052  &   0.00015  &  5.969 \\
${4f_0-f{\rm _m}}$   &    6.571116  &   0.00013  &  1.999 \\
${5f_0-f{\rm _m}}$   &    8.224252  &   0.00007  &  3.496 \\
${6f_0-f{\rm _m}}$   &    9.877031  &   0.00007  &  5.558 \\
${7f_0-f{\rm _m}}$   &   11.529681  &   0.00004  &  0.594 \\
${8f_0-f{\rm _m}}$   &   13.182138  &   0.00004  &  3.375 \\
${10f_0-f{\rm _m}}$  &   16.487827  &   0.00004  &  0.997 \\
${f'}$         &    2.389287  &   0.00015  &  4.079 \\
\noalign{\smallskip}
\hline
\end{tabular}
\end{flushleft}
\end{table}
\end{center}

Despite the heavy blending, the spectrum shows the frequency corresponding to the 
fundamental mode pulsation (${f_0}$), and ten harmonics. In addition, the  
triplet components of the modulation found around ${f_0}$ are revealed 
around most of the harmonics after pre-whitening. The right side-lobes (${k*f_0+f_m}$) are present with higher amplitudes than their left counterparts. The remaining spectrum consists of the 
known orbital frequencies of CoRoT, and their linear combinations with the sideral 
day, many peaks due to low-frequency variations in the frequency interval ${\rm 0.2-0.8\,d^{-1}}$, some residuals around the main frequency and low-order harmonics, and a few remaining peaks close to or below the significance level between ${f_0}$ and 
${2f_0}$. The remaining grass is at the level of 28 $\mu$mag. The amplitude and 
frequency variation due to the Blazhko-modulation was presented in \citet{szabo2009}.

\section{Frequencies of the CoRoT RRc star 0105036241}\label{app241}

\begin{center}
\begin{table}
\begin{flushleft}
\caption[]{Frequency table of the CoRoT RRc star, {\tt 241}.}
\label{tab241}
\begin{tabular}{rrcr}
\hline \hline
\noalign{\smallskip}
ID        & freq. & ampl. & phase   \\
\hline
          & [$\rm d^{-1}$] & [mag] & [rad]   \\
\noalign{\smallskip}
\hline
\noalign{\smallskip}
${f_1}$  &  2.68153  & 0.19610 &  0.337    \\
${2f_1}$ &  5.36271	 & 0.01186 &  4.704    \\
${3f_1}$ &	8.04458	 & 0.01409 &  2.349	 \\
${4f_1}$ & 10.72611	 & 0.01120 &  5.662	 \\
${5f_1}$ & 13.40729	 & 0.00728 &  2.976	 \\
${6f_1}$ & 16.08882	 & 0.00454 &  6.118	 \\
${7f_1}$ & 18.77035	 & 0.00288 &  2.850	 \\
${8f_1}$ & 21.45187	 & 0.00177 &  5.771	 \\
${9f_1}$ & 24.13375	 & 0.00105 &  1.973	 \\
${10f_1}$ & 26.81493 & 0.00065 &  5.299	 \\
${11f_1}$ & 29.49611 & 0.00044 &  2.292	 \\
${12f_1}$ & 32.17764 & 0.00027 &  5.439	 \\
${f'}$   &  4.37783 & 0.00332 &  2.953    \\

${f' - f_1}$  &  1.69630 & 0.00052 & 3.322  \\
${f' + f_1}$  &  7.05901 & 0.00304 & 5.886  \\
${2f_1 + f'}$ &  9.74054 & 0.00147 & 1.980  \\
${2f_1 - f'}$ &  0.98488 & 0.00039 & 5.187  \\
${3f_1 + f'}$ & 12.42241 & 0.00035 & 4.613  \\

${f_1 - f_{\rm m}}$  &  2.67601  & 0.00415   & 0.369   \\
${f_1 + f_{\rm m}}$  &  2.69084  & 0.00151   & 5.743   \\
${f_1 - 2f_{\rm m}}$ &  2.67015  & 0.00147   & 4.535   \\
${f_1 + 2f_{\rm m}}$ &  2.69946  & 0.00055   & 5.690   \\
${f_1 - 3f_{\rm m}}$ &  2.66256  & 0.00026   & 3.728   \\

${2f_1 + f_{\rm m}}$ &  5.35823  & 0.00060   & 2.256   \\
${2f_1 - f_{\rm m}}$ &  5.36823  & 0.00036   & 4.349   \\
${3f_1 + f_{\rm m}}$ &  8.03906  & 0.00054   & 1.887   \\
${3f_1 - f_{\rm m}}$ &  8.05286  & 0.00025   & 2.841   \\
${4f_1 + f_{\rm m}}$ & 10.72025  & 0.00063   & 5.756   \\
${4f_1 - f_{\rm m}}$ & 10.73232  & 0.00025   & 1.842   \\
${5f_1 - f_{\rm m}}$ & 13.40143  & 0.00043   & 2.244   \\
${6f_1 - f_{\rm m}}$ & 16.08296  & 0.00029   & 5.635   \\
${7f_1 - f_{\rm m}}$ & 18.76414  & 0.00026   & 2.625   \\
${8f_1 - f_{\rm m}}$ & 20.44610  & 0.00032   & 5.533   \\

${f' - f_{\rm m}}$  &  4.37266  & 0.00035   & 1.050   \\
${f' + f_{\rm m}}$  &  4.38472  & 0.00186   & 2.572   \\

${f_1 - f_{\rm b}}$  &  2.64290  & 0.00044   & 1.131   \\
${f_1 - 2f_{\rm b}}$  & 2.60014  & 0.00038   & 3.784   \\
${2f_1 - f_{\rm b}}$  & 9.69881  & 0.00039   & 4.512   \\

${f' - f_{\rm b}}$  & 4.33852    & 0.00172   & 1.475   \\
${f' + f_{\rm b}}$  & 4.41852    & 0.00026   & 2.532   \\
${f' - 2f_{\rm b}}$  & 4.30265   & 0.00074   & 5.346   \\
${f' - 3f_{\rm b}}$  & 4.25989   & 0.00024   & 5.469   \\

${f' + f_1 + f_{\rm m}}$  &  7.06625  & 0.00159   & 5.015  \\
${f' + f_1 - f_{\rm m}}$  &  7.05418  & 0.00062   & 3.355   \\

${2f_1 + f' + f_{\rm m}}$  & 9.74743  & 0.00084   & 1.622   \\

${f' + f_1 - 2f_{\rm b}}$  & 6.98383 & 0.00036   & 2.248   \\
${f' + f_1 + 2f_{\rm b}}$  & 7.14379 & 0.00043   & 3.320   \\

${f' - f_{\rm b} - f_{\rm m}}$  & 4.33369  & 0.00069   &  2.981  \\
${f' + f_{\rm b} - f_{\rm m}}$  & 4.34472  & 0.00043   &  4.744  \\
${f' + f_{\rm b} + f_{\rm m}}$  & 4.42611  & 0.00053   &  4.089  \\
${f' + 2f_{\rm b} + f_{\rm m}}$ & 4.46197  & 0.00098   &  5.216  \\

\noalign{\smallskip}
\hline
\end{tabular}
\end{flushleft}
\end{table}
\end{center}

Besides the dominant first overtone frequency ($f_1 = {\rm 2.68153\,d^{-1}}$) and its harmonics, we see ${f'}$ in the spectrum with a characteristic 0.613 frequency ratio with the first overtone radial pulsation (Table~\ref{tab241}). In addition, a bunch of frequencies were found in the [0.5;1.5]\,${\rm d^{-1}}$ frequency range. Upon inspection of the data we found that their origin can be traced back to two remaining discontinuities. Therefore these portions of the data set CJD [3046.0-3048.0] and [3156.5-3157.5] were removed. In the following we analyze the remaining data set. 

We found high left side peaks around the main frequency and its harmonics. If we suppose that $f_m={\rm 0.00585\,d^{-1}}$ is a modulation frequency, the period of the modulation would be longer than the data set. According to that this star may show a long-period Blazhko-modulation, but more data would be necessary to confirm this finding. 
 Another set of modulation-like frequency difference appears in the data set (Fig.~\ref{241_frsp}, Table~\ref{tab241}). We denote the corresponding frequency ${f_b}$. Neither ${f_m}$, nor ${f_b}$ can be found in the frequency spectrum, they appear only through combination frequencies. Even combination frequencies involving both ${f_m}$ and ${f_b}$ can be identified. We note here that as we demonstrated in Sec.~\ref{rrc} ${f'}$ has temporal amplitude variation, 
and this is the most probable culprit to cause the appearance of the ${f_b}$ modulation frequencies. The strongest argument favoring this scenario is that ${f_b}$ appears only close to and in combination with  ${f'}$, and is not seen around the main pulsation frequency, ${f_1}$. In addition, frequencies associated with the orbital period of the satellite and its daily aliases are seen at $f={\rm 13.967924, 14.974027, 12.969585\,d^{-1}}$ as usual in CoRoT data. We omitted these peaks from Table~\ref{tab241}. The following epoch for maxima was found:  

\begin{equation}
$$2454572.6300(7) \thinspace {\rm HJD} + 0\dotd3729214(2) \cdot {\rm E} $$
\end{equation}

or taking into account a gradual period change:

\begin{equation}
$$2454572.6300(7) \thinspace {\rm HJD} + 0\dotd3729214(2) \cdot {\rm E} - 1\dotd4(7)\cdot 10^{-7} \cdot {\rm E^2}$$
\end{equation}

\section{Frequencies of the CoRoT RRc star 0105735652}\label{app652}

We detect the main frequency $f_1 = {\rm 3.58218\,d^{-1}}$ and its harmonics,  but also many other frequencies with lower amplitude. Among them we found a highly significant frequency at $f'= {\rm 5.82484\,d^{-1}}$ with several peaks around it, then additional peaks around ${f_1}$. The frequencies are available in Table~\ref{652tab}. The following epoch for maxima and period were found:

\begin{equation}
$$2454572.7323\thinspace {\rm HJD} + 0\dotd2791596(38) \cdot {\rm E} $$
\end{equation}

After pre-whitening with the frequencies enumerated in Table~\ref{652tab}, a dense forest of fre\-quen\-cies remains around ${f_1}$. We also see similar residual power around the second and third harmonics, frequencies around ${\rm 5.9, 9.4, 16.5, 20.1\,d^{-1}}$, and frequencies related to the orbital frequency of CoRoT.

The large number of side frequencies seen around frequencies ${f'}$ and
${\rm 9.470874\,d^{-1}}$ may be the result of their non-stationary nature (modulation). We gave an example in \citet{szabo2010} where the frequency forest found around the the half-integer frequencies was modeled and explained by the varying amplitude of these frequencies. We see a very similar situation here.

\begin{center}
\begin{table}
\begin{flushleft}
\caption[]{Frequency table of the CoRoT RRc star, {\tt 652}.}
\label{652tab}
\begin{tabular}{rrcr}
\hline \hline
\noalign{\smallskip}
ID        & freq. & ampl. & phase   \\
\hline
          & [$\rm d^{-1}$] & [mag] & [rad]   \\
\noalign{\smallskip}
\hline
\noalign{\smallskip}
${f_1}$        &  3.58218  & 0.20403    &  0.548    \\
${2f_1}$       &  7.16435  & 0.03209    &  4.309    \\
${3f_1}$       & 10.74653  & 0.01666    &  1.496    \\
${4f_1}$       & 14.32871  & 0.01356    &  5.530    \\
${5f_1}$       & 17.91088  & 0.00992    &  2.685    \\
${6f_1}$       & 21.49306  & 0.00671    &  5.825    \\
${7f_1}$       & 25.07541  & 0.00420    &  2.693    \\
${8f_1}$       & 28.65754  & 0.00240    &  5.670    \\
${9f_1}$       & 32.23960  & 0.00145    &  2.291    \\
${10f_1}$      & 35.82177  & 0.00074    &  5.136    \\
${11f_1}$      & 39.40394  & 0.00054    &  1.507    \\
${12f_1}$      & 42.98612  & 0.00039    &  4.327    \\
${13f_1}$      & 46.56830  & 0.00033    &  0.575    \\
${14f_1}$      & 50.15047  & 0.00023    &  0.712    \\

${f'}$       &    5.82484  &    0.00216  &    5.731   \\
${f'+f_{\rm m}}$   &    5.88925  &    0.00144  &    5.090   \\
${f'-f_{\rm m}}$   &    5.76587  &    0.00084  &    3.699   \\
${f'+2f_{\rm m}}$  &    6.00000  &    0.00073  &    2.904   \\ 
${f'-2f_{\rm m}}$  &    5.71282  &    0.00035  &    1.072   \\
${}$   &    5.93520  &    0.00043  &    6.179   \\
${}$   &    5.87227  &    0.00054  &    2.259   \\
${}$   &    5.77659  &    0.00075  &    0.001   \\
  &  9.47087    &   0.00089   &  2.479     \\
  &  9.40747    &   0.00064   &  0.961     \\
  &  9.58200    &   0.00048   &  5.982     \\
  &  9.34758    &   0.00037   &  0.292     \\
  &  9.29643    &   0.00023   &  3.506     \\ 
  &  9.62932    &   0.00017   &  1.606     \\
${f_1+f_{\rm m}}$ &  3.64669    &   0.00087   &  2.133     \\
            &  2.66379    &   0.00064   &  3.466     \\

\noalign{\smallskip}
\hline
\end{tabular}
\end{flushleft}
\end{table}
\end{center}

\end{appendix}
\end{document}